%% file: main.tex
\documentclass[twocolumn]{aastex63}

\usepackage{rotating}
\usepackage{acronym}
\usepackage{xspace}
\usepackage{multirow}
\usepackage{mathtools}
\usepackage{amsmath}
\usepackage{hyperref}
\usepackage{units}

\usepackage{xcolor}

\graphicspath{{./}{figures/}}

\input{commands}
\input{values}
\input{acronyms}

\newcommand{\KICP}{\affiliation{Kavli Institute for Cosmological Physics, The University of Chicago, 5640 South Ellis Avenue, Chicago, IL 60637, USA}}
\newcommand{\EFI}{\affiliation{Enrico Fermi Institute, The University of Chicago, 933 East 56th Street, Chicago, IL 60637, USA}}
\newcommand{\Geneva}{\affiliation{Département d’Astronomie, Université de Genève, Chemin Pegasi 51, CH-1290 Versoix, Switzerland}}

\shorttitle{Suspicious Siblings}
\shortauthors{Zevin \& Bavera 2022}

\begin{document}

%\title{Suspicious Siblings: Can Isolated Binary Evolution Produce\\
%Asymmetric Binary Black Holes with Spinning Primaries?}
\title{Suspicious Siblings: The Distribution of Mass and Spin across\\ Component Black Holes in Isolated Binary Evolution}

\author[0000-0002-0147-0835]{Michael Zevin}\email{michaelzevin@uchicago.edu}\thanks{NASA Hubble Fellow}
\KICP \EFI
\author[0000-0002-3439-0321]{Simone S. Bavera}
\Geneva

\begin{abstract}
The LIGO and Virgo gravitational-wave detectors have uncovered binary black hole systems with definitively nonzero spins, as well as systems with significant spin residing in the more massive black hole of the pair. 
We investigate the ability of isolated binary evolution in forming such highly spinning, asymmetric-mass systems through both accretion onto the first-born black hole and tidal spin-up of the second-born black hole using a rapid population synthesis approach with detailed considerations of spin-up through tidal interactions. 
Even with the most optimistic assumptions regarding the efficiency at which an accreting star receives material from a donor, we find that it is difficult to form systems with significant mass asymmetry and moderate or high spins in the primary black hole component. 
Assuming efficient angular momentum transport within massive stars and Eddington-limited accretion onto black holes, we find that $< \UpperLimitQbGreaterTwoSpinGreaterZeropTwo$ of systems in the underlying binary black hole population have a primary black hole spin greater than $0.2$ and a mass asymmetry of greater than 2:1 in our most optimistic models, with most models finding that this criteria is only met in $\sim \ApproxQbGreaterTwoSpinGreaterZeropTwo$ of systems. 
The production of systems with significant mass asymmetries and spin in the primary black hole component is thus an unlikely byproduct of isolated evolution unless highly super-Eddington accretion is invoked or angular momentum transport in massive stars is less efficient than typically assumed.

\end{abstract}

%\keywords{space!}

\section{Introduction}\label{sec:intro}

Prior to the direct observation of compact binary coelescences via \acp{GW}, the expected birth properties of \acp{BH} relied primarily on highly uncertain predictions from stellar population modeling and a limited number of indirect observations. 
The past half-decade has brought an observational sample of \ac{BBH} mergers that has not only provided invaluable insights into the birth properties of \acp{BH}, but also unveiled a number of unexpected systems that are in tension with the theoretical expectations of \ac{BH} formation channels. 
Trends and features of the population as a whole are also becoming apparent, such as structure beyond a sharp cutoff in the \ac{BH} mass distribution~\citep[e.g.,][]{GWTC2_pops} and potential correlations between intrinsic parameters of \ac{BBH} systems~\citep[e.g.,][]{Callister2021a}. 

Certain \ac{BBH} systems observed by the \ac{LVK} Collaboration show unexpected features not just in their masses or spins individually, but their distribution of mass and spin across the two binary components. 
For example, GW190412 was the first \ac{BBH} to have been measured with definitively asymmetric component masses ($\q \simeq 0.28$, where $\q \equiv m_2/m_1$ with $m_2 \leq m_1$) and nonzero spin ($\chieff > 0.14$ at 90\% credibility, where $\chieff \equiv (m_1 a_{1z} + m_2 a_{2z}) / (m_1+m_2)$ and $a_{z}$ is the projection of the \ac{BH} dimensionless spin aligned with the \ac{AM};~\citealt{GW190412}). 
Because of the mass asymmetry in the signal, the component spin of the more massive primary \ac{BH} was able to be disentangled from the leading-order spin term \chieff, with the more massive \ac{BH} having a dimensionless spin of $a_1 > 0.22$ at 95\% credibility. 
More \ac{BBH} systems with spinning component \acp{BH} have been identified in recent \ac{GW} catalogs~\citep[e.g.,][]{GWTC2.1,GWTC3,Nitz2021a,Olsen2022}. 

The combination of \ac{BH} masses and spins across both binary components holds important clues about both the formation environment of \ac{BBH} systems as well as the intricate physical processes occurring within and between their progenitor stars. 
In massive stars, \ac{AM} transport between the stellar core and envelope significantly impacts the expected spin of the resultant \ac{BH}. 
Models of \ac{AM} transport in massive stars, such as the Taylor--Spruit magnetic dynamo~\citep{Spruit2002}, indicate \ac{AM} transport to be highly efficient~\citep{Heger2005,Fuller2019a}. 
Under this paradigm, \ac{AM} would thus be transported to the outer layers of the star during its giant phase and lost due to wind mass loss or \ac{RLO} \ac{MT} onto a binary companion, resulting in a slowly spinning core and upon collapse a \ac{BH} with a dimensionless spin extremely close to zero (\citealt{Qin2018,Fuller2019b}; though see \citealt{Belczynski2020} for variations in this mechanism that can lead to slightly larger spins of $a \simeq 0.1$). 
The low effective spins in most \acp{BBH} also observationally hint at efficient \ac{AM} transport in their progenitors~\citep{GWTC2_pops,GWTC3_pops,Miller2020,Zevin2021}, although certain systems are beginning to challenge the universality of quasi-isolated \acp{BH} having low spins at birth~\citep[e.g.,][]{Zevin2020a,Qin2022a}. 

Even in the efficient \ac{AM} transport paradigm, binary processes such as tides can induce high spins on stellar cores that can be preserved in their remnants. 
This can be accomplished in three general ways: (a) chemically homogeneous evolution, where two massive stars with near-equal masses in a tight binary system at \ac{ZAMS} tidally interact on the main sequence, which induces strong rotational mixing that prevents expansion and \ac{AM} loss, resulting in two massive, spinning \acp{BH}~\citep{DeMink2016,Mandel2016b,Marchant2016}; (b) Case A \ac{MT} (i.e., while the donor is on the main sequence) for binaries on tight orbits of about a day or less, where the donor and accretor are tidally locked during \ac{MT} and the envelope of the donor is stripped, leading the donor to never expand during its evolution into a \ac{WR} star~\citep{Qin2019}; (c) tidal spin-up of a \ac{WR} star (i.e., a naked helium star), either by an already-formed compact object following a stable or unstable \ac{MT} event that hardens the binary to subday orbital periods~\citep{Detmers2008,Kushnir2017,Hotokezaka2017,Zaldarriaga2018,Qin2018,Bavera2020,Olejak2021a,Steinle2021,Fuller2022} or following a double-core \ac{CE} scenario where two helium cores of supergiant stars are enveloped in the envelope of one or both of the supergiants and hardened through a \ac{CE} phase to a point where the two cores can tidally interact~\citep{Brown1995,Dewi2006,Hotokezaka2017,Neijssel2019,Olejak2021a}. 
Extremely metal-poor stars born in the early universe may also be able to retain a substantial hydrogen envelope and collapse into high-spinning \acp{BH}, though their contribution to the local merger population is likely small~\citep{Cruz-Osorio2021,Tanikawa2022b}. 
Following formation, a \ac{BH} can also gain \ac{AM} through accretion, though any appreciable spin-up would require the accretion to either be highly super-Eddington or transpire over timescales far longer than the evolutionary timescales of massive stars~\citep{VanSon2020,Bavera2021,Qin2022}. 

Systems with unequal masses and spinning primaries provide a challenge to the isolated binary evolution scenario. 
\acp{BBH} that result from chemically homogeneous evolution strongly favor near-equal-mass systems~\citep{Marchant2016,Mandel2016b,duBuisson2020}. 
The Case A \ac{MT} scenario has been invoked for explaining the large inferred spins of \acp{BH} in high-mass X-ray binaries~\citep{Qin2019}, though binary modeling finds that the high-mass X-ray binaries in the Milky Way are unlikely to form merging compact binary systems~\citep{Belczynski2011,Belczynski2012,Neijssel2021}. 
The \ac{BH}--\ac{WR} tidal spin-up scenario is predicted to be common for post-\ac{CE} binaries~\citep{Bavera2021}, though the efficiency of this pathway is highly dependent on \ac{CE} ejection efficiency and can only impart spin on the second-born \ac{BH} progenitor. 
Though the tidal spin-up of two \ac{WR} stars following a double-core \ac{CE} may lead to significant spins in both \acp{BH}~\citep{Olejak2021a}, it typically leads to near-equal-mass mergers and operates at a much lower rate than the \ac{BH}--\ac{WR} tidal spin-up scenario~\citep{Neijssel2019}. 
Lastly, although spinning up the first-born \ac{BH} through accretion can be accomplished via highly super-Eddington \ac{MT}, if accretion is pushed too high, the rate of mergers from this channel can drop by orders of magnitude~\citep{Bavera2021}. 

The \ac{BH}--\ac{WR} tidal spin-up scenario described above may be a valid explanation for the observed spins of primary \acp{BH} if the second-born \ac{BH} can oftentimes be more massive than the first-born \ac{BH}. 
In isolation, a more massive star at \ac{ZAMS} will typically lead to a more massive remnant (though see, e.g., \citealt{Patton2022}). 
However, binary interactions throughout the coevolution of the two stars can alter this picture. 
In particular, since the more massive star at \ac{ZAMS} will almost always evolve off the main sequence first, it will be the first to overflow its Roche lobe and transfer material onto its companion. 
The fraction of transferred mass deposited onto the companion depends on an uncertain \ac{MT} accretion efficiency~\citep{Bouffanais2021a}, and can potentially lead to a \ac{MRR}, where the star that was originally less massive at \ac{ZAMS} has its mass inflated from the accreted material and leads to a more massive remnant~\citep[e.g.,][]{Olejak2021a,Broekgaarden2022,Hu2022}. 
The originally more massive star will still proceed through the remainder of its evolution quicker, and create the first compact object in the binary. 
Then, the originally less massive star can be stripped of its envelope and harden its orbit with the compact object companion during the second \ac{MT} episode, and be tidally spun up as a \ac{WR} star. 

In this paper, we explore the viability of the \ac{MRR}/tidal interaction and \ac{BH} accretion spin-up scenarios for forming the asymmetric-mass, spinning-primary \ac{BBH} systems observed by the \ac{LVK}. 
In Section~\ref{sec:models} we overview our population models and the determination of \ac{BH} spins through tidal spin-up and accretion. 
Analysis of these models, with a particular focus on mass ratios, \ac{MRR}, and \ac{BH} spins, is in Section~\ref{sec:results}. 
We contextualize our results with \ac{GW} events and population properties in Section~\ref{sec:gw_results}, and discuss the broader implications and caveats in Section~\ref{sec:discussion}. 
Throughout this work, we assume solar metallicity of ${Z_\odot=0.017}$~\citep{Grevesse1998} and Planck 2018 cosmological parameters~\citep{PlanckCollaboration2018}.

\section{Population Models}\label{sec:models}

We use the open-source binary population synthesis code \texttt{COSMIC}\footnote{\url{cosmic-popsynth.github.io}, Version 3.4}~\citep{Breivik2020} for modeling the populations of \ac{BBH} mergers. 
\texttt{COSMIC} is based on single-star evolutionary tracks from the \texttt{SSE} code~\citep{Hurley2000} and the binary star implementations of \texttt{BSE}~\citep{Hurley2002}. 
A large number of updates have been made to the physical prescriptions used for initial conditions, winds, the onset of \ac{MT} and system evolution during \ac{RLO}, remnant formation, and natal kicks. 
Rather than simulating a fixed number of systems with specific physical assumptions at a given metallicity, \texttt{COSMIC} samples the systems until user-specified convergence criteria have been reached in the population.\footnote{We set the maximum number of simulated binaries for all runs to $10^8$, and thus the specified convergence criteria are sometimes not reached, especially for high-metallicity populations where the \ac{BH} formation efficiency is extremely low.} 
Although we highlight physical assumptions pertinent to this work in the following section, we refer the reader to \cite{Breivik2020} for further details. 

\subsection{Physical Assumptions}\label{subsec:assumptions}

A large number of physical uncertainties embed models of massive-star binary evolution, which have significant impacts on the population properties of their \ac{BBH} remnants~\citep[e.g.,][]{Giacobbo2018a,Giacobbo2018b,Kruckow2018,Bouffanais2019,Stevenson2019,VanSon2020,Bavera2021,Belczynski2022a,Santoliquido2021,Zevin2021}. 
Although recent studies have become increasingly thorough in their coverage of this highly uncertain parameter space (see \citealt{Broekgaarden2021} for a recent overview), it is computationally expensive and can make interpretation difficult. 
We instead narrow our parameter space coverage to uncertainties that have the strongest effect on \ac{MRR} and \ac{BH} spin-up. 

The main parameter varied between models is the accretion efficiency: 
\begin{equation}
    \facc = \frac{\dot{M}_\mathrm{acc}}{\dot{M}_\mathrm{don}} \in [0,1]
\end{equation}
where $\dot{M}_\mathrm{don}$ is the mass-loss rate of the donor during \ac{RLO} and $\dot{M}_\mathrm{acc}$ is the mass accepted by the accretor. 
The additional mass $(1-\facc)\dot{M}_\mathrm{don}$ is lost from the system with an \ac{AM} as if it were a wind from the accretor. 
We simulate five variations for \facc: $[0.0, 0.25, 0.5, 0.75, 1.0]$ where $\facc = 0.0$ means that the accretor accepts no mass from the donor, and $\facc = 1.0$ means that the accretor accepts all of the mass transferred by the donor. 
Although the timescales pertinent to the response of the envelope to \ac{MT} may provide a more realistic description of the amount of material a star can accrete in a given amount of time, this simple parameterization can capture the extreme limits of accretion efficiency, which has an important impact on possible \ac{MRR} during stable \ac{MT} and subsequent tidal spin-up of the second-born \ac{BH}. 
This parameter also impacts the compact binary merger rates and mass spectra~\citep[see, e.g.,][]{Kruckow2018}, and has the promise of being constrained with \ac{GW} data~\citep{Bouffanais2021a}. 
The mass-loss rate from the donor during \ac{RLO}, $\dot{M}_\mathrm{don}$, follows the prescription of \cite{Hurley2002}. 

In addition, we simulate two separate assumptions for the amount at which \acp{BH} can accrete material above the Eddington rate, which impacts the accretion-induced spin-up of the first-born \ac{BH}: 
\begin{equation}
    \dot{M}_\mathrm{BH}^\mathrm{max} \simeq \gammaEdd \times R_\mathrm{s} \times 2.08 \times 10^{-3}~\Msun~\mathrm{yr}^{-1}
\end{equation}
where $R_\mathrm{s}$ is the Schwarzschild radius of the \ac{BH} in units of solar radii and $\gammaEdd$ is a multiplicative factor that allows for super-Eddington accretion (i.e., $\gammaEdd = 1$ would limit accretion onto a \ac{BH} to the Eddington rate). 
The maximum rate at which a \ac{BH} can accrete is thus given by
\begin{equation}
    \dot{M}_\mathrm{BH} = \mathrm{MIN}(\facc \dot{M}_\mathrm{don}, \dot{M}_\mathrm{BH}^\mathrm{max})
\end{equation}
Note that we allow \facc to alter the accretion efficiency of compact objects as well as (nondegenerate) stars in our models, and thus at $\facc=0$, \acp{BH} will not gain mass through accretion. 
We choose two values for $\gammaEdd$: $[1, 10^5]$. 
The highly super-Eddington parameterization of $\gammaEdd = 10^5$ is extreme but chosen because it can lead to a significant number of \acp{BH} to be spun up through stable \ac{MT} onto the \ac{BH}, whereas lower values for $\gammaEdd$ are much less capable~\citep[see, e.g.,][]{Bavera2021}. 

The final parameter variations we explore govern the evolution through a \ac{CE} phase. 
These affect both the survival of systems that evolve through a \ac{CE} phase and post-\ac{CE} orbital separations, thereby impacting the ability of the progenitors of the second-born \ac{BH} to tidally spin up. 
We model three variations for the efficiency at which orbital energy of the inspiraling binary is transmitted into the energy needed to eject the envelope, assuming the $\alpha - \lambda$ formalism for energetics during \ac{CE} evolution~\citep{Webbink1984}: $\alphaCE = [0.5, 1.0, 2.0]$. 
We use the variable prescription in \cite{Claeys2014} for determining the value of $\lambda$. 
Stellar type-dependent mass ratios that determine whether \ac{MT} proceeds stably or unstably follow \cite{Neijssel2019}, though we also run a subset of models that follow \cite{Belczynski2008} to investigate the impact this parameterization has on \ac{MRR} and tidal spin-up (see Appendix~\ref{app:qcrit}). 

All models sample the binary initial conditions independently following \cite{Kroupa2001} and \cite{Sana2012}, and assume a pessimistic \ac{CE} scenario in which Hertzprung gap stars that experience unstable \ac{RLO} merge in the \ac{CE} and do not form compact binaries~\citep[see][]{Belczynski2008}. 
We assume a fixed binary fraction of $f_\mathrm{bin} = 0.7$, which only affects the arbitrary normalization during the resampling described in Section~\ref{subsec:star_formation}. 
Wind mass loss follows \cite{Hurley2002} with updates for O and B stars~\citep{Vink2001} and metallicity-dependent \ac{WR} winds~\citep{Vink2005}. 
Remnant masses are determined assuming the delayed supernova engine prescription of \cite{Fryer2012} with neutrino mass loss as implemented in \cite{Zevin2020b}. 
A maximum neutron star mass (and therefore minimum \ac{BH} mass) of $3~\Msun$ is assumed. 
Supernova kicks, which can tilt the orbital plane and lead to \ac{BH} spins that are misaligned from the orbital \ac{AM}, are drawn from a Maxwellian
with a dispersion of $265~\mathrm{km\,s}^{-1}$~\citep{Hobbs2005} and are fallback modulated, leading to a suppression of kick strength as a function of \ac{BH} mass~\citep[e.g.,][]{Rodriguez2016}. 
\ac{BH} masses as a result of pulsational pair instability and pair instability supernovae are treated using fits to the results of \cite{Marchant2019}. 

Population models only retain systems that result in a \ac{BBH} that merges within a Hubble time. 
For each population model, we simulate 12 fixed metallicities spaced uniformly in log between $Z = [10^{-4}, 0.03]$. 
The permutation of all these model variations leads to $5\,[\facc] \times 2\,[\gammaEdd] \times 3\,[\alphaCE] \times 12\,[Z] = 360$ individual population sequences. 
Throughout this work, we present the results that combine the $12$ discrete metallicities for each model set according to the joint star formation and metallicity evolution of the universe (see Section~\ref{subsec:star_formation}).

\subsection{BH Spins}\label{subsec:spins}

Tracking the spin evolution of \ac{BH} progenitors is difficult in rapid population synthesis, and typically relies on simplified prescriptions. 
This is because rapid population synthesis codes lack information about the internal structure of a star, and thus cannot accurately model the \ac{AM} transport between binary components or back-reaction on the structure and evolution of each individual star. 
We assign spins to the first-born \ac{BH} ($a_\mathrm{1b}$) and second-born \ac{BH} ($a_\mathrm{2b}$) in our population during post-processing, assuming that \ac{AM} transport is highly efficient~\citep{Spruit2002,Fuller2019b} such that the first-born \ac{BH} can only attain nonzero spin through accretion after it becomes a compact object, and the second-born \ac{BH} can only attain a nonzero spin through tidal spin-up of its \ac{WR} progenitor. 
Thus, \acp{BH} born in quasi-isolation where their progenitors are not susceptible to tidal spin-up are assumed to be Schwarzschild \acp{BH}, though we examine the impact of relaxing this assumption on our results to account for other possible mechanisms for inducing spin in Section~\ref{sec:discussion}. 
We note that post-processing spins in this way leads to a slight inconsistency in our determination of inspiral times through \ac{GW} emission, as spins aligned with the orbital \ac{AM} will slightly expedite the eventual merger. 
However, this has a minor effect for the timescales considered, because spin effects enter at a higher post-Newtonian order and do not impact the inspiral rate significantly until the binary is close to merger~\citep{Damour2001}. 

The spin of the first-born \ac{BH} is calculated assuming stable \ac{MT} from a disk of material being accreted from the innermost stable circular orbit of the \ac{BH} as in \cite{Thorne1974}: 
\begin{equation}
    a_\mathrm{1b} = 
    \begin{cases}
        \sqrt{\frac{2}{3}} \frac{M_\mathrm{i}}{M_\mathrm{f}} \left[ 4 - \sqrt{\left(\frac{18 M_\mathrm{i}^2}{M_f^2} - 2\right)}\,\right] & \,\, 1 \leq \frac{M_\mathrm{f}}{M_\mathrm{i}} \leq \sqrt{6} \\
        1 & \,\, \frac{M_\mathrm{f}}{M_\mathrm{i}} > \sqrt{6}
    \end{cases}
\end{equation}
where $M_\mathrm{i}$ is the mass of the \ac{BH} prior to \ac{MT} and $M_\mathrm{f}$ is the mass of the \ac{BH} after \ac{MT} has ceased. 
For the determination of $a_\mathrm{1b}$, we only consider the mass gained from stable \ac{RLO} \ac{MT} and neglect any potential mass gain from wind accretion. 
We also assume that the \acp{BH} gain no mass when inspiraling through a \ac{CE}~\citep[though see, e.g.,][]{MacLeod2015a,Cruz-Osorio2020}. 
The spin of the second-born \ac{BH} is determined using the semianalytic fits from \cite{Bavera2021b}, which are based on the detailed spin evolution of \ac{BH}--\ac{WR} systems during tidal spin-up using the \texttt{MESA} simulations~\citep{Paxton2011,Paxton2013,Paxton2015,Paxton2018,Paxton2019} under the \texttt{POSYDON}\footnote{\url{posydon.org}} framework~\citep{Fragos2022}. 
\cite{Bavera2021b} found the spin of the second-born \ac{BH} to be well approximated by a quadratic function dependent on \ac{BH}--\ac{WR} log-orbital period $\log_{10}(p/\mathrm{day})$, which implicitly is dependent on the mass of the \ac{WR} star $M_\mathrm{WR}$ at helium depletion: 
\begin{equation}\label{eq:tidal_spinup}
    a_\mathrm{2b} = 
    \begin{cases}
        f^\alpha \log_{10}\left(\frac{p}{\mathrm{day}}\right)^2 + f^\beta \log_{10}\left(\frac{p}{\mathrm{day}}\right) & \,\, 0.1 \leq \frac{p}{\mathrm{day}} \leq 1 \\
        0 & \,\, \frac{p}{\mathrm{day}} > 1
    \end{cases}
\end{equation}
where $f^{(\alpha,\beta)} = -c_1^{(\alpha,\beta)} / [c_2^{(\alpha,\beta)} + \mathrm{exp}(-c_3^{(\alpha,\beta)} M_\mathrm{WR} / \Msun)]$ with coefficients $c_1^{(\alpha,\beta)}$, $c_2^{(\alpha,\beta)}$, and  $c_3^{(\alpha,\beta)}$ determined through least-square minimization; see \cite{Bavera2021b}. 
Since the grid of systems used for the fit only extended down to orbital periods of $0.1~\mathrm{day}$ (given by the physical limit of RLO at zero-age helium main sequence in the \texttt{MESA} simulations), for systems in our population with $p < 0.1~\mathrm{day}$ we assume their orbital periods are $0.1~\mathrm{day}$ when computing the fit. 
Across our population models, \PercentageSystemsPorbLessZeropOne of \ac{BH}--\ac{WR} systems have orbital periods of $p < 0.1~\mathrm{day}$ and are extrapolated in this manner. 

\begin{figure}[t]
\includegraphics[width=0.48\textwidth]{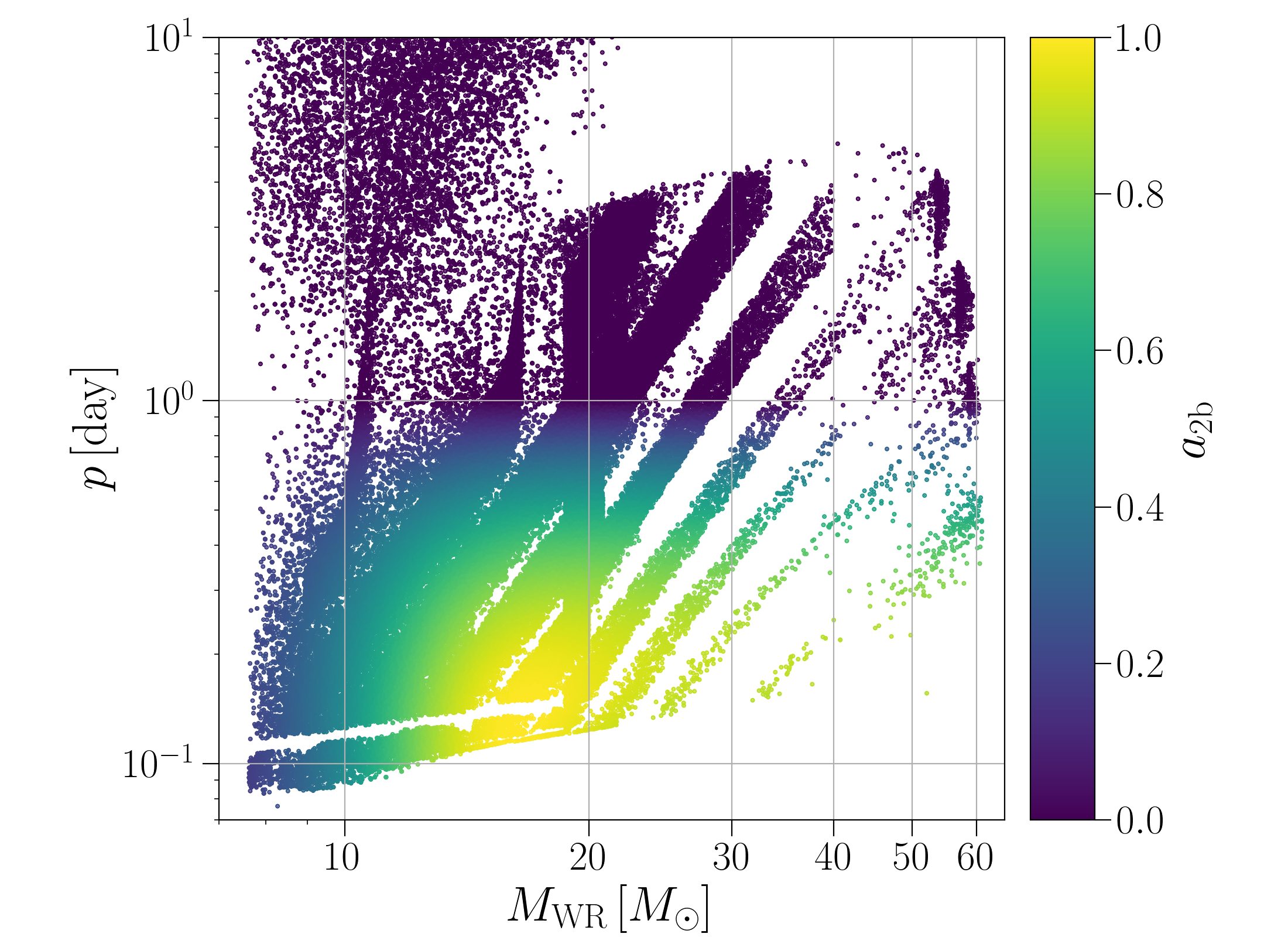}
\caption{Dimensionless spin magnitudes of the second-born \ac{BH} ($a_\mathrm{2b}$) determined using the fits from \cite{Bavera2021b} for a single population model as a function of the orbital period $p$ and the \ac{WR}-star mass at He-depletion $M_\mathrm{WR}$. 
This model assumes $\facc = 0.5$, $\gammaEdd = 1$, and $\alphaCE = 1$, with other model assumptions as specified in Section~\ref{subsec:assumptions}. 
Diagonal line-like features are the result of the set of discrete metallicities run for each population model, which are resampled according to the star-formation history and metallicity evolution of the universe as described in Section~\ref{subsec:star_formation}. 
}
\label{fig:spin_fits}
\end{figure}

We show the second-born \ac{BH} spins for a single population model in Figure~\ref{fig:spin_fits}. 
For $M_\mathrm{WR} < 14~\Msun$ at low periods ($p < 0.3~\mathrm{day}$), the resultant \ac{BH} spin decays as a function of \ac{WR} star mass. 
Such a trend is dictated by the core-collapse mechanism. 
For stars with carbon-oxygen core masses $m_\mathrm{CO-core} < 11~M_\odot$, the \citet{Fryer2012} delayed prescription predicts mass ejection in the supernova event. 
The ejected mass will carry away the \ac{AM} stored in the outer layers, lowering the fraction of \ac{AM} transferred to the \ac{BH} from the \ac{WR} star. 
In the simulations of \cite{Bavera2021b}, we also see a similar \ac{BH} spin decay for \ac{BH}--\ac{WR} systems with $M_\mathrm{WR}>40~M_\odot$ and low orbital periods ($p < 0.3~\mathrm{day}$). 
Such suppression is due to strong \ac{WR} stellar winds and pulsations due to the pair instability process, which both cause the star to lose a nonnegligible fraction of material from its envelope, depleting the \ac{WR} \ac{AM} reservoir.

One additional subchannel of \ac{BBH} formation from isolated binary evolution that is present in our populations is the double-core \ac{CE} channel, in which the helium cores of two supergiant stars proceed through a \ac{CE} in one or both of the envelopes of the stars, leading to a tight \ac{WR}--\ac{WR} system where both binary components can be tidally spun up. 
Since this channel is subdominant and we currently lack the simulation grids and fits to monitor this subdominant tidal spin-up scenario, we exclude such systems from our population and focus solely on the standard \ac{CE} and stable \ac{MT} channels that lead to tidal spin-up. 
This channel also typically forms near-equal-mass \ac{BBH} systems and therefore does not strongly affect our key results. 
We comment more on our exclusion of this subchannel and implications in Section~\ref{sec:discussion}.

%%% FIGURE 1

\subsection{Star Formation and Metallicity Evolution}\label{subsec:star_formation}

As described in Section~\ref{subsec:assumptions}, for each model (parameterized by \facc, \gammaEdd, and \alphaCE), we simulate $12$ discrete metallicities. 
Each binary is evolved for a full Hubble time $t_\mathrm{H}$. 
To get a population of merging \acp{BBH} that is representative of the underlying population in the universe, we resample $2 \times 10^5$ binaries from the metallicity runs and populate the binaries in redshift according to star formation history and metallicity evolution as predicted by the \texttt{Illustris-TNG} simulations~\citep{Nelson2015}.

From \texttt{Illustris-TNG}, we have a grid of stellar mass formed over metallicity and lookback time in a 100 comoving Mpc$^3$ cube. 
Marginalizing over the metallicity axis provides the star formation rate density as a function of redshift, $\psi(z)$. 
However, although our population modeling tracks the total stellar mass sampled across all single stars and binaries, our population models only retain systems of interest (i.e., those that form \acp{BBH}). 
Therefore, drawing the birth redshifts for our target population according to this distribution does not account for the relative formation efficiency of our target population at each redshift, which is dependent on the metallicity distribution at each redshift. 
For each metallicity model $Z_\mathrm{sim}$ we have a formation efficiency 
\begin{equation}
    \zeta(Z_\mathrm{sim}) = \frac{N_\mathrm{BBH}(Z_\mathrm{sim})}{M_\mathrm{sample}(Z_\mathrm{sim})}, 
\end{equation}
where $N_\mathrm{BBH}$ is the number of \acp{BBH} formed and $M_\mathrm{sample}$ is the total mass sampled in the particular metallicity model, accounting for the entirety of the initial mass function and binary fraction. 
At a given redshift $z$, we can construct a distribution of metallicities from the cosmological model. 
The cumulative distribution function of this distribution can be used to determine the relative contribution of each discrete metallicity at a given redshift. 
We define $p_Z(Z|z)$ as the support of the metallicity cumulative distribution function at redshift $z$ closest to a metallicity model $Z_\mathrm{sim}$ in log-space, such that $\sum_{Z\mathrm{sim}} p_Z(Z|z) = 1$. 
Combined, $\zeta(Z_\mathrm{sim}) p_Z(Z_\mathrm{sim}|z)$ gives the number of systems per sampled stellar mass that should be drawn from each metallicity model at a given redshift. 

The number of \ac{BBH} systems formed per comoving volume at a particular discrete metallicity across all redshifts is thus
\begin{equation}
    \frac{\mathrm{d}N}{\mathrm{d}V_\mathrm{c}}(Z_\mathrm{sim}) = \int_{z=0}^{z=\infty} \frac{\psi(z) \zeta(Z_\mathrm{sim}) p_Z(Z_\mathrm{sim}|z)}{H_0(1+z)E(z)} \mathrm{d} z. 
\end{equation}
where we perform the change of variable $\mathrm{d}t = \frac{\mathrm{d}t}{\mathrm{d}z} \mathrm{d}z = (H_0(1+z)E(z))^{-1}\mathrm{d}z$ with $E(z)=\sqrt{\Omega_m(1+z)^3+\Omega_k(1+z)^2+\Omega_\Lambda}$. 
This integral gives us the total number of \ac{BBH} systems formed per source-frame year in a comoving box at a discrete metallicity over the entire history of the universe. 
At a given metallicity, we densely discretize redshifts and evaluate this integral between the bounds of our redshift bins, which gives a relative weight at a particular redshift and metallicity that we use when resampling our population models. 

With the relative probability of drawing a system from our target population at a given redshift and metallicity in hand, we randomly draw the birth redshifts and metallicities of $2 \times 10^5$ systems for each population model set. 
We remove the systems from our resampled population that merge after the present day; that is, systems that satisfy the condition $\tau(z) - t_\mathrm{delay} < 0$, where $t_\mathrm{delay}$ is the delay time (defined as the time between \ac{ZAMS} and \ac{BBH} merger) and $\tau(z)$ is the lookback time of a system born at redshift $z$. 
This eliminates \PercentageSystemsMergingAfterTodayRange of systems from our populations of merging \acp{BBH}. 
The codebase used for generating a metallicity and redshift resampled population based on the \texttt{COSMIC} models using various star formation history and metallicity evolution assumptions is available on Github.\footnote{\url{https://github.com/michaelzevin/resample-population}}

%%% FIGURE 2
\begin{figure}[b]
\includegraphics[width=0.48\textwidth]{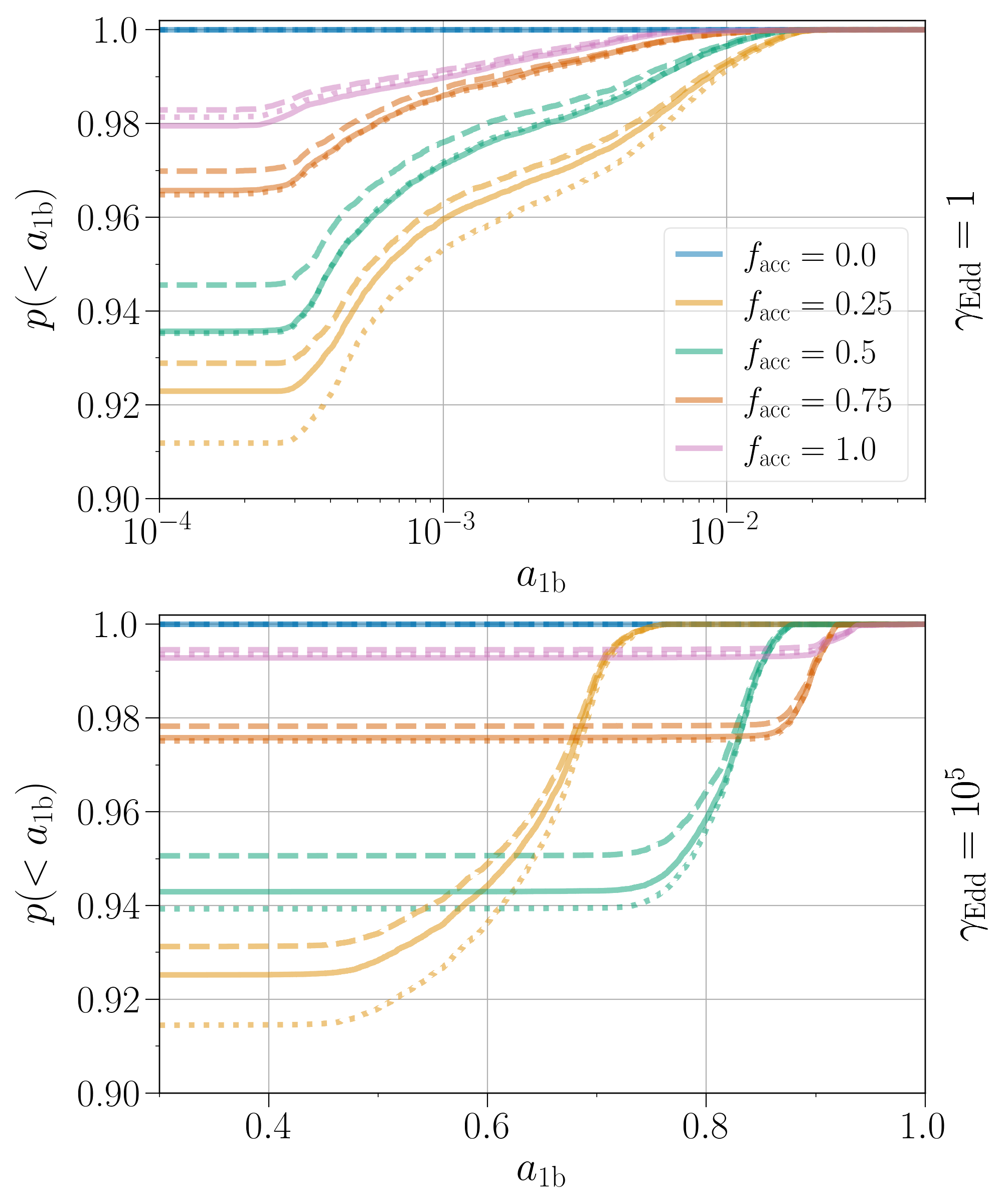}
\caption{Spin magnitude distributions for first-born \acp{BH}. 
Nonzero spin magnitudes for the first-born \ac{BH} population are the result of accretion onto the already-formed \ac{BH} from stellar companions. 
Colors show distributions for different assumptions of the accretion efficiency \facc, with rows showing variations in the super-Eddington accretion parameter \gammaEdd. 
Dotted, dashed, and solid lines show the distributions for \alphaCE values of 0.5, 1, and 2, respectively. 
}
\label{fig:spinmag_firstborn}
\end{figure}

\section{Results}\label{sec:results}

We now investigate the viability of isolated evolution at forming systems with asymmetric masses and spinning primaries. 
We first look at the component spin distributions and mass ratio distributions individually for our array of population models to identify trends across our physical assumption variations in Sections~\ref{subsec:spin_magnitudes} and \ref{subsec:mass_ratios}. 
We then look at broader population properties in mass ratio--spin space in Section~\ref{subsec:pop_props}.

\subsection{Spin Magnitudes of the First- and Second-born BH}~\label{subsec:spin_magnitudes}

Under the assumption of efficient \ac{AM} transport in massive stars, spin in the first-born \ac{BH} can be achieved through accretion via stable \ac{MT} onto the already-formed \ac{BH} by a stellar companion. 
Figure~\ref{fig:spinmag_firstborn} shows the distribution of first-born \ac{BH} spins $a_\mathrm{1b}$ across our various physical assumptions. 
The distributions transition from a flat region, which indicates the branching fraction of stable \ac{MT} systems in each model, to a monotonically increasing behavior; the transition at $a_\mathrm{1b} \sim 3 \times 10^{-4}$ in all models corresponds to the minimum accretion timescale found in our populations of $\sim 0.01~\mathrm{Myr}$. 
Appreciable spin in the first-born \ac{BH} can only be attained via accretion if the \ac{BH} can accrete far above the Eddington limit. 
For Eddington-limited accretion onto the first-born \ac{BH}, we find that no more than $\FirstBornSpinGreaterZeropZeroOneGammaEddOne$ of systems are spun up beyond $a_\mathrm{1b} > 0.01$ across all population models. 
This is expected due to the short evolutionary timescales of the massive stars that are \ac{BH} progenitors, which have post-main-sequence lifetimes of $\lesssim 1~\mathrm{Myr}$ and thus can only accrete at most a fraction of a solar mass if accretion is limited to the Eddington rate. 
More systems proceed through this stable \ac{MT} channel at lower accretion efficiencies ($\simeq \SMTPrecentageFaccZeropTwoFiveGammaEddOne$ at $\facc=0.25$ compared to $\simeq \SMTPrecentageFaccOneGammaEddOne$ at $\facc=1$ in the Eddington-limited models), though higher values for the accretion efficiency lead to larger typical first-born spins when only considering the stable \ac{MT} channel. 
Variations in the \ac{CE} efficiency, shown with different linestyles in Figure~\ref{fig:spinmag_firstborn}, have a minor and indirect impact on the first-born spin distributions; these slight variations are due to the changes in the relative fraction of merging \acp{BBH} that proceed through a \ac{CE} phase rather than stable \ac{MT}.

%%% FIGURE 3
\begin{figure}[t]
\includegraphics[width=0.48\textwidth]{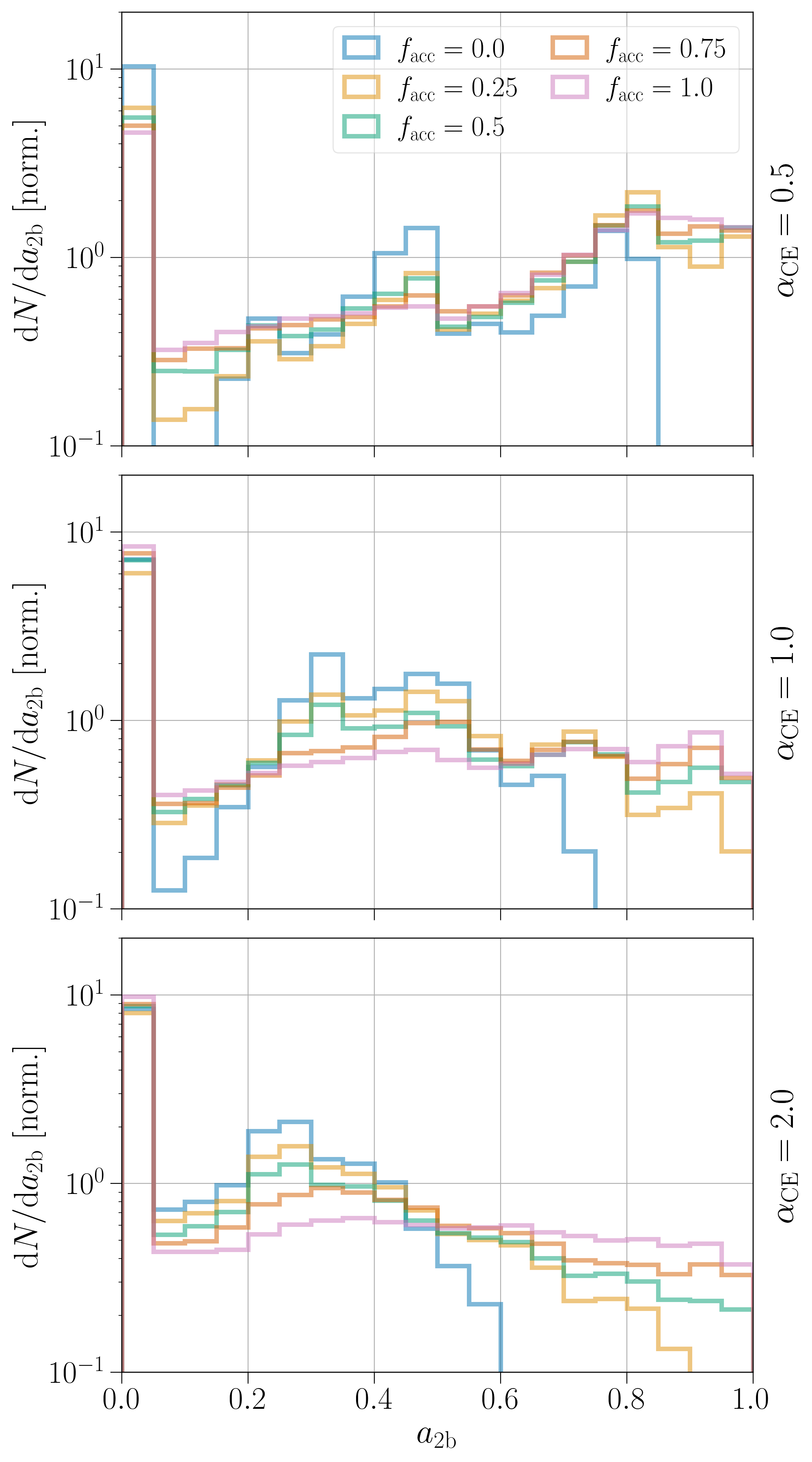}
\caption{Spin magnitude distributions for the second-born \ac{BH}. 
Nonzero spin magnitudes for the first-born \ac{BH} population are the result of tidal spin-up of the \ac{WR} progenitor. 
Colors show distributions for different assumptions of the accretion efficiency \facc, with rows showing variations in \ac{CE} efficiencies $\alphaCE$. 
Accretion is limited to the Eddington rate ($\gammaEdd = 1$) for all plotted models. 
}
\label{fig:spinmag_secondborn}
\end{figure}

Increasing the accretion limit onto \acp{BH} to $10^5$ times the Eddington rate drives the first-born \ac{BH} spins to more extreme values. 
The differences in the transition between a flat behavior and monotonically increasing behavior for different accretion efficiency models in Figure~\ref{fig:spinmag_firstborn} indicates that the maximum \ac{MT} rate is now often imposed by the $\facc$ rather than the $\gammaEdd$, with the transition happening at larger spins as $\facc$ increases. 
For the nonzero accretion efficiencies $\facc>0$, a sizeable fraction of systems in our populations have first-born spins with $a_\mathrm{1b}>0.5$:  $\approx \FirstBornSpinGreaterZeropFiveFaccZeropFiveGammaEddOneEFive$ in our $\facc=0.5$ models. 
Almost no systems have first-born spins of $a_\mathrm{1b} < 0.5$ in any of our super-Eddington accretion models. 
Though we do not explicitly consider \ac{BBH} merger rates here, we note that significantly increasing the possible accretion rate onto \acp{BH} drives down the expected merger rate of systems with highly spinning first-born \acp{BH} due to the conservative \ac{MT} not shrinking the orbit as efficiently as nonconservative \ac{MT}~\citep{Bavera2021}. 
This is seen in the bottom panel of Figure~\ref{fig:spinmag_firstborn}; though increasing the accretion efficiency does indeed lead to more highly spinning \acp{BH} from the stable \ac{MT} channel, the relative contribution of this channel compared to the full \ac{BBH} population decreases.

%%% FIGURE 4
\begin{figure*}[t]
\includegraphics[width=1.0\textwidth]{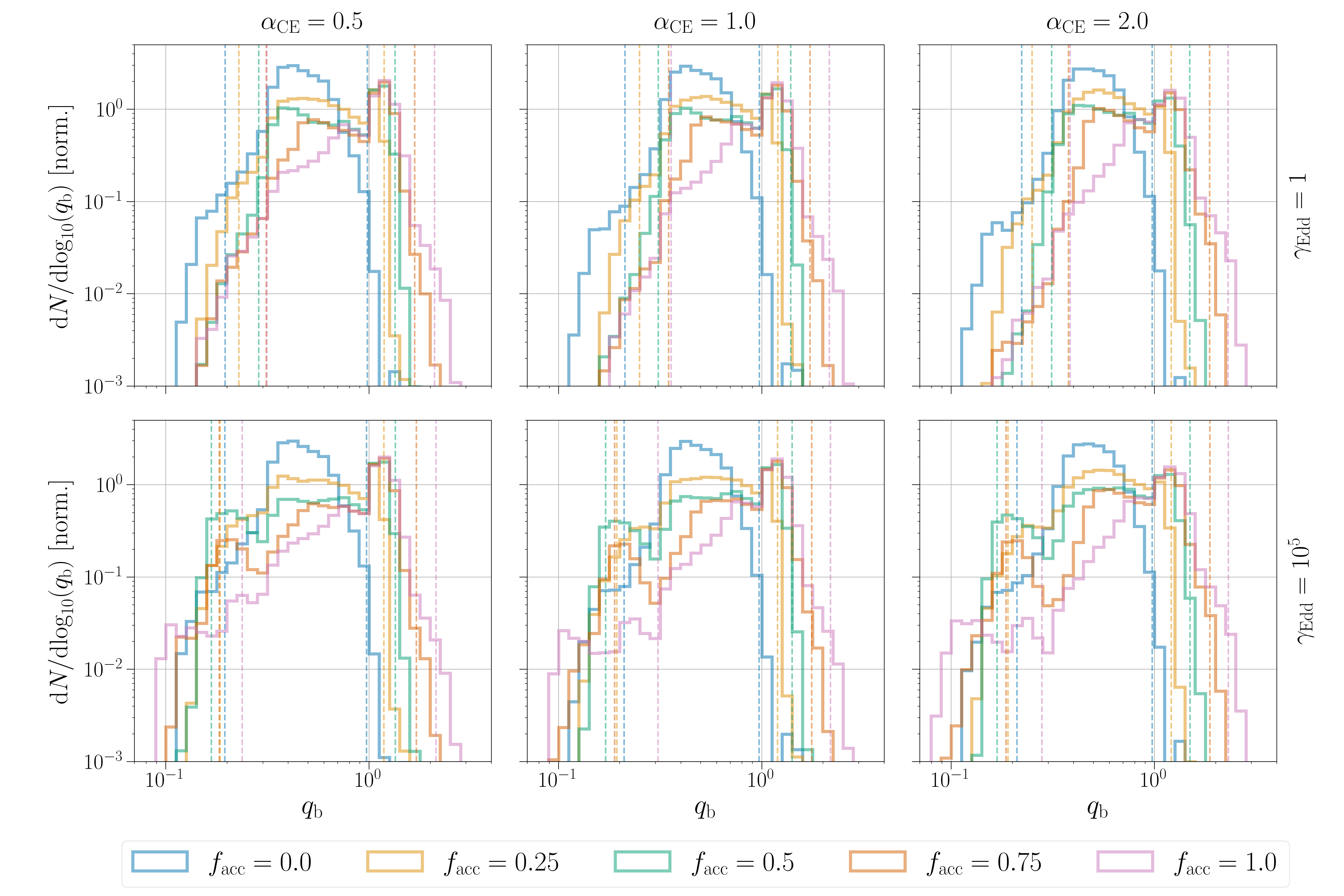}
\caption{Distribution of mass ratios between the first-born \ac{BH} and second-born \ac{BH}. 
Values with $\qb<1$ indicate that the first-born \ac{BH} is the more massive component, whereas values with $\qb>1$ indicate that the second-born \ac{BH} is the more massive component. 
As in Figures~\ref{fig:spinmag_firstborn} and \ref{fig:spinmag_secondborn}, colors show distributions for different assumptions of the accretion efficiency \facc. 
Rows and columns show variations in \ac{CE} efficiency \alphaCE and super-Eddington accretion \gammaEdd, respectively. 
Vertical dashed lines mark the region in which 99\% of all systems in the respective populations reside. 
}
\label{fig:qb_dist}
\end{figure*}

In Figure~\ref{fig:spinmag_secondborn}, we show the spin distributions for the second-born \acp{BH} $a_\mathrm{2b}$ in our models, which is driven by tidal spin-up of He cores by the first-born \ac{BH}. 
For all models, we find a broad range of spin magnitudes ranging from nonspinning to maximally spinning, with little dependence on \gammaEdd because the \ac{CE} channel dominates over the stable \ac{MT} channel in our models. 
The peak at $a_\mathrm{2b} \simeq 0$ in all models is the result of systems that have orbital periods outside of the spin-up regime at \ac{WR}--\ac{BH} formation; see Figure~\ref{fig:spin_fits}. 
Across our models, we find that \SecondBornVanishingSpin of secondary spin magnitudes are $a_\mathrm{2b} < 0.05$. 

We find a general trend of second-born spin distributions pushing to larger values with increasing accretion efficiency and decreasing \ac{CE} efficiency, in agreement with \cite{Bavera2021}. 
For $\facc = 0.5$ and $\alphaCE = 1$, \SecondBornSpinGreaterZeropFiveFaccZeropFiveAlphaCEOne of systems have second-born \ac{BH} spins of $a_\mathrm{2b} > 0.5$. 
This fraction drops to \SecondBornSpinGreaterZeropFiveFaccZeropFiveAlphaCETwo for $\alphaCE = 2$ and increases to \SecondBornSpinGreaterZeropFiveFaccZeropFiveAlphaCEZeropFive for $\alphaCE = 0.5$. 
The increase in high-spinning \acp{BH} with decreasing \ac{CE} efficiency is due to the less efficient \ac{CE} phases leading to more hardened orbits post-\ac{CE}, which makes the stripped He core more susceptible to tidal spin-up. 
At $\alphaCE=1$, the fraction of systems with $a_\mathrm{2b} > 0.5$ drops to \SecondBornSpinGreaterZeropFiveFaccZeroAlphaCEOne for $\facc=0$ whereas at $\facc=1$ it increases to \SecondBornSpinGreaterZeropFiveFaccOneAlphaCEOne. 
This trend of increasingly high-spinning second-born \acp{BH} with increasing accretion efficiency is due to the progenitor of the second-born \ac{BH} gaining more mass during the first stable \ac{MT} phase; the post-\ac{CE} separation scales inversely with the product of the total mass and envelope mass of the donor pre-\ac{CE}.

\subsection{Mass Ratios and Mass Ratio Reversal}~\label{subsec:mass_ratios}

We now turn to the mass ratio distributions in our populations. 
Since the order at which \acp{BH} are born in a binary determines the mechanism responsible for their potential spin, we choose to define the birth-mass-ratio parameter $\qb = m_\mathrm{2b} / m_\mathrm{1b}$, where $m_\mathrm{1b}$ and $m_\mathrm{2b}$ are the masses of the first-born \ac{BH} and second-born \ac{BH}, respectively. 
Thus, for values $\qb > 1$, the system proceeded through a \ac{MRR} that led to the less massive star at \ac{ZAMS} forming in the more massive \ac{BH} in the binary. 
Note that in our models the only mechanisms for inducing the spin on \ac{BH} remnants are accretion onto the \ac{BH} and tidal spin-up, and thus the first-born \ac{BH} is only susceptible to the former, and the second-born \ac{BH} is only susceptible to the latter of these processes. 

We show the population distributions of the mass ratio parameter \qb in Figure~\ref{fig:qb_dist}. 
The accretion efficiency \facc has the largest impact in determining whether \ac{MRR} ensues, as this parameters governs the amount of material the less massive star at \ac{ZAMS} can accept from the initially more massive donor after it evolves off the main sequence and overflows its Roche lobe. 
For example, in our default model with $\alphaCE = 1$ and $\gammaEdd = 1$, we find \QbGreaterOneFaccZeroAlphaCEOne of merging \ac{BBH} systems to have $\qb > 1$ for $\facc = 0$, which rises to \QbGreaterOneFaccZeropFiveAlphaCEOne at $\facc = 0.5$ and \QbGreaterOneFaccOneAlphaCEOne at $\facc = 1$. 
Despite this, the second-born \ac{BH} rarely reaches masses of more than $\simeq 3$ times the mass of the first-born \ac{BH}; in our most optimistic \ac{MRR} models, we find the birth-mass-ratio parameter to be constrained to $\qb < \QbNinetyNinePercentileFaccOne$ at 99\% confidence, and it never reaches a value larger than $\qb = \QbHighExtreme$. 

Increasing the accretion limit onto \acp{BH} leads to more systems occupying the low-end tail of the \qb distribution. 
This is due to the first-born \ac{BH} being able to gain significantly more mass from accretion relative to the Eddington-limited case, and can generate systems with mass ratios of $\qb \simeq \QbLowEndExtremeEstimate$ in the most extreme cases. 
However, these extreme mass ratio systems are still rare in the context of the full population; at $\facc = 0.5$ and $\alphaCE = 1$, we find that 99\% of systems have $\qb > \QbOnePercentileFaccZeropFiveAlphaCEOneGammaEddOneEFive$ for $\gammaEdd = 10^5$, which raises to $\qb > \QbOnePercentileFaccZeropFiveAlphaCEOneGammaEddOne$ for the Eddington-limited case with $\gammaEdd = 1$. 
Thus, when considering all systems regardless if they proceeded through a \ac{MRR} or not, we find that forming binaries with mass asymmetries of more than $3$:$1$ is a rare occurrence unless super-Eddington accretion is invoked.

%%% FIGURE 5
\begin{figure}[t]
\includegraphics[width=0.5\textwidth]{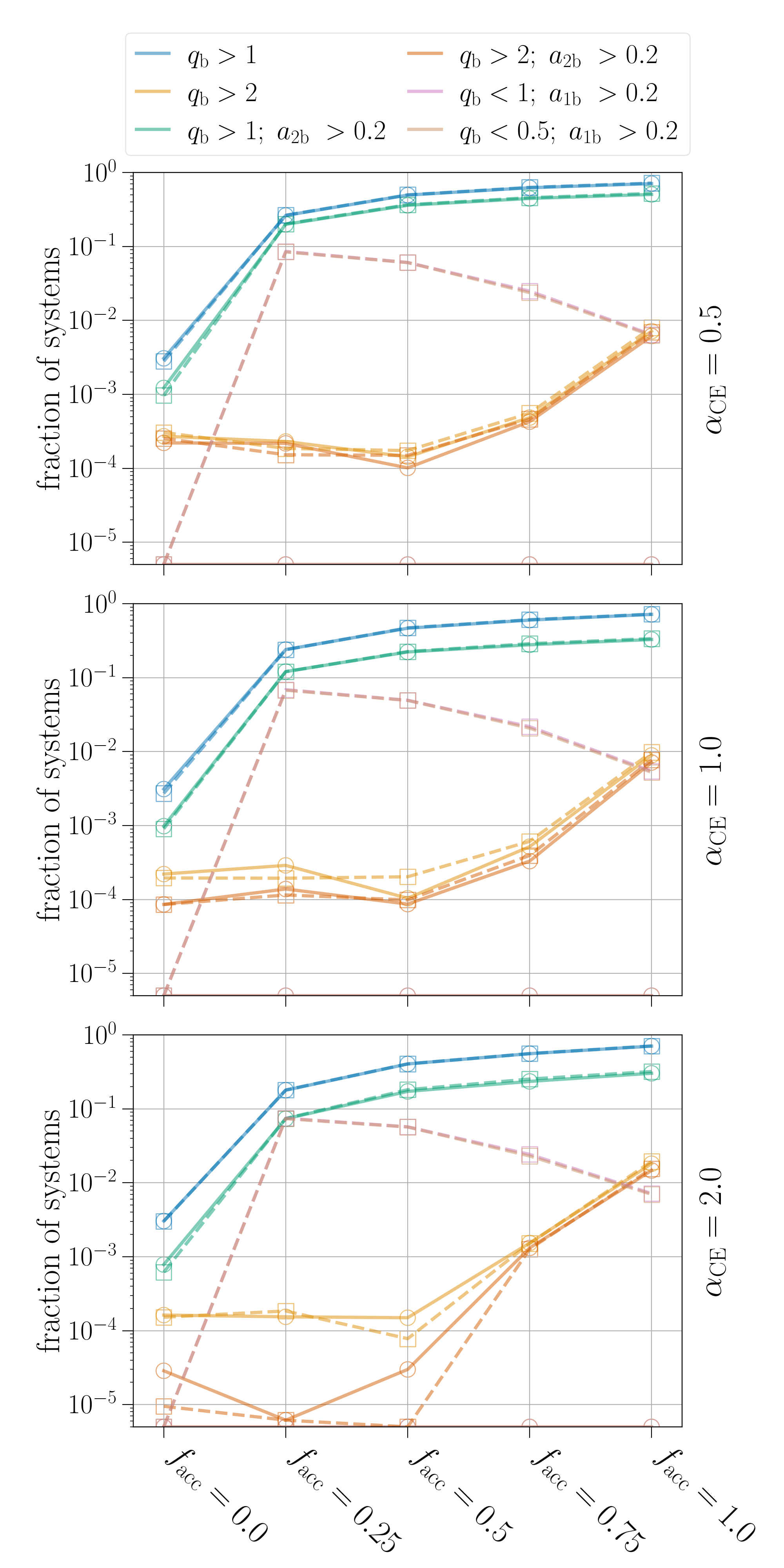}
\caption{Fraction of systems that satisfy criteria based on mass ratios and component spins as a function of accretion efficiency \facc. 
Different criteria are denoted with different colored lines as indicated in the legend, with solid lines with circular markers (dashed lines with square markers) showing the $\gammaEdd=1$ ($\gammaEdd=10^5$) populations and rows showing different values of $\alphaCE$. 
Points for models where no systems in the population satisfy a particular criteria are plotted on the horizontal axis. 
Note that the points for the pink ($\qb < 1;\,a_\mathrm{1b} > 0.2$) and brown ($\qb < 0.5;\,a_\mathrm{1b} > 0.2$) conditions are mostly overlapping. 
}
\label{fig:mass_spin_criteria}
\end{figure}

\subsection{Population Properties}~\label{subsec:pop_props}

We now examine both component spins and mass ratios simultaneously to make broader statements about the viability of generating asymmetric-mass systems with spinning primaries. 
Figure~\ref{fig:mass_spin_criteria} marks the fraction of systems in our full \ac{BBH} population that satisfy particular criteria of interest. 
We specifically highlight the systems that proceed through \ac{MRR} ($\qb > 1$), systems that have significant asymmetries in their \ac{BH} masses ($\qb > 2$ or $\qb < 0.5$), and systems that achieve a significant primary \ac{BH} spin ($a > 0.2$). 

As shown in Figure~\ref{fig:qb_dist}, the number of systems that proceed through a \ac{MRR} is significantly enhanced with increasing accretion efficiency (blue and green lines). 
However, even for perfectly conservative \ac{MT} where all of the material from the donor is accepted by the accretor, the number of systems that have significantly asymmetric masses and nonnegligible spins in the primary \ac{BH} is small (yellow and orange lines). 
Across our Eddington-limited population models, \PercentageFaccOneQbGreaterTwoSpinGreaterZeropTwo satisfy this criteria at $\facc = 1$; this number drops by an order of magnitude or more for lower values of $\facc$. 
Therefore, in the efficient \ac{AM} transport paradigm, we find the formation of systems with asymmetric masses and spinning primary \acp{BH} to be inefficient even for the most optimistic assumptions about \ac{MT} physics. 

For our models that allow for highly super-Eddington accretion ($\gammaEdd = 10^5$), the percentage of systems with nonnegligible primary \ac{BH} spins and mass ratios greater than $2$:$1$ can exceed \PercentageEddfacOneEFiveQbLessZeropFiveSpinGreaterZeropTwoHigh and drop to less than \PercentageEddfacOneEFiveQbLessZeropFiveSpinGreaterZeropTwoLow depending on the choice of accretion efficiency, so long as $\facc > 0$. 
When the accretion onto \acp{BH} is Eddington limited, no systems in our population meet this criteria since the first-born \ac{BH} spin-up is highly suppressed (see Figure~\ref{fig:spinmag_firstborn}). 
Thus, in either spin-up scenario, we find the fraction of systems that have significant mass asymmetries ($> 2$:$1$) and significant primary spins ($> 0.2$) to be percent-level at best, and requires highly conservative \ac{MT} or highly super-Eddington accretion to achieve these numbers.

\section{Contextualizing with the Gravitational-wave Population}\label{sec:gw_results}

In the previous section, we examined the efficiency of forming systems with asymmetric masses and spinning primary components, and found that such systems are relatively rare occurrences in the underlying population even when taking optimistic assumptions about \ac{MT} efficiency. 
We now turn to the implications of these results, in the context of explaining both the individual events and general trends that are observed in the \ac{BBH} population.

%%% FIGURE 6
\begin{figure*}[t]
\includegraphics[width=1.0\textwidth]{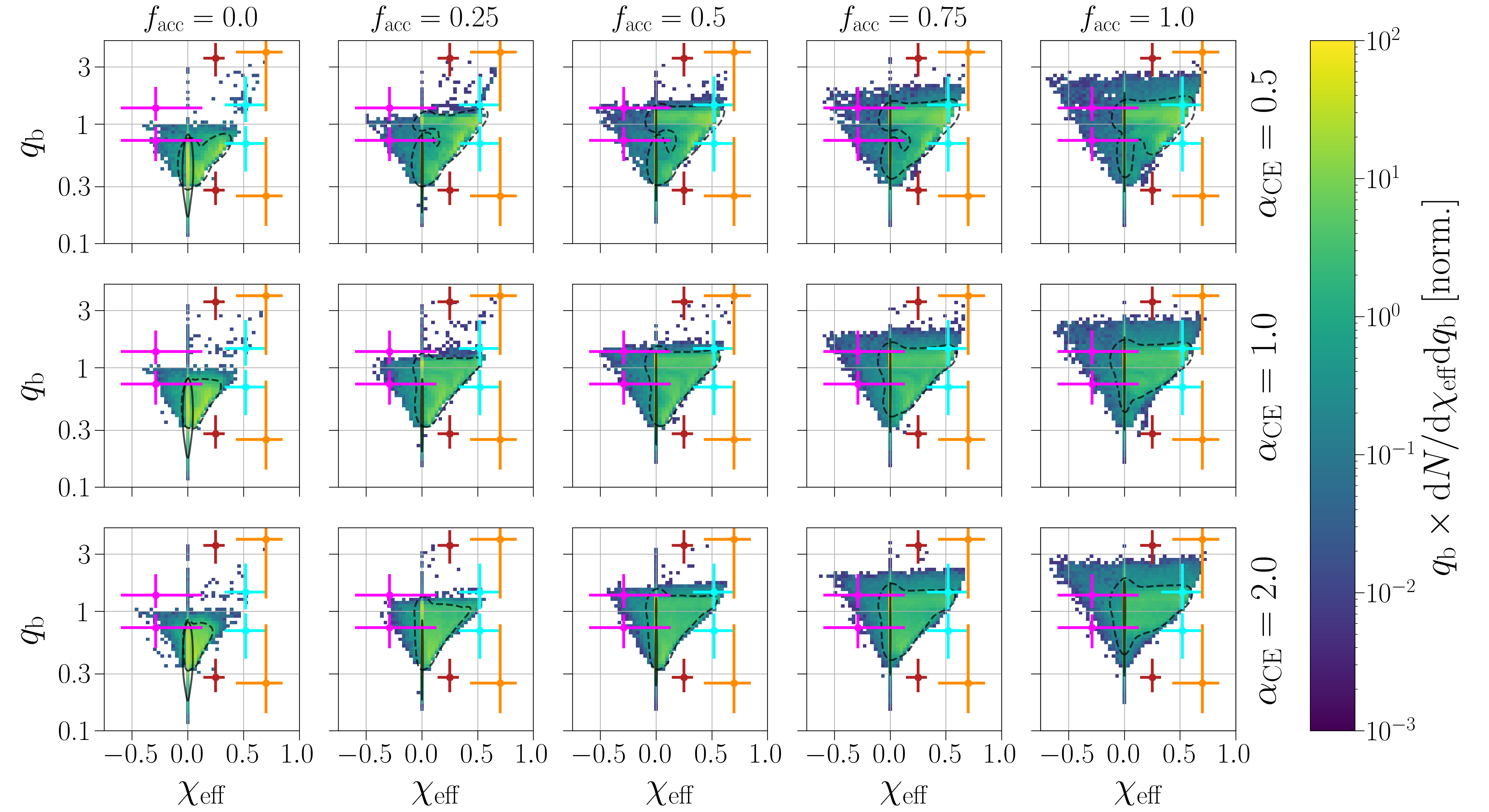}
\caption{Joint distribution of mass ratio between the first-born and second-born \ac{BH}, $\qb$, and effective spin, $\chieff$, for various assumptions regarding the accretion efficiency, \facc (columns), and \ac{CE} efficiency, \alphaCE (rows). 
Values of $\qb<1$ indicate that the first-born \ac{BH} is the more massive component, whereas $\qb>1$ indicate that the second-born \ac{BH} is the more massive component. 
For all distributions in this figure, we show results for models that assume Eddington-limited accretion, with $\gammaEdd=1$. 
Black dashed (solid) lines surround 90\% of systems from the stable \ac{MT} (\ac{CE}) channel. 
Colored points show the 90\% credible intervals for mass ratio and effective spin for a selection of \ac{GW} events: GW190403\_051519 (orange), GW190412 (red), GW190517\_055101 (cyan), and GW191109\_010717 (pink). 
Two points are shown for each event that is reflected over the boundary at $\qb = 1$. 
}
\label{fig:q_chieff_eddfac1e0}
\end{figure*}

A number of systems in the current population of \acp{BBH} exhibit interesting combinations of mass ratio and spin. 
For example, GW190412 and GW190517\_055101 are both high-confidence events with positive effective spins of $\chieff = \GWNineteenZeroFourTwelveChiEff$ and $\chieff = \GWNineteenZeroFiveSeventeenChiEff$, respectively, at 90\% credibility~\citep{GWTC2}. 
The primary \ac{BH} in GW190412 is 3--4 time more massive than the secondary, with a mass ratio of $q = \GWNineteenZeroFourTwelveMassRatio$~\citep{GW190412}, whereas the mass ratio of GW190517\_055101 is more consistent with unity ($q = \GWNineteenZeroFiveSeventeenMassRatio$, \citealt{GWTC2}). 
Because of the large measured effective spins and/or mass asymmetry, the spin magnitudes of the primary \ac{BH} can be constrained to $a_1 = \GWNineteenZeroFourTwelvePrimarySpin$ and $a_1 = \GWNineteenZeroFiveSeventeenPrimarySpin$ for GW190412 and GW190517\_055101, respectively, whereas the spin of the secondary \ac{BH} is unconstrained.\footnote{Though strong astrophysical priors have been shown to lead to alternate interpretation in the distribution of spin between the primary and secondary components of GW190412~\citep{Mandel2020}, this interpretation is statistically disfavored relative to the system having a spinning primary~\citep{Zevin2020a}.} 
The extended catalog for the first half of the third observing run uncovered the system GW190403\_051519, and although it has a relatively low astrophysical probability compared to GW190412 and GW190517\_055101 and a primary mass that is in tension with the maximum \ac{BH} mass expected from the pair instability process, its significantly asymmetric masses ($q = \GWNineteenZeroFourZeroThreeMassRatio$) and high primary spin ($a_1 = \GWNineteenZeroFourZeroThreePrimarySpin$) make it a system of interest. 
In the most recent \ac{GW} catalog published by the \ac{LVK}, the high-confidence event GW191109\_010717 also has intriguing spin signatures with the primary \ac{BH} spinning at $a_1 = \GWNineteenElevenZeroNinePrimarySpin$~\citep{GWTC3}, although its high mass and in-plane spin may be indicative of a dynamical formation scenario. 
Other \ac{GW} catalogs outside the \ac{LVK} have also uncovered additional systems with interesting configurations of spin and mass ratio that may help to probe the viable spin-up mechanisms~\citep[e.g.,][]{Nitz2021a,Olsen2022}. 

These \acp{BBH}, particularly GW190412, are potential exemplars of systems whose progenitors underwent a \ac{MRR} where the more massive \ac{BH} was born second and its progenitor was tidally spun up~\citep{GW190412,Zevin2020a,Olejak2020,Olejak2021a}, or alternatively having the first-born primary \ac{BH} spun up by super-Eddington accretion during the second stable \ac{MT} episode~\citep{Bavera2021}. 
As we showed in Figure~\ref{fig:spinmag_secondborn}, population modeling predicts a significant number of systems that can have appreciable spins across the entire range of physical spin magnitudes in the underlying population. 
However, the formation of systems that can also achieve spinning primary components drops precipitously as the mass of the primary relative to the mass of the secondary increases, as shown in Figure~\ref{fig:mass_spin_criteria}. 
Although the mass ratio distribution of GW190517\_055101 is too broad to definitively rule out this formation mechanism, we find the formation of GW190412 through \ac{MRR} and tidal spin-up to be improbable. 
Taking the upper and lower values of GW190412's 90\% symmetric credible interval for mass ratio and primary spin, respectively (i.e., taking values closest to nonspinning and equal mass), we find that $< \PercentageGWNineteenZeroFourTwelveMassRatioSpinOptimisticMassInversion$ of systems satisfy this criteria in our Eddington-limited accretion model even when we take the optimistic assumption of $\facc = 1$. 
For our highly super-Eddington models, we find the formation of such systems to be more likely, with up to \PercentageGWNineteenZeroFourTwelveMassRatioSpinOptimisticSuperEddington of systems satisfying this criteria. 

We find the detection of asymmetric-mass, spinning primary \ac{BH} systems through the tidal spin-up scenario to be less likely when selection effects are accounted for, consistent with the findings of \cite{Bavera2021}. 
Since the tidal spin-up of a \ac{WR} star by a \ac{BH} requires subday orbital periods (see Figure~\ref{fig:spin_fits}), these systems have short inspiral times when they form \acp{BBH}. 
Systems will more readily proceed through tidal spin-up at lower metallicities (and therefore higher redshifts) because the \ac{WR} stellar winds are weaker, which lessens orbital expansion and keeps the \ac{BH}--\ac{WR} system in the regime where tidal spin-up can still be efficient~\citep{Bavera2020}. 
Therefore, many tidal spin-up systems will merge outside of the horizon of current ground-based \ac{GW} detectors, and a lower fraction of systems in the detectable population is predicted to have high spins from tidal spin-up relative to the underlying population~\citep[e.g.,][]{Safarzadeh2020c,Bavera2021}. 
Applying selection effects to our population causes the probability of detecting systems similar to GW190412 through the \ac{MRR} and tidal spin-up channel to further decrease by more than an order of magnitude. 
However, we find the super-Eddington accretion spin-up scenario to be less impacted by selection effects, still accounting for a percent-level number of systems in the detectable population. 
We describe our implementation of selection effects in Appendix~\ref{app:selection_effects}.

%%% FIGURE 7
\begin{figure*}[t]
\includegraphics[width=1.0\textwidth]{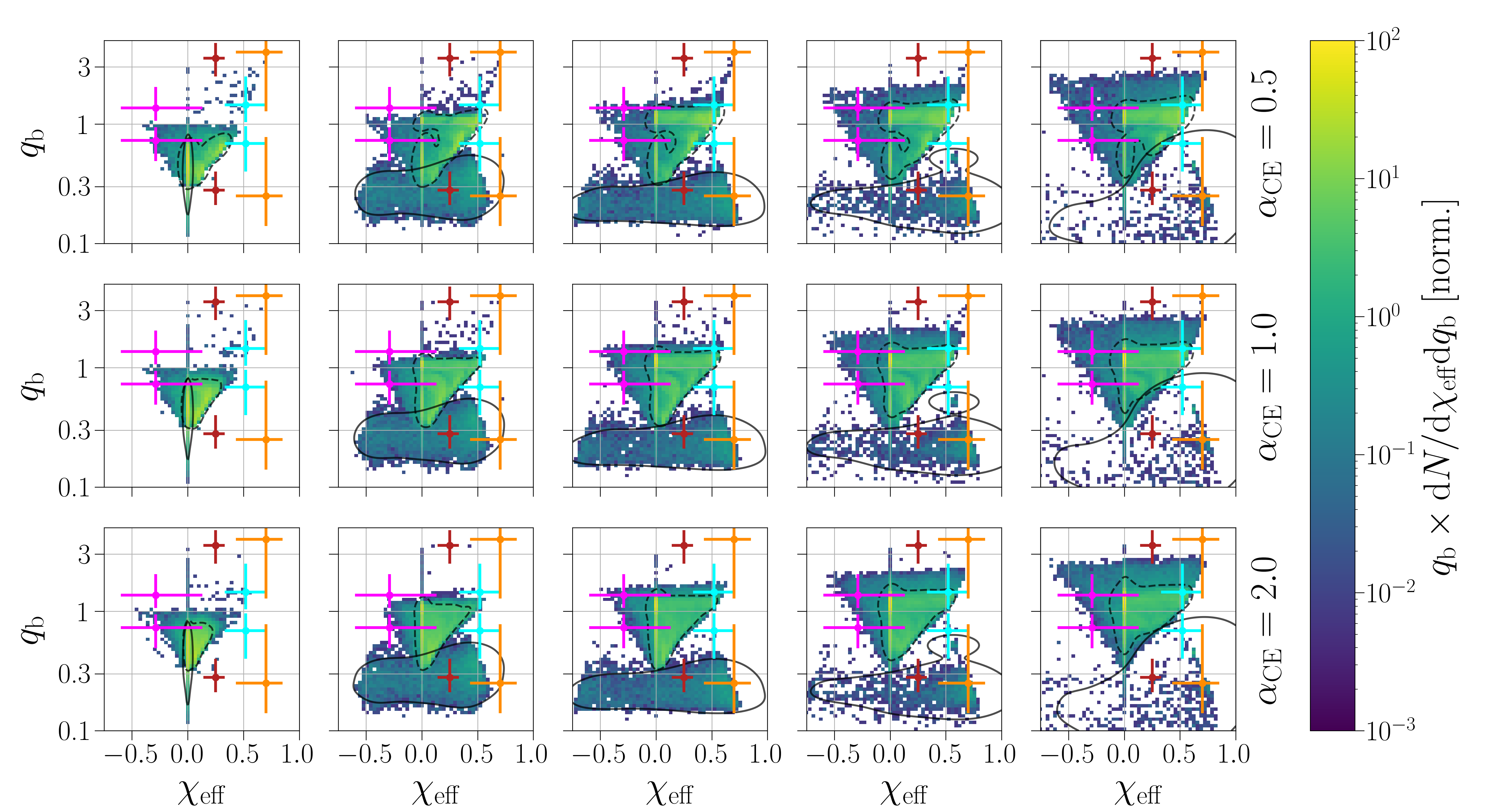}
\caption{Same as Figure~\ref{fig:q_chieff_eddfac1e0} except for models with a super-Eddington accretion parameter of $\gammaEdd=10^5$. 
}
\label{fig:q_chieff_eddfac1e5}
\end{figure*}

In addition to the individual \ac{BBH} systems in the \ac{GW}-detected population having interesting parameter estimates, populations analyses are useful for uncovering trends and features in the data as a whole~\citep[see, e.g.,][]{Mandel2019,Vitale2020a}. 
One example that is particularly pertinent to mass ratios and spin configurations is that the population of \acp{BBH} shows tantalizing signs of (anti)correlations between mass ratio and effective spin~\citep{Callister2021a,GWTC3_pops}. 
In Figures~\ref{fig:q_chieff_eddfac1e0} and \ref{fig:q_chieff_eddfac1e5}, we show joint distributions of the birth-mass-ratio parameter, $\qb$, and effective spin, $\chieff$, in our \ac{BBH} populations for $\gammaEdd=1$ and $\gammaEdd=10^5$, respectively. 
A few \ac{GW} events with interesting mass ratio and spin measurements are overplotted for comparison. 

We find no strong indication of trends in effective spins as a function of mass ratio (for either $\qb > 1$ or $\qb < 1$) in our Eddington-limited population models shown in Figure~\ref{fig:q_chieff_eddfac1e0}. 
The effective spin distributions tend to narrow closer to $\chieff \simeq 0$ as mass ratios decrease from $\qb = 1$ to $\qb = 0$. 
For values of $\qb > 1$, the effective spin distributions remain broad, but retain a peak in the distribution at $\chieff \simeq 0$. 

This situation differs in the super-Eddington models, as shown in Figure~\ref{fig:q_chieff_eddfac1e5}. 
Significantly increasing the accretion limit onto \acp{BH} leads to the stable \ac{MT} channel populating the region of parameter space with large effective spins and small mass ratios ($\qb \ll 1$). 
The combination of the bulk of systems dominating the low-spin and equal-mass regime, \ac{MRR} and tidal spin-up accounting for mildly asymmetric masses and moderate spins, and stable \ac{MT} with super-Eddington accretion producing low mass ratios and high spins can lead to an apparent trend between mass ratio and effective spin, and a potential means of generating the anticorrelation observed in \ac{GW} data. 
Visually, the super-Eddington stable \ac{MT} channel can populate the regions of parameter space occupied by unique systems such as GW190412. 
We also see an increased number of systems with $q \ll 1$ and antialigned spins, which result from the second-born \ac{BH} being lower mass and more susceptible to strong natal kicks that can significantly tilt the orbital plane. 
Although the relative fraction of \acp{BBH} formed via \ac{CE} and stable \ac{MT} channels may not be accurately captured by rapid population synthesis approaches (see Section~\ref{sec:discussion} for further discussion), the combination of these isolated evolution subchannels may provide one explanation for anticorrelations between mass ratio and effective spin. 
However, our models find that extreme assumptions regarding the accretion rate onto \acp{BH} are required for such an interpretation.

It is interesting to note that certain analysis techniques can have a significant impact on measurements of spins in individual \ac{BH} systems and the population of component spins. 
For example, rather than inferring the properties of the most massive and least massive \acp{BH} in a particular system, \cite{Biscoveanu2021a} show that, by instead inferring the parameters of the highest-spinning and lowest-spinning components of the system, spin magnitude measurements can be more informative especially for near-equal-mass systems. 
Such variations to the standard analyses may be useful in better constraining the component spin magnitudes, though it yields less differences in parameter estimates for unequal-mass systems such as those considered in this work, due to the spin of the heavier \ac{BH} playing a larger role in measured spin quantities such as the effective spin. 
Nonetheless, such techniques may uncover more component \acp{BH} with significant spin magnitudes in the broader population~\citep{GWTC3_pops}.

\section{Discussion and Conclusions}\label{sec:discussion}

In this work, we simulated a suite of population models to investigate the joint distribution of masses and spins across an astrophysical population of \acp{BBH}. 
In particular, we focus on the ability of stellar progenitors to proceed through a \ac{MRR} where the lighter star at \ac{ZAMS} leads to the more massive compact object, and the potential means of generating spin in \acp{BH} even with efficient \ac{AM} transport in their stellar progenitors. 
Our key results are as follows. 

\begin{enumerate}
    \item Although the distribution is peaked near zero, the second-born \ac{BH} can attain a range of spin magnitudes from near zero to maximally spinning through tidal spin-up, with less efficient \ac{CE} phases leading to larger typical spins. 
    Across our model variations, $\SecondarySpinGreaterZeropFiveRange$ of systems have second-born \acp{BH} with spin magnitudes $> 0.5$ (Figure~\ref{fig:spinmag_secondborn}). 
    \item Assuming highly efficient \ac{AM} transport and Eddington-limited accretion, the spin of the first-born \ac{BH} will naturally be small, with typical values $\ll 10^{-2}$. 
    However, increasing the accretion rate to many orders of magnitude above the Eddington limit will lead to a sizeable fraction of first-born \acp{BH} with significant spin. 
    Though higher accretion efficiencies lead to larger typical spins from this subchannel, it causes the contribution of this subchannel to drop relative to the full population (Figure~\ref{fig:spinmag_firstborn}). 
    \item Naturally, the number of systems that proceed through \ac{MRR} is highly sensitive to the assumed accretion efficiency at which the accreting star accepts material from the donor star. 
    Our parameterization choices for the accretion efficiency bracket the range of systems that have a more massive second-born \ac{BH} to $\MassInversionRange$, though even at a low accretion efficiency of $\facc = 0.25$, more than $\MassInversionZeropTwoFiveLow$ of systems have a more massive second-born \ac{BH}. 
    \item It is rare for systems to form with asymmetric masses and spinning primary \acp{BH} through \ac{MRR} and tidal spin-up; across our various model assumptions with $\facc > 0$, \MassInversionSpinningRange of systems in the underlying population have second-born \acp{BH} that are twice as massive as their first-born counterparts and have spin magnitudes $> 0.2$. 
    These numbers decrease by an order of magnitude or more when \ac{GW}-detector selection effects are considered due to the short delay times of this channel and lower efficiency of tidal spin-up at lower redshifts and higher metallicities. 
    The number of systems where the first-born \ac{BH} is twice as massive as the second-born \ac{BH} and spinning at $> 0.2$ is negligible for our Eddington-limited models, though can account for \NoMassInversionSpinningRangeSuperEddington of systems in the underlying population of our models that assume highly super-Eddington accretion. 
    We therefore find that the formation of \ac{BBH} systems with asymmetric masses and spinning primary \acp{BH}, such as GW190412, is unlikely through the channels considered in this work unless highly super-Eddington accretion is invoked. 
    \item We find no indication of a correlation between mass ratio and effective spin in our model variations other than in the models that assume highly super-Eddington accretion onto \acp{BH}. 
\end{enumerate}

Though we aim to survey the parameterizations of binary stellar evolution that most impact \ac{MRR} and tidal spin-up, the vast array of physical uncertainties that embed binary population modeling prevent us from being completely exhaustive in our coverage of parameter space. 
We anticipate that the criteria determining the onset of unstable \ac{MT} could lead to potential changes in our results. 
This is typically encoded in rapid population modeling via an array of stellar-type-specific critical mass ratio values, $\vec{q}_\mathrm{crit}$, where \ac{MT} will be unstable if the donor mass is sufficiently more massive than the accretor mass. 
One variation in this parameter for a subset of our other model variations is described in Appendix~\ref{app:qcrit}. 
However, it is possible that the values typically used in population synthesis are not representative of the true physical picture. 
For example, \cite{Gallegos-Garcia2021} found that rapid population synthesis may severely overpredict the number of systems that proceed through a successful \ac{CE} phase and form \ac{BBH} systems. 
In Figure~\ref{fig:q_chieff_eddfac1e5}, the high-density region at low mass ratios ($\qb \ll 1$) and large, positive effective spins is populated almost exclusively through systems that do not proceed through \ac{CE} evolution, and could potentially lead to a mass ratio--effective spin correlation in the full \ac{BBH} population if the \ac{CE} channel is suppressed, though this feature is only apparent in the models that assume super-Eddington accretion onto \acp{BH} (see also \citealt{Bavera2021}). 
In our Eddington-limited models, we find that the stable \ac{MT} channel typically does not harden \ac{BH}--\ac{WR} binaries to orbital periods of less than 1 day, and thus is inefficient at spinning up the second-born \ac{BH} progenitor through tides. 
However, we note that the work investigating tidal spin-up in isolated evolution with other population synthesis codes, such as \cite{Olejak2021a}, finds that evolutionary sequences that do not involve a \ac{CE} phase can still lead to an appreciable fraction of systems that are spun up via tides. 

One subchannel present in our models, though excluded in our analysis, is the double-core \ac{CE} channel, in which both helium cores can tidally interact and spin up before collapsing into \acp{BH}. 
This channel is excluded (a) because we lack detailed simulations grids following tidal spin-up in this scenario, and (b) so our results are less opaque and the first-born \ac{BH} can only attain spin through accretion following \ac{BH} formation, and the second-born \ac{BH} can only attain spin through \ac{BH}--\ac{WR} tidal spin-up. 
Although subdominant, this channel contributes up to $\sim 10\%$ of our underlying population of merging \acp{BBH} in certain models, similar to the underlying populations of \cite{Neijssel2019}. 
It does, however, contribute to a much smaller fraction of the detectable population (see, e.g., Figure 12 of \citealt{Neijssel2019}). 
We anticipate that proper treatment of this spin-up process would lead to spinning first-born and second-born \acp{BH} and large, positive effective spins. 
However, in our models, this channel favors the formation of near-equal-mass binaries, with $\sim 95\%$ of systems having mass ratios of $q > 0.75$. 
Therefore, although the inclusion and proper treatment of this channel would affect the formation of near-equal-mass systems with high effective spins, we find it to be inefficient at generating systems with spinning primaries and significantly asymmetric masses. 

Another evolutionary scenario not explored in detail here that can induce the spins in \ac{BH} progenitors is through \ac{MT} that occurs during the main sequence. 
In this Case A \ac{MT} scenario, the primary star is in a tight binary with an orbital period at \ac{ZAMS} of $\sim 0.5-1.2~\mathrm{day}$ and is tidally locked when it overfills its Roche lobe on the main sequence. 
This strips the primary star of its envelope and causes it to become a fast-spinning \ac{WR} star without any significant expansion. 
Although this channel has been shown to lead to the formation of high-mass X-ray binaries with appreciable \ac{BH} spin, it will typically not lead to a merging \ac{BBH}~\citep{Belczynski2012,Qin2019,Neijssel2021}. 

Throughout this work, we have assumed that \ac{AM} transport in massive stars is highly efficient, leading to \acp{BH} formed from progenitors that do not undergo substantial tidal interaction to have near-zero spins. 
Although the numerical simulations of \ac{AM} transport in high-mass stars through Taylor--Spruit dynamo and the majority of spins in \ac{BBH} systems being small hint at efficient \ac{AM} transport, the astroseismic measurements that encode this process are typically for low-mass stars. 
Furthermore, efficient \ac{AM} transport may still lead to \ac{BH} spins that are low but nonnegligible ($a \simeq 0.1$, \citealt{Belczynski2020}), and processes such as supernova fallback may induce some spin on the resultant \ac{BH} even if \ac{AM} transport is highly efficient~\citep{Schroder2018}. 
As a simple approximation for less efficient \ac{AM} transport, we set a floor on the minimum spin of \acp{BH} at birth: 
\begin{equation}
    a_{\mathrm{1b,2b}} = 
    \begin{cases}
        a_{\mathrm{1b,2b}} & \,\, a_{\mathrm{1b,2b}} \geq a_\mathrm{min} \\
        a_\mathrm{min} & \,\, a_{\mathrm{1b,2b}} < a_\mathrm{min}
    \end{cases}
\end{equation}
where $a_\mathrm{min}$ is the minimum spin magnitude of quasi-isolated \acp{BH} at birth. 
With $a_\mathrm{min} = 0.2$, we find only minor changes in our main results regarding \ac{MRR} and spinning primaries. 
For example, with all second-born \acp{BH} satisfying the criteria $a_\mathrm{2b} \geq 0.2$, we find that the percentage of systems in our most optimistic \ac{MT} efficiency models with $\qb > 2$ and $a_\mathrm{2b} \geq 0.2$ to raise from $\sim \PercentageFaccOneQbGreaterTwoSpinGreaterZeropTwo$ to $\sim \PercentageFaccOneQbGreaterTwoSpinGreaterZeropTwoSpinFloor$. 
However, setting the minimum natal spin magnitude of \acp{BH} has a larger impact on the systems that do not proceed through a \ac{MRR}, with upwards of \PercentageQbLessZeropFiveSpinGreaterZeropTwoSpinFloorHigh of systems satisfying $\qb < 0.5$ and $a_\mathrm{1b} \geq 0.2$. 

Of course, isolated evolution is only one of many proposed formation scenarios for generating \ac{BBH} mergers, and the full population of \ac{BBH} mergers may be the result of a combination of formation pathways~\citep{Bouffanais2021,Wong2021a,Zevin2021}. 
Although we will not cover alternative potential formation pathways in detail here (see, e.g., \citealt{Mandel2022} for a review), we note that the formation of systems with spinning primaries and asymmetric masses may be the result of hierarchical mergers in dense stellar environments. 
However, the formation of GW190412-like systems still proves to be enigmatic in the standard hierarchical merger paradigm, as its primary \ac{BH} component is unlikely to be the product of a second-generation merger in an environment such as a classical globular cluster because its primary spin is lower than what one would expect for the merger product of two low-spinning \acp{BH}~\citep{GW190412}. 
A number of other formation scenarios have been proposed such as third-generation mergers~\citep{Rodriguez2020}, hierarchical systems~\citep{Hamers2020}, and formation in the disks of active galactic nuclei~\citep{McKernan2021,Tagawa2021}. 
If systems such as GW190412 and GW190517\_055101 result from subdominant formation scenarios, it may impact the astrophysical interpretation of the apparent anticorrelation between mass ratio and effective spin, as the exclusion of these events in such analyses (slightly) reduces the statistical significance of such a correlation~\citep{Callister2021a}, and different formation pathways do not necessarily need to share correlations in their mass and spin parameters. 

The population of \ac{BBH} mergers observed via \ac{GW} emission has already led to many astrophysical lessons. 
In addition to the properties of systems as a whole, the distribution of mass and spin across the two components of the merging binary is enlightening when considering formation scenarios and physical processes. 
As always, the hundreds of anticipated \ac{BBH} detections in the upcoming observing runs of the \ac{LVK} interferometer network will help to determine whether such asymmetric-mass, highly spinning systems are commonplace. 
Given the quantity and flexibility of the proposed formation scenarios for compact binary mergers, we stress the importance of determining the regions of parameter space that cannot be easily populated by astrophysical formation models in addition to regions that can. 
This will be equally as important for constraining the merger rates and uncertain physical processes that embed the myriad of potential formation pathways.

\acknowledgments
We thank Mario Spera for insightful discussion and a detailed read of this manuscript, Luke Kelley for providing processed Illustris-TNG data used in our population resampling, and Katie Breivik for useful conversations as well as pertinent developments to \texttt{COSMIC}. 
We also thank the anonymous referee whose comments improved this manuscript. 
Support for this work and M.Z. was provided by NASA through the NASA Hubble Fellowship grant HST-HF2-51474.001-A awarded by the Space Telescope Science Institute, which is operated by the Association of Universities for Research in Astronomy, Incorporated, under NASA contract NAS5-26555. 
S.S.B. is supported by a Swiss National Science Foundation professorship grant (project number PP00P2\_176868; PI Fragos)

\software{\texttt{Astropy}~\citep{TheAstropyCollaboration2013,TheAstropyCollaboration2018}; 
\texttt{iPython}~\citep{ipython}; 
\texttt{Matplotlib}~\citep{matplotlib}; 
\texttt{NumPy}~\citep{numpy,numpy2}; 
\texttt{Pandas}~\citep{pandas};
\texttt{SciPy}~\citep{scipy}.}

\appendix

\section{Incorporating Selection Effects}\label{app:selection_effects}

The observation of compact binary coalescences via \acp{GW} is prone to selection effects, most notably that more massive systems are more luminous and thus can be seen out to higher luminosity distances. 
Current ground-based detectors are most sensitive at frequencies of $\sim 100~\mathrm{Hz}$, and systems with redshifted total masses $\gtrsim 500~\Msun$ become unobservable since they merge at frequencies below the seismic floor~\citep[e.g.,][]{Mehta2022}. 
Mass ratios and spins also impact the observability of compact binary mergers, with the detectors being most sensitive to systems with near-equal masses and high, aligned spins. 
Selection effects must be accounted for in our population models to yield a fair comparison against the observed \ac{BBH} population. 

We use a semianalytic treatment to incorporate selection effects in our population models, which relies on precomputed detection probabilities over a grid of chirp masses $\mchirp = (m_1 m_2)^{3/5} / (m_1+m_2)^{1/5}$, mass ratios $\q = m_2/m_1$ with $m_2 \leq m_1$, and redshifts \z where $m_1$ and $m_2$ are the primary and secondary masses. 
We assume a three-detector network consisting of LIGO--Hanford, LIGO--Livingston, and Virgo operating at either \texttt{midhighlatelow} or \texttt{design} sensitivity~\citep{LVC_ObservingScenarios}. 
Extrinsic parameters are sampled over, and systems with network signal-to-noise ratio of $\rho_\mathrm{thresh} = 10$ are considered detectable~\citep{Nitz2020b,GWTC2}, such that the detection probability \pdet is given 
\begin{equation}
    \pdet = \frac{1}{N} \sum_j^N \mathcal{H} \left[ \left( \sum_i  \rho_i(\psi_j) \right)^{1/2} - \rho_\mathrm{thresh} \right]
\end{equation}
where $\psi_j$ are the sampled extrinsic parameters, $i$ defines the detector, and $\mathcal{H}$ is a Heaviside step function. 
Spin effects are ignored, as they have a minor effect on detection probabilities~\citep{Ng2018}. 
The relative weight of each system in the population thus becomes
\begin{equation}
    w = \pdet \frac{dV_\mathrm{c}}{dz} (1+z)^{-1}
\end{equation}
where $\frac{dV_\mathrm{c}}{dz}$ is the differential comoving volume at the merger redshift and $dt_\mathrm{src}/dt_\mathrm{obs}(z) = (1+z)^{-1}$ is the time dilation between the merger and detectors. 
Detection probabilities for each system are determined using a k-nearest neighbors regressor trained on our detection probability grid. 

\begin{figure*}[t]
\includegraphics[width=1.0\textwidth]{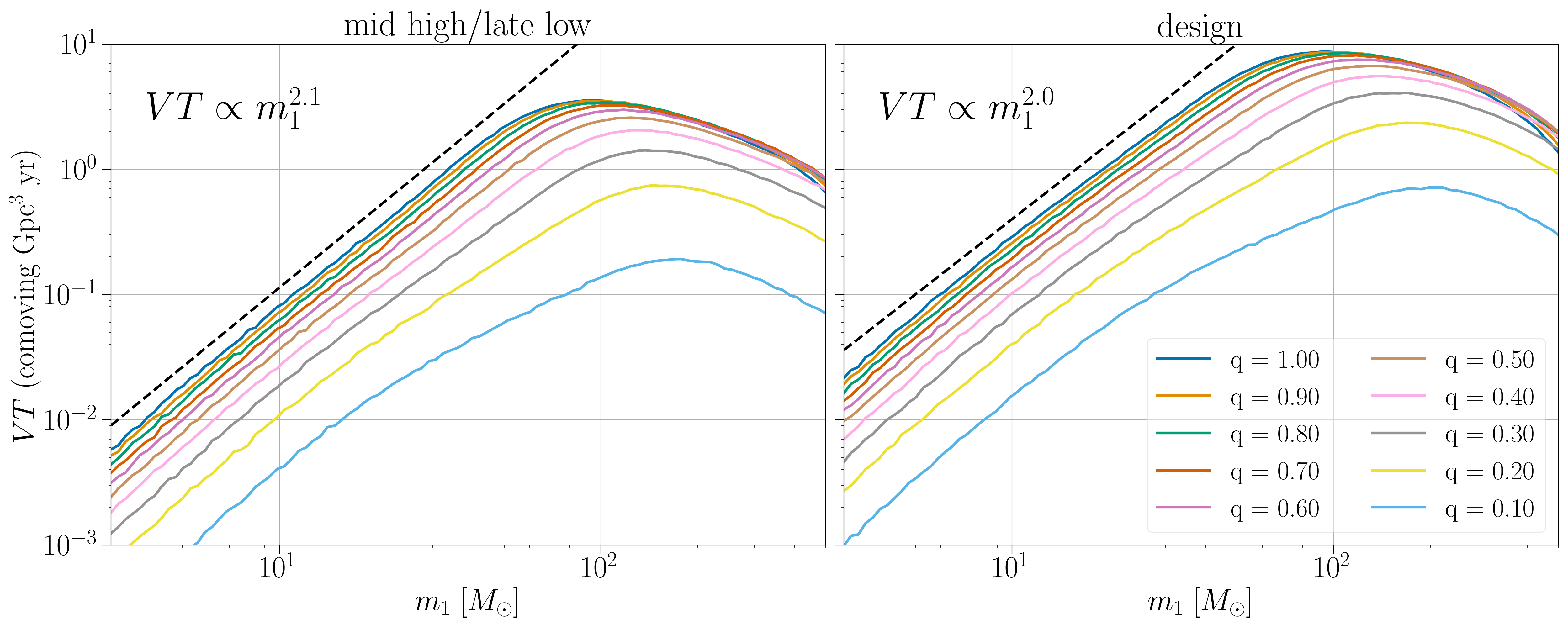}
\caption{Sensitive redshifted spacetime volume, $VT$, as a function of primary mass $m_1$ and mass ratio $q$ for a three-detector network consisting of LIGO-Hanford, LIGO-Livingston, and Virgo with fixed power spectral densities representative of the third observing run O3 (left) and at design sensitivity (right). 
Power spectral densities for all three interferometers are taken from \cite{LVC_ObservingScenarios}. 
Colors show $VT$ for mass ratios ranging from 10:1 to 1:1. 
Until the turnover in sensitivity at $m_1 \simeq 100~\Msun$, we find the scaling in $VT$ to be well approximated by a power law in $m_1$ with $VT \propto m_1^{2.1}$ for mid high/late low sensitivity and $VT \propto m_1^{2.0}$ for design sensitivity, similar to \cite{Fishbach2017a}. 
}
\label{fig:VT_grid}
\end{figure*}

Figure~\ref{fig:VT_grid} shows the sensitive comoving spacetime volume, $VT$, for which a given search is sensitive to a \ac{BBH} system with a particular primary mass and mass ratio, defined as 
\begin{equation}
    VT(m_1, q) = T \int \frac{dV_\mathrm{c}}{dz} \frac{1}{1+z} p_\mathrm{det}(m_1, q, z) \mathrm{d}z
\end{equation}
where $T$ is the observing time, which is fixed to 1 yr, and $p_\mathrm{det}(m_1, q, z)$ is the detection probability for a system merging at redshift $z$ with primary mass $m_1$ and mass ratio $q$~\citep[see, e.g.,][]{Fishbach2017a}. 
Our grid of precomputed detection probabilities is publicly available on Zenodo~\citep{Zenodo_selection_effects}, and the associated codebase for applying the selection effects is available on Github.\footnote{\url{https://github.com/michaelzevin/selection-effects}}

%%% FIGURE 8

\section{Varying the Onset of Unstable MT}\label{app:qcrit}

\begin{figure*}[ht]
\includegraphics[width=1.0\textwidth]{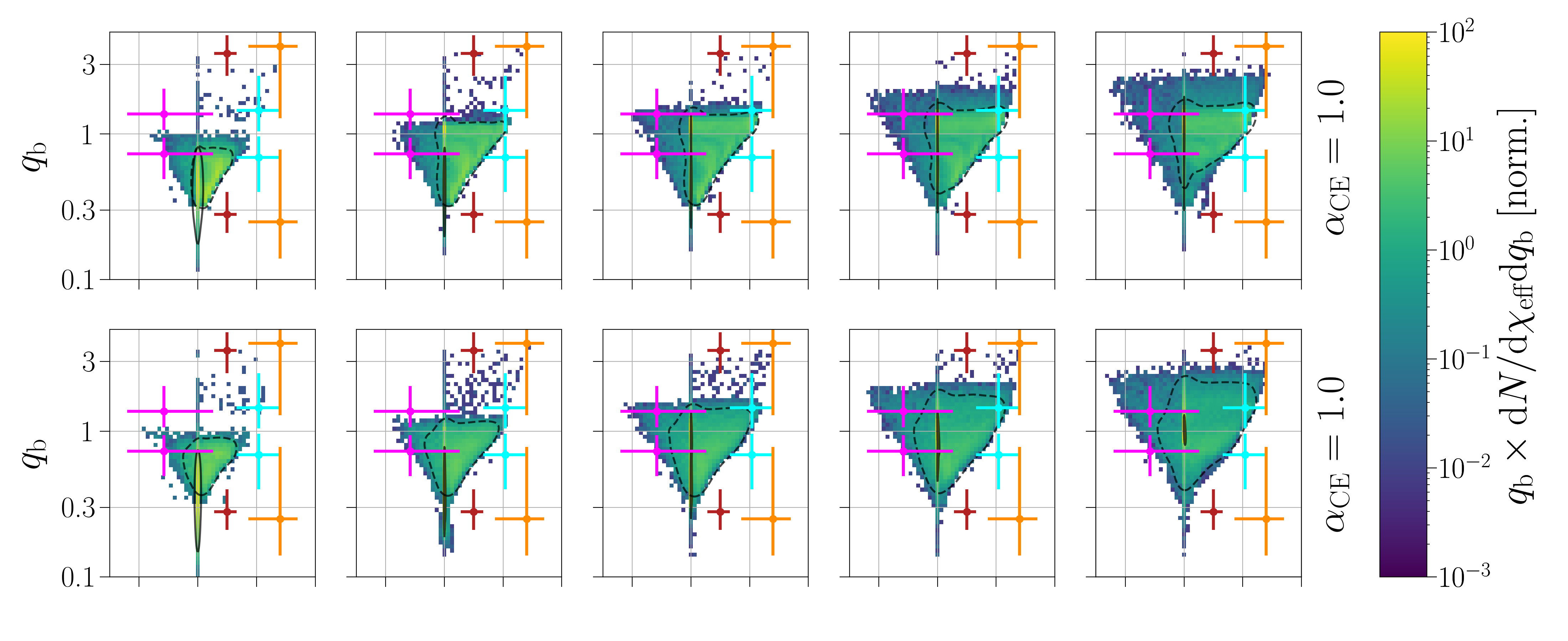}
\caption{Same as Figure~\ref{fig:q_chieff_eddfac1e0} except comparing the stellar-type-specific critical mass ratios of \cite{Neijssel2019} (N+19, used throughout the rest of this work) and \cite{Belczynski2008} (B+08). 
For the models shown here, the \ac{CE} efficiency and super-Eddington factor are $\alphaCE = 1.0$ and $\gammaEdd = 1.0$, respectively. 
}
\label{fig:qcrit_Startrack}
\end{figure*}

Throughout this work, we have assumed that the critical mass ratios that govern the onset of unstable \ac{MT} follow the prescription in \cite{Neijssel2019}. 
Here, we briefly discuss how key results change with an alternative choice of critical mass ratios by adopting the values from \cite{Belczynski2008}. 
Figure~\ref{fig:qcrit_Startrack} shows the birth mass ratio versus effective spin distribution across our five simulated accretion efficiency ($\facc$) values, for a single model variation of \ac{CE} efficiency ($\alphaCE = 1$) and super-Eddington accretion factor ($\gammaEdd = 1$). 
Overall, the mass ratio and effective spin distributions are similar between the two critical mass ratio models. 
The critical mass ratio parameterization of \cite{Belczynski2008} leads to slightly less systems proceeding through \ac{MRR} and tidal spin-up. 
Compared to the critical mass ratios of \cite{Neijssel2019}, at $\facc = 0.5$, the \cite{Belczynski2008} model decreases the percentage of systems with $\qb > 1$ from $\QbGreaterOneFaccZeropFiveAlphaCEOne$ to $\QbGreaterOneFaccZeropFiveAlphaCEOneBelczynski$. 
The fraction of systems that proceed through \ac{MRR} and have second-born spins of $a_\mathrm{2b} > 0.2$ decreases by about a factor of $2$.

%%% FIGURE 9

\pagebreak
\bibliography{library}{}
\bibliographystyle{aasjournal}

\end{document}

%% file: commands.tex
\definecolor{chmagenta}{rgb}{0.54, 0.17, 0.88}

\def\mchirp{\ensuremath{\mathcal{M}_\mathrm{c}}\xspace}
\def\chieff{\ensuremath{\chi_\mathrm{eff}}\xspace}
\def\q{\ensuremath{q}\xspace}
\def\z{\ensuremath{z}\xspace}

\def\Msun{\ensuremath{\mathit{M_\odot}}\xspace}

\def\pdet{\ensuremath{p_\mathrm{det}}\xspace}

\def\facc{\ensuremath{f_\mathrm{acc}}\xspace}
\def\gammaEdd{\ensuremath{\gamma_\mathrm{Edd}}\xspace}

\def\alphaCE{\ensuremath{\alpha_\mathrm{CE}}\xspace}
\def\qb{\ensuremath{\q_\mathrm{b}}\xspace}

%% file: values.tex
\def\UpperLimitQbGreaterTwoSpinGreaterZeropTwo{\ensuremath{1.5\%}\xspace}
\def\ApproxQbGreaterTwoSpinGreaterZeropTwo{\ensuremath{0.01\%}\xspace}

\def\PercentageSystemsMergingAfterTodayRange{\ensuremath{4\mbox{--}13\%}\xspace}
\def\PercentageSystemsPorbLessZeropOne{\ensuremath{0.2\mbox{--}2.9\%}\xspace}

\def\FirstBornSpinGreaterZeropZeroOneGammaEddOne{\ensuremath{0.9\%}\xspace}
\def\SMTPrecentageFaccZeropTwoFiveGammaEddOne{\ensuremath{8\%}\xspace}
\def\SMTPrecentageFaccOneGammaEddOne{\ensuremath{2\%}\xspace}
\def\FirstBornSpinGreaterZeropFiveFaccZeropFiveGammaEddOneEFive{\ensuremath{6\%}\xspace}

\def\SecondBornVanishingSpin{\ensuremath{22\%\mbox{--}52\%}\xspace}
\def\SecondBornSpinGreaterZeropFiveFaccZeropFiveAlphaCEOne{\ensuremath{31\%}\xspace}
\def\SecondBornSpinGreaterZeropFiveFaccZeropFiveAlphaCETwo{\ensuremath{18\%}\xspace}
\def\SecondBornSpinGreaterZeropFiveFaccZeropFiveAlphaCEZeropFive{\ensuremath{52\%}\xspace}
\def\SecondBornSpinGreaterZeropFiveFaccZeroAlphaCEOne{\ensuremath{18\%}\xspace}
\def\SecondBornSpinGreaterZeropFiveFaccOneAlphaCEOne{\ensuremath{33\%}\xspace}

\def\QbGreaterOneFaccZeroAlphaCEOne{\ensuremath{0.3\%}\xspace}
\def\QbGreaterOneFaccZeropFiveAlphaCEOne{\ensuremath{47\%}\xspace}
\def\QbGreaterOneFaccOneAlphaCEOne{\ensuremath{72\%}\xspace}

\def\QbNinetyNinePercentileFaccOne{\ensuremath{2.2}\xspace}
\def\QbHighExtreme{\ensuremath{3.7}\xspace}

\def\QbLowEndExtremeEstimate{\ensuremath{0.09}\xspace}
\def\QbOnePercentileFaccZeropFiveAlphaCEOneGammaEddOneEFive{\ensuremath{0.18}\xspace}
\def\QbOnePercentileFaccZeropFiveAlphaCEOneGammaEddOne{\ensuremath{0.33}\xspace}

\def\PercentageFaccOneQbGreaterTwoSpinGreaterZeropTwo{\ensuremath{0.6\%\mbox{--}1.5\%}\xspace}

\def\PercentageEddfacOneEFiveQbLessZeropFiveSpinGreaterZeropTwoLow{\ensuremath{0.5\%}\xspace}
\def\PercentageEddfacOneEFiveQbLessZeropFiveSpinGreaterZeropTwoHigh{\ensuremath{8\%}\xspace}
\def\PercentageGWNineteenZeroFourTwelveMassRatioSpinOptimisticMassInversion{\ensuremath{0.1\%}\xspace}
\def\PercentageGWNineteenZeroFourTwelveMassRatioSpinOptimisticSuperEddington{\ensuremath{8\%}\xspace}

\def\PercentageFaccOneQbGreaterTwoSpinGreaterZeropTwoSpinFloor{\ensuremath{0.7\%\mbox{--}1.9\%}\xspace}
\def\PercentageQbLessZeropFiveSpinGreaterZeropTwoSpinFloorHigh{\ensuremath{52\%}\xspace}

\def\SecondarySpinGreaterZeropFiveRange{\ensuremath{4\mbox{--}57\%}\xspace}
\def\MassInversionRange{\ensuremath{0.3\mbox{--}72\%}\xspace}
\def\MassInversionZeropTwoFiveLow{\ensuremath{18\%}\xspace}
\def\MassInversionSpinningRange{\ensuremath{10^{-5}\mbox{--}10^{-2}}\xspace}
\def\NoMassInversionSpinningRangeSuperEddington{\ensuremath{5 \times 10^{-3}\mbox{--}8 \times 10^{-2}}\xspace}

\def\QbGreaterOneFaccZeropFiveAlphaCEOneBelczynski{\ensuremath{22\%}\xspace}

\def\GWNineteenZeroFourTwelveChiEff{\ensuremath{0.25^{+0.08}_{-0.11}}\xspace}
\def\GWNineteenZeroFourTwelveMassRatio{\ensuremath{0.28^{+0.12}_{-0.07}}\xspace}
\def\GWNineteenZeroFourTwelvePrimarySpin{\ensuremath{0.44^{+0.16}_{-0.22}}\xspace}

\def\GWNineteenZeroFiveSeventeenChiEff{\ensuremath{0.52^{+0.19}_{-0.19}}\xspace}
\def\GWNineteenZeroFiveSeventeenMassRatio{\ensuremath{0.69^{+0.27}_{-0.29}}\xspace}
\def\GWNineteenZeroFiveSeventeenPrimarySpin{\ensuremath{0.86^{+0.13}_{-0.35}}\xspace}

\def\GWNineteenZeroFourZeroThreeMassRatio{\ensuremath{0.25^{+0.53}_{-0.11}}\xspace}
\def\GWNineteenZeroFourZeroThreePrimarySpin{\ensuremath{0.92^{+0.07}_{-0.22}}\xspace}

\def\GWNineteenElevenZeroNinePrimarySpin{\ensuremath{0.83^{+0.15}_{-0.58}}\xspace}

%% file: acronyms.tex
\acrodef{GW}{gravitational wave}
\acrodef{BH}{black hole}
\acrodef{BBH}{binary black hole}
\acrodef{NS}{black hole}
\acrodef{BNS}{binary neutron star}

\acrodef{LIGO}{Laser Interferometer Gravitational-wave Observatory}
\acrodef{LVK}{LIGO--Virgo--KAGRA}
\acrodef{O1}{first observing run}
\acrodef{O2}{second observing run}
\acrodef{O3}{third observing run}
\acrodef{O3a}{first half of the third observing run}

\acrodef{AM}{angular momentum}
\acrodef{CE}{common envelope}
\acrodef{ZAMS}{zero-age main sequence}
\acrodef{MT}{mass transfer}
\acrodef{RLO}{Roche-lobe overflow}
\acrodef{WR}{Wolf--Rayet}
\acrodef{MRR}{mass ratio reversal}

%% file: main.bbl
\begin{thebibliography}{}
\expandafter\ifx\csname natexlab\endcsname\relax\def\natexlab#1{#1}\fi
\providecommand{\url}[1]{\href{#1}{#1}}
\providecommand{\dodoi}[1]{doi:~\href{http://doi.org/#1}{\nolinkurl{#1}}}
\providecommand{\doeprint}[1]{\href{http://ascl.net/#1}{\nolinkurl{http://ascl.net/#1}}}
\providecommand{\doarXiv}[1]{\href{https://arxiv.org/abs/#1}{\nolinkurl{https://arxiv.org/abs/#1}}}

\bibitem[{Abbott {et~al.}(2018)Abbott, Abbott, Abbott, Abernathy, Acernese,
  Ackley, Adams, Adams, Addesso, Adhikari, Adya, Affeldt, Agathos, Agatsuma,
  Aggarwal, Aguiar, Aiello, Ain, Ajith, Akutsu, Allen, Allocca, Altin,
  Ananyeva, Anderson, Anderson, Ando, Appert, Arai, Araya, Araya, Areeda,
  Arnaud, Arun, Asada, Ascenzi, Ashton, Aso, Ast, Aston, Astone, Atsuta,
  Aufmuth, Aulbert, Avila-Alvarez, Awai, Babak, Bacon, Bader, Baiotti, Baker,
  Baldaccini, Ballardin, Ballmer, Barayoga, Barclay, Barish, Barker, Barone,
  Barr, Barsotti, Barsuglia, Barta, Bartlett, Barton, Bartos, Bassiri, Basti,
  Batch, Baune, Bavigadda, Bazzan, B{\'{e}}csy, Beer, Bejger, Belahcene,
  Belgin, Bell, Berger, Bergmann, Berry, Bersanetti, Bertolini, Betzwieser,
  Bhagwat, Bhandare, Bilenko, Billingsley, Billman, Birch, Birney, Birnholtz,
  Biscans, Bisht, Bitossi, Biwer, Bizouard, Blackburn, Blackman, Blair, Blair,
  Blair, Bloemen, Bock, Boer, Bogaert, Bohe, Bondu, Bonnand, Boom, Bork,
  Boschi, Bose, Bouffanais, Bozzi, Bradaschia, Brady, Braginsky, Branchesi,
  Brau, Briant, Brillet, Brinkmann, Brisson, Brockill, Broida, Brooks, Brown,
  Brown, Brown, Brunett, Buchanan, Buikema, Bulik, Bulten, Buonanno, Buskulic,
  Buy, Byer, Cabero, Cadonati, Cagnoli, Cahillane, {Calder{\'{o}}n Bustillo},
  Callister, Calloni, Camp, Cannon, Cao, Cao, Capano, Capocasa, Carbognani,
  Caride, {Casanueva Diaz}, Casentini, Caudill, Cavagli{\`{a}}, Cavalier,
  Cavalieri, Cella, Cepeda, {Cerboni Baiardi}, Cerretani, Cesarini, Chamberlin,
  Chan, Chao, Charlton, Chassande-Mottin, Cheeseboro, Chen, Chen, Cheng,
  Chincarini, Chiummo, Chmiel, Cho, Cho, Chow, Christensen, Chu, Chua, Chua,
  Chung, Ciani, Clara, Clark, Cleva, Cocchieri, Coccia, Cohadon, Colla,
  Collette, Cominsky, Constancio, Conti, Cooper, Corbitt, Cornish, Corsi,
  Cortese, Costa, Coughlin, Coughlin, Coulon, Countryman, Couvares, Covas,
  Cowan, Coward, Cowart, Coyne, Coyne, Creighton, Creighton, Cripe, Crowder,
  Cullen, Cumming, Cunningham, Cuoco, Canton, Danilishin, D'Antonio, Danzmann,
  Dasgupta, {Da Silva Costa}, Dattilo, Dave, Davier, Davies, Davis, Daw, Day,
  Day, De, DeBra, Debreczeni, Degallaix, {De Laurentis}, Del{\'{e}}glise, {Del
  Pozzo}, Denker, Dent, Dergachev, {De Rosa}, DeRosa, DeSalvo, Devine,
  Dhurandhar, D{\'{i}}az, Fiore, Giovanni, Girolamo, Lieto, Pace, Palma,
  Virgilio, Doctor, Doi, Dolique, Donovan, Dooley, Doravari, Dorrington,
  Douglas, {Dovale {\'{A}}lvarez}, Downes, Drago, Drever, Driggers, Du, Ducrot,
  Dwyer, Eda, Edo, Edwards, Effler, Eggenstein, Ehrens, Eichholz, Eikenberry,
  Eisenstein, Essick, Etienne, Etzel, Evans, Evans, Everett, Factourovich,
  Fafone, Fair, Fairhurst, Fan, Farinon, Farr, Farr, Fauchon-Jones, Favata,
  Fays, Fehrmann, Fejer, {Fern{\'{a}}ndez Galiana}, Ferrante, Ferreira,
  Ferrini, Fidecaro, Fiori, Fiorucci, Fisher, Flaminio, Fletcher, Fong,
  Forsyth, Fournier, Frasca, Frasconi, Frei, Freise, Frey, Frey, Fries,
  Fritschel, Frolov, Fujii, Fujimoto, Fulda, Fyffe, Gabbard, Gadre, Gaebel,
  Gair, Gammaitoni, Gaonkar, Garufi, Gaur, Gayathri, Gehrels, Gemme, Genin,
  Gennai, George, Gergely, Germain, Ghonge, Ghosh, Ghosh, Ghosh, Giaime,
  Giardina, Giazotto, Gill, Glaefke, Goetz, Goetz, Gondan, Gonz{\'{a}}lez,
  {Gonzalez Castro}, Gopakumar, Gorodetsky, Gossan, Gosselin, Gouaty, Grado,
  Graef, Granata, Grant, Gras, Gray, Greco, Green, Groot, Grote, Grunewald,
  Guidi, Guo, Gupta, Gupta, Gushwa, Gustafson, Gustafson, Hacker, Hagiwara,
  Hall, Hall, Hammond, Haney, Hanke, Hanks, Hanna, Hannam, Hanson, Hardwick,
  Harms, Harry, Harry, Hart, Hartman, Haster, Haughian, Hayama, Healy,
  Heidmann, Heintze, Heitmann, Hello, Hemming, Hendry, Heng, Hennig, Henry,
  Heptonstall, Heurs, Hild, Hirose, Hoak, Hofman, Holt, Holz, Hopkins, Hough,
  Houston, Howell, Hu, Huerta, Huet, Hughey, Husa, Huttner, Huynh-Dinh, Indik,
  Ingram, Inta, Ioka, Isa, Isac, Isi, Isogai, Itoh, Iyer, Izumi, Jacqmin, Jani,
  Jaranowski, Jawahar, Jim{\'{e}}nez-Forteza, Johnson, Jones, Jones, Jonker,
  Ju, Junker, Kagawa, Kajita, Kakizaki, Kalaghatgi, Kalogera, Kamiizumi, Kanda,
  Kandhasamy, Kanemura, Kaneyama, Kang, Kanner, Karki, Karvinen, Kasprzack,
  Kataoka, Katsavounidis, Katzman, Kaufer, Kaur, Kawabe, Kawai, Kawamura,
  K{\'{e}}f{\'{e}}lian, Keitel, Kelley, Kennedy, Key, Khalili, Khan, Khan,
  Khan, Khazanov, Kijbunchoo, Kim, Kim, Kim, Kim, Kim, Kim, Kimbrell, Kimura,
  King, King, Kirchhoff, Kissel, Klein, Kleybolte, Klimenko, Koch, Koehlenbeck,
  Kojima, Kokeyama, Koley, Komori, Kondrashov, Kontos, Korobko, Korth, Kotake,
  Kowalska, Kozak, Kr{\"{a}}mer, Kringel, Krishnan, Kr{\'{o}}lak, Kuehn, Kumar,
  Kumar, Kumar, Kuo, Kuroda, Kutynia, Kuwahara, Lackey, Landry, Lang, Lange,
  Lantz, Lanza, Lartaux-Vollard, Lasky, Laxen, Lazzarini, Lazzaro, Leaci,
  Leavey, Lebigot, Lee, Lee, Lee, Lee, Lee, Lehmann, Lenon, Leonardi, Leong,
  Leroy, Letendre, Levin, Li, Libson, Littenberg, Liu, Lockerbie, Lombardi,
  London, Lord, Lorenzini, Loriette, Lormand, Losurdo, Lough, Lousto, Lovelace,
  L{\"{u}}ck, Lundgren, Lynch, Ma, Macfoy, Machenschalk, MacInnis, Macleod,
  Maga{\~{n}}a-Sandoval, Majorana, Maksimovic, Malvezzi, Man, Mandic, Mangano,
  Mano, Mansell, Manske, Mantovani, Marchesoni, Marchio, Marion, M{\'{a}}rka,
  M{\'{a}}rka, Markosyan, Maros, Martelli, Martellini, Martin, Martynov, Mason,
  Masserot, Massinger, Masso-Reid, Mastrogiovanni, Matichard, Matone,
  Matsumoto, Matsushima, Mavalvala, Mazumder, McCarthy, McClelland, McCormick,
  McGrath, McGuire, McIntyre, McIver, McManus, McRae, McWilliams, Meacher,
  Meadors, Meidam, Melatos, Mendell, Mendoza-Gandara, Mercer, Merilh,
  Merzougui, Meshkov, Messenger, Messick, Metzdorff, Meyers, Mezzani, Miao,
  Michel, Michimura, Middleton, Mikhailov, Milano, Miller, Miller, Miller,
  Miller, Millhouse, Minenkov, Ming, Mirshekari, Mishra, Mitrofanov,
  Mitselmakher, Mittleman, Miyakawa, Miyamoto, Miyamoto, Miyoki, Moggi, Mohan,
  Mohapatra, Montani, Moore, Moore, Moraru, Moreno, Morii, Morisaki, Moriwaki,
  Morriss, Mours, Mow-Lowry, Mueller, Muir, Mukherjee, Mukherjee, Mukherjee,
  Mukund, Mullavey, Munch, Muniz, Murray, Mytidis, Nagano, Nakamura, Nakamura,
  Nakano, Nakano, Nakano, Nakao, Napier, Nardecchia, Narikawa, Naticchioni,
  Nelemans, Nelson, Neri, Nery, Neunzert, Newport, Newton, Nguyen, Ni, Nielsen,
  Nissanke, Nitz, Noack, Nocera, Nolting, Normandin, Nuttall, Oberling,
  Ochsner, Oelker, Ogin, Oh, Oh, Ohashi, Ohishi, Ohkawa, Ohme, Okutomi, Oliver,
  Ono, Ono, Oohara, Oppermann, Oram, O'Reilly, O'Shaughnessy, Ottaway,
  Overmier, Owen, Pace, Page, Pai, Pai, Palamos, Palashov, Palomba, Pal-Singh,
  Pan, Pankow, Pannarale, Pant, Paoletti, Paoli, Papa, Paris, Parker, Pascucci,
  Pasqualetti, Passaquieti, Passuello, Patricelli, Pearlstone, Pedraza,
  Pedurand, Pekowsky, Pele, {Pe{\~{n}}a Arellano}, Penn, Perez, Perreca, Perri,
  Pfeiffer, Phelps, Piccinni, Pichot, Piergiovanni, Pierro, Pillant, Pinard,
  Pinto, Pitkin, Poe, Poggiani, Popolizio, Post, Powell, Prasad, Pratt, Predoi,
  Prestegard, Prijatelj, Principe, Privitera, Prodi, Prokhorov, Puncken,
  Punturo, Puppo, P{\"{u}}rrer, Qi, Qin, Qiu, Quetschke, Quintero,
  Quitzow-James, Raab, Rabeling, Radkins, Raffai, Raja, Rajan, Rakhmanov,
  Rapagnani, Raymond, Razzano, Re, Read, Regimbau, Rei, Reid, Reitze, Rew,
  Reyes, Rhoades, Ricci, Riles, Rizzo, Robertson, Robie, Robinet, Rocchi,
  Rolland, Rollins, Roma, Romano, Romie, Rosi{\'{n}}ska, Rowan, R{\"{u}}diger,
  Ruggi, Ryan, Sachdev, Sadecki, Sadeghian, Sago, Saijo, Saito, Sakai,
  Sakellariadou, Salconi, Saleem, Salemi, Samajdar, Sammut, Sampson, Sanchez,
  Sandberg, Sanders, Sasaki, Sassolas, Sathyaprakash, Sato, Sato, Saulson,
  Sauter, Savage, Sawadsky, Schale, Scheuer, Schmidt, Schmidt, Schmidt,
  Schnabel, Schofield, Sch{\"{o}}nbeck, Schreiber, Schuette, Schutz, Schwalbe,
  Scott, Scott, Sekiguchi, Sekiguchi, Sellers, Sengupta, Sentenac, Sequino,
  Sergeev, Setyawati, Shaddock, Shaffer, Shahriar, Shapiro, Shawhan, Sheperd,
  Shibata, Shikano, Shimoda, Shoda, Shoemaker, Shoemaker, Siellez, Siemens,
  Sieniawska, Sigg, Silva, Singer, Singer, Singh, Singh, Singhal, Sintes,
  Slagmolen, Smith, Smith, Smith, Somiya, Son, Sorazu, Sorrentino, Souradeep,
  Spencer, Srivastava, Staley, Steinke, Steinlechner, Steinlechner, Steinmeyer,
  Stephens, Stevenson, Stone, Strain, Straniero, Stratta, Strigin, Sturani,
  Stuver, Sugimoto, Summerscales, Sun, Sunil, Sutton, Suzuki, Swinkels,
  Szczepa{\'{n}}czyk, Tacca, Tagoshi, Takada, Takahashi, Takahashi, Takamori,
  Talukder, Tanaka, Tanaka, Tanaka, Tanner, T{\'{a}}pai, Taracchini, Tatsumi,
  Taylor, Telada, Theeg, Thomas, Thomas, Thomas, Thorne, Thrane, Tippens,
  Tiwari, Tiwari, Tokmakov, Toland, Tomaru, Tomlinson, Tonelli, Tornasi,
  Torrie, T{\"{o}}yr{\"{a}}, Travasso, Traylor, Trifir{\`{o}}, Trinastic,
  Tringali, Trozzo, Tse, Tso, Tsubono, Tsuzuki, Turconi, Tuyenbayev, Uchiyama,
  Uehara, Ueki, Ueno, Ugolini, Unnikrishnan, Urban, Ushiba, Usman, Vahlbruch,
  Vajente, Valdes, van Bakel, van Beuzekom, van~den Brand, {Van Den Broeck},
  Vander-Hyde, van~der Schaaf, van Heijningen, van Putten, van Veggel, Vardaro,
  Varma, Vass, Vas{\'{u}}th, Vecchio, Vedovato, Veitch, Veitch, Venkateswara,
  Venugopalan, Verkindt, Vetrano, Vicer{\'{e}}, Viets, Vinciguerra, Vine,
  Vinet, Vitale, Vo, Vocca, Vorvick, Voss, Vousden, Vyatchanin, Wade, Wade,
  Wade, Wakamatsu, Walker, Wallace, Walsh, Wang, Wang, Wang, Wang, Ward,
  Warner, Was, Watchi, Weaver, Wei, Weinert, Weinstein, Weiss, Wen, We{\ss}els,
  Westphal, Wette, Whelan, Whiting, Whittle, Williams, Williams, Williamson,
  Willis, Willke, Wimmer, Winkler, Wipf, Wittel, Woan, Woehler, Worden, Wright,
  Wu, Wu, Yam, Yamamoto, Yamamoto, Yamamoto, Yancey, Yano, Yap, Yokoyama,
  Yokozawa, Yoon, Yu, Yu, Yuzurihara, Yvert, Zadro{\.{z}}ny, Zangrando,
  Zanolin, Zeidler, Zendri, Zevin, Zhang, Zhang, Zhang, Zhang, Zhao, Zhou,
  Zhou, Zhu, Zhu, Zucker, \& Zweizig}]{LVC_ObservingScenarios}
Abbott, B.~P., Abbott, R., Abbott, T.~D., {et~al.} 2018, Living Reviews in
  Relativity, 21, 3, \dodoi{10.1007/s41114-018-0012-9}

\bibitem[{Abbott {et~al.}(2020)Abbott, Abbott, Abraham, Acernese, Ackley,
  Adams, Adhikari, Adya, Affeldt, Agathos, Agatsuma, Aggarwal, Aguiar, Aich,
  Aiello, Ain, Ajith, Akcay, Allen, Allocca, Altin, Amato, Anand, Ananyeva,
  Anderson, Anderson, Angelova, Ansoldi, Antier, Appert, Arai, Araya, Areeda,
  Ar{\`{e}}ne, Arnaud, Aronson, Arun, Asali, Ascenzi, Ashton, Aston, Astone,
  Aubin, Aufmuth, Aultoneal, Austin, Avendano, Babak, Bacon, Badaracco, Bader,
  Bae, Baer, Baird, Baldaccini, Ballardin, Ballmer, Bals, Balsamo, Baltus,
  Banagiri, Bankar, Bankar, Barayoga, Barbieri, Barish, Barker, Barkett,
  Barneo, Barone, Barr, Barsotti, Barsuglia, Barta, Bartlett, Bartos, Bassiri,
  Basti, Bawaj, Bayley, Bazzan, B{\'{e}}csy, Bejger, Belahcene, Bell, Beniwal,
  Benjamin, Benkel, Bentley, Bergamin, Berger, Bergmann, Bernuzzi, Berry,
  Bersanetti, Bertolini, Betzwieser, Bhandare, Bhandari, Bidler, Biggs,
  Bilenko, Billingsley, Birney, Birnholtz, Biscans, Bischi, Biscoveanu, Bisht,
  Bissenbayeva, Bitossi, Bizouard, Blackburn, Blackman, Blair, Blair, Blair,
  Bobba, Bode, Boer, Boetzel, Bogaert, Bondu, Bonilla, Bonnand, Booker, Boom,
  Bork, Boschi, Bose, Bossilkov, Bosveld, Bouffanais, Bozzi, Bradaschia, Brady,
  Bramley, Branchesi, Brau, Breschi, Briant, Briggs, Brighenti, Brillet,
  Brinkmann, Brito, Brockill, Brooks, Brooks, Brown, Brunett, Bruno, Bruntz,
  Buikema, Bulik, Bulten, Buonanno, Buskulic, Byer, Cabero, Cadonati, Cagnoli,
  Cahillane, {Calder{\'{o}}n Bustillo}, Callaghan, Callister, Calloni, Camp,
  Canepa, Cannon, Cao, Cao, Carapella, Carbognani, Caride, Carney, Carullo,
  {Casanueva Diaz}, Casentini, Casta{\~{n}}eda, Caudill, Cavagli{\`{a}},
  Cavalier, Cavalieri, Cella, Cerd{\'{a}}-Dur{\'{a}}n, Cesarini, Chaibi,
  Chakravarti, Chan, Chan, Chao, Charlton, Chase, Chassande-Mottin, Chatterjee,
  Chaturvedi, Chatziioannou, Chen, Chen, Chen, Cheng, Cheong, Chia, Chiadini,
  Chierici, Chincarini, Chiummo, Cho, Cho, Cho, Christensen, Chu, Chua, Chung,
  Chung, Ciani, Ciecielag, Cie{\'{s}}lar, Ciobanu, Ciolfi, Cipriano, Cirone,
  Clara, Clark, Clearwater, Clesse, Cleva, Coccia, Cohadon, Cohen, Colleoni,
  Collette, Collins, Colpi, Constancio, Conti, Cooper, Corban, Corbitt,
  Cordero-Carri{\'{o}}n, Corezzi, Corley, Cornish, Corre, Corsi, Cortese,
  Costa, Cotesta, Coughlin, Coughlin, Coulon, Countryman, Couvares, Covas,
  Coward, Cowart, Coyne, Coyne, Creighton, Creighton, Cripe, Croquette,
  Crowder, Cudell, Cullen, Cumming, Cummings, Cunningham, Cuoco, Curylo, {Dal
  Canton}, D{\'{a}}lya, Dana, Daneshgaran-Bajastani, D'Angelo, Danilishin,
  D'Antonio, Danzmann, Darsow-Fromm, Dasgupta, Datrier, Dattilo, Dave, Davier,
  Davies, Davis, Daw, Debra, Deenadayalan, Degallaix, {De Laurentis},
  Del{\'{e}}glise, Delfavero, {De Lillo}, {Del Pozzo}, Demarchi, D'Emilio,
  Demos, Dent, {De Pietri}, {De Rosa}, {De Rossi}, Desalvo, {De Varona},
  Dhurandhar, Di{\'{a}}z, Diaz-Ortiz, Dietrich, {Di Fiore}, {Di Fronzo}, {Di
  Giorgio}, {Di Giovanni}, {Di Giovanni}, {Di Girolamo}, {Di Lieto}, Ding, {Di
  Pace}, {Di Palma}, {Di Renzo}, Divakarla, Dmitriev, Doctor, Donovan, Dooley,
  Doravari, Dorrington, Downes, Drago, Driggers, Du, Ducoin, Dupej, Durante,
  D'Urso, Dwyer, Easter, Eddolls, Edelman, Edo, Edy, Effler, Ehrens, Eichholz,
  Eikenberry, Eisenmann, Eisenstein, Ejlli, Errico, Essick, Estelles, Estevez,
  Etienne, Etzel, Evans, Evans, Ewing, Fafone, Fairhurst, Fan, Farinon, Farr,
  Farr, Fauchon-Jones, Favata, Fays, Fazio, Feicht, Fejer, Feng, Fenyvesi,
  Ferguson, Fernandez-Galiana, Ferrante, Ferreira, Ferreira, Fidecaro, Fiori,
  Fiorucci, Fishbach, Fisher, Fittipaldi, Fitz-Axen, Fiumara, Flaminio, Floden,
  Flynn, Fong, Font, Forsyth, Fournier, Frasca, Frasconi, Frei, Freise, Frey,
  Frey, Fritschel, Frolov, Fronz{\`{e}}, Fulda, Fyffe, Gabbard, Gadre, Gaebel,
  Gair, Galaudage, Ganapathy, Ganguly, Gaonkar, Garci{\'{a}}-Quir{\'{o}}s,
  Garufi, Gateley, Gaudio, Gayathri, Gemme, Genin, Gennai, George, George,
  Gergely, Ghonge, Ghosh, Ghosh, Ghosh, Giacomazzo, Giaime, Giardina, Gibson,
  Gier, Gill, Glanzer, Gniesmer, Godwin, Goetz, Goetz, Gohlke, Goncharov,
  Gonz{\'{a}}lez, Gopakumar, Gossan, Gosselin, Gouaty, Grace, Grado, Granata,
  Grant, Gras, Grassia, Gray, Gray, Greco, Green, Green, Gretarsson, Griggs,
  Grignani, Grimaldi, Grimm, Grote, Grunewald, Gruning, Guidi, Guimaraes,
  Guix{\'{e}}, Gulati, Guo, Gupta, Gupta, Gupta, Gustafson, Gustafson, Haegel,
  Halim, Hall, Hamilton, Hammond, Haney, Hanke, Hanks, Hanna, Hannam,
  Hannuksela, Hansen, Hanson, Harder, Hardwick, Haris, Harms, Harry, Harry,
  Hasskew, Haster, Haughian, Hayes, Healy, Heidmann, Heintze, Heinze, Heitmann,
  Hellman, Hello, Hemming, Hendry, Heng, Hennes, Hennig, Heurs, Hild, Hinderer,
  Hoback, Hochheim, Hofgard, Hofman, Holgado, Holland, Holt, Holz, Hopkins,
  Horst, Hough, Howell, Hoy, Huang, H{\"{u}}bner, Huerta, Huet, Hughey, Hui,
  Husa, Huttner, Huxford, Huynh-Dinh, Idzkowski, Iess, Inchauspe, Ingram,
  Intini, Isac, Isi, Iyer, Jacqmin, Jadhav, Jadhav, James, Jani, Janthalur,
  Jaranowski, Jariwala, Jaume, Jenkins, Jiang, Johns, Johnson-Mcdaniel, Jones,
  Jones, Jones, Jones, Jones, Jonker, Ju, Junker, Kalaghatgi, Kalogera, Kamai,
  Kandhasamy, Kang, Kanner, Kapadia, Karki, Kashyap, Kasprzack, Kastaun,
  Katsanevas, Katsavounidis, Katzman, Kaufer, Kawabe, K{\'{e}}f{\'{e}}lian,
  Keitel, Keivani, Kennedy, Key, Khadka, Khalili, Khan, Khan, Khan, Khazanov,
  Khetan, Khursheed, Kijbunchoo, Kim, Kim, Kim, Kim, Kim, Kim, Kim, Kimball,
  King, Kinley-Hanlon, Kirchhoff, Kissel, Kleybolte, Klimenko, Knowles,
  Knyazev, Koch, Koehlenbeck, Koekoek, Koley, Kondrashov, Kontos, Koper,
  Korobko, Korth, Kovalam, Kozak, Kringel, Krishnendu, Kr{\'{o}}lak, Krupinski,
  Kuehn, Kumar, Kumar, Kumar, Kumar, Kumar, Kuo, Kutynia, Lackey, Laghi,
  Lalande, Lam, Lamberts, Landry, Lane, Lang, Lange, Lantz, Lanza, {La Rosa},
  Lartaux-Vollard, Lasky, Laxen, Lazzarini, Lazzaro, Leaci, Leavey, Lecoeuche,
  Lee, Lee, Lee, Lee, Lee, Lehmann, Leroy, Letendre, Levin, Li, Li, Li, Li, Li,
  Linde, Linker, Linley, Littenberg, Liu, Liu, Llorens-Monteagudo, Lo,
  Lockwood, London, Longo, Lorenzini, Loriette, Lormand, Losurdo, Lough,
  Lousto, Lovelace, L{\"{u}}ck, Lumaca, Lundgren, Ma, MacAs, MacFoy, MacInnis,
  MacLeod, MacMillan, MacQuet, {Magan{\~{a}} Hernandez}, Magan{\~{a}}-Sandoval,
  Magee, Majorana, Maksimovic, Malik, Man, Mandic, Mangano, Mansell, Manske,
  Mantovani, Mapelli, Marchesoni, Marion, M{\'{a}}rka, M{\'{a}}rka, Markakis,
  Markosyan, Markowitz, Maros, Marquina, Marsat, Martelli, Martin, Martin,
  Martinez, Martynov, Masalehdan, Mason, Massera, Masserot, Massinger,
  Masso-Reid, Mastrogiovanni, Matas, Matichard, Mavalvala, Maynard, McCann,
  McCarthy, McClelland, McCormick, McCuller, McGuire, McIsaac, McIver, McManus,
  McRae, McWilliams, Meacher, Meadors, Mehmet, Mehta, {Mejuto Villa}, Melatos,
  Mendell, Mercer, Mereni, Merfeld, Merilh, Merritt, Merzougui, Meshkov,
  Messenger, Messick, Metzdorff, Meyers, Meylahn, Mhaske, Miani, Miao,
  Michaloliakos, Michel, Middleton, Milano, Miller, Miller, Millhouse, Mills,
  Milotti, Milovich-Goff, Minazzoli, Minenkov, Mishkin, Mishra, Mistry, Mitra,
  Mitrofanov, Mitselmakher, Mittleman, Mo, Mogushi, Mohapatra, Mohite,
  Molina-Ruiz, Mondin, Montani, Moore, Moraru, Morawski, Moreno, Morisaki,
  Mours, Mow-Lowry, Mozzon, Muciaccia, Mukherjee, Mukherjee, Mukherjee,
  Mukherjee, Mukund, Mullavey, Munch, Mu{\~{n}}iz, Murray, Nagar, Nardecchia,
  Naticchioni, Nayak, Neil, Neilson, Nelemans, Nelson, Nery, Neunzert, Ng, Ng,
  Nguyen, Nguyen, Nichols, Nichols, Nissanke, Nocera, Noh, North, Nothard,
  Nuttall, Oberling, O'Brien, Oganesyan, Ogin, Oh, Oh, Ohme, Ohta, Okada,
  Oliver, Olivetto, Oppermann, Oram, O'Reilly, Ormiston, Ortega, O'Shaughnessy,
  Ossokine, Osthelder, Ottaway, Overmier, Owen, Pace, Pagano, Page, Pagliaroli,
  Pai, Pai, Palamos, Palashov, Palomba, Pan, Panda, Pang, Pankow, Pannarale,
  Pant, Paoletti, Paoli, Parida, Parker, Pascucci, Pasqualetti, Passaquieti,
  Passuello, Patricelli, Payne, Pearlstone, Pechsiri, Pedersen, Pedraza, Pele,
  Penn, Perego, Perez, P{\'{e}}rigois, Perreca, Perri{\`{e}}s, Petermann,
  Pfeiffer, Phelps, Phukon, Piccinni, Pichot, Piendibene, Piergiovanni, Pierro,
  Pillant, Pinard, Pinto, Piotrzkowski, Pirello, Pitkin, Plastino, Poggiani,
  Pong, Ponrathnam, Popolizio, Porter, Powell, Prajapati, Prasai, Prasanna,
  Pratten, Prestegard, Principe, Prodi, Prokhorov, Punturo, Puppo,
  P{\"{u}}rrer, Qi, Quetschke, Quinonez, Raab, Raaijmakers, Radkins, Radulesco,
  Raffai, Rafferty, Raja, Rajan, Rajbhandari, Rakhmanov, Ramirez, Ramos-Buades,
  Rana, Rao, Rapagnani, Raymond, Razzano, Read, Regimbau, Rei, Reid, Reitze,
  Rettegno, Ricci, Richardson, Richardson, Ricker, Riemenschneider, Riles,
  Rizzo, Robertson, Robinet, Rocchi, Rodriguez-Soto, Rolland, Rollins, Roma,
  Romanelli, Romano, Romel, Romero-Shaw, Romie, Rose, Rose, Rose,
  Rosi{\'{n}}ska, Rosofsky, Ross, Rowan, Rowlinson, Roy, Roy, Roy, Ruggi,
  Rutins, Ryan, Sachdev, Sadecki, Sakellariadou, Salafia, Salconi, Saleem,
  Samajdar, Sanchez, Sanchez, Sanchis-Gual, Sanders, Santiago, Santos, Sarin,
  Sassolas, Sathyaprakash, Sauter, Savage, Savant, Sawant, Sayah, Schaetzl,
  Schale, Scheel, Scheuer, Schmidt, Schnabel, Schofield, Sch{\"{o}}nbeck,
  Schreiber, Schulte, Schutz, Schwarm, Schwartz, Scott, Scott, Seidel, Sellers,
  Sengupta, Sennett, Sentenac, Sequino, Sergeev, Setyawati, Shaddock, Shaffer,
  Shahriar, Sharifi, Sharma, Sharma, Shawhan, Shen, Shikauchi, Shink,
  Shoemaker, Shoemaker, Shukla, Shyamsundar, Siellez, Sieniawska, Sigg, Singer,
  Singh, Singh, Singha, Singhal, Sintes, Sipala, Skliris, Slagmolen,
  Slaven-Blair, Smetana, Smith, Smith, Somala, Son, Soni, Sorazu, Sordini,
  Sorrentino, Souradeep, Sowell, Spencer, Spera, Srivastava, Srivastava,
  Staats, Stachie, Standke, Steer, Steinke, Steinlechner, Steinlechner,
  Steinmeyer, Stevenson, Stocks, Stops, Stover, Strain, Stratta, Strunk,
  Sturani, Stuver, Sudhagar, Sudhir, Summerscales, Sun, Sunil, Sur, Suresh,
  Sutton, Swinkels, Szczepa{\'{n}}czyk, Tacca, Tait, Talbot, Tanasijczuk,
  Tanner, Tao, T{\'{a}}pai, Tapia, {Tapia San Martin}, Tasson, Taylor, Tenorio,
  Terkowski, Thirugnanasambandam, Thomas, Thomas, Thompson, Thondapu, Thorne,
  Thrane, Tinsman, Saravanan, Tiwari, Tiwari, Tiwari, Toland, Tonelli, Tornasi,
  Torres-Forn{\'{e}}, Torrie, {Tosta E Melo}, To{\"{y}}r{\"{a}}, Trail,
  Travasso, Traylor, Tringali, Tripathee, Trovato, Trudeau, Tsang, Tse, Tso,
  Tsukada, Tsuna, Tsutsui, Turconi, Ubhi, Udall, Ueno, Ugolini, Unnikrishnan,
  Urban, Usman, Utina, Vahlbruch, Vajente, Valdes, Valentini, {Van Bakel}, {Van
  Beuzekom}, {Van Den Brand}, {Van Den Broeck}, Vander-Hyde, {Van Der Schaaf},
  {Van Heijningen}, {Van Veggel}, Vardaro, Varma, Vass, Vas{\'{u}}th, Vecchio,
  Vedovato, Veitch, Veitch, Venkateswara, Venugopalan, Verkindt, Veske,
  Vetrano, Vicer{\'{e}}, Viets, Vinciguerra, Vine, Vinet, Vitale, Vivanco, Vo,
  Vocca, Vorvick, Vyatchanin, Wade, Wade, Wade, Walet, Walker, Wallace,
  Wallace, Walsh, Wang, Wang, Wang, Ward, Warden, Warner, Was, Watchi, Weaver,
  Wei, Weinert, Weinstein, Weiss, Wellmann, Wen, We{\ss}els, Westhouse, Wette,
  Whelan, Whiting, Whittle, Wilken, Williams, Willis, Willke, Winkler, Wipf,
  Wittel, Woan, Woehler, Wofford, Wong, Wright, Wu, Wysocki, Xiao, Yamamoto,
  Yang, Yang, Yang, Yap, Yazback, Yeeles, Yu, Yu, Yuen, Zadrozny, Zadrozny,
  Zanolin, Zelenova, Zendri, Zevin, Zhang, Zhang, Zhang, Zhao, Zhao, Zhou,
  Zhou, Zhu, Zimmerman, Zucker, \& Zweizig}]{GW190412}
Abbott, R., Abbott, T.~D., Abraham, S., {et~al.} 2020, Physical Review D, 102,
  43015, \dodoi{10.1103/PhysRevD.102.043015}

\bibitem[{Abbott {et~al.}(2021{\natexlab{a}})Abbott, Abbott, Abraham, Acernese,
  Ackley, Adams, Adams, Adhikari, Adya, Affeldt, Agathos, Agatsuma, Aggarwal,
  Aguiar, Aiello, Ain, Ajith, Allen, Allocca, Altin, Amato, Anand, Ananyeva,
  Anderson, Anderson, Angelova, Ansoldi, Antelis, Antier, Appert, Arai, Araya,
  Areeda, Ar{\`{e}}ne, Arnaud, Aronson, Arun, Asali, Ascenzi, Ashton, Aston,
  Astone, Aubin, Aufmuth, AultONeal, Austin, Avendano, Babak, Badaracco, Bader,
  Bae, Baer, Bagnasco, Baird, Ball, Ballardin, Ballmer, Bals, Balsamo, Baltus,
  Banagiri, Bankar, Bankar, Barayoga, Barbieri, Barish, Barker, Barneo, Barnum,
  Barone, Barr, Barsotti, Barsuglia, Barta, Bartlett, Bartos, Bassiri, Basti,
  Bawaj, Bayley, Bazzan, Becher, B{\'{e}}csy, Bedakihale, Bejger, Belahcene,
  Beniwal, Benjamin, Bennett, Bentley, Bergamin, Berger, Bergmann, Bernuzzi,
  Berry, Bersanetti, Bertolini, Betzwieser, Bhandare, Bhandari, Bhattacharjee,
  Bidler, Bilenko, Billingsley, Birney, Birnholtz, Biscans, Bischi, Biscoveanu,
  Bisht, Bitossi, Bizouard, Blackburn, Blackman, Blair, Blair, Blair, Blanch,
  Bobba, Bode, Boer, Boetzel, Bogaert, Boldrini, Bondu, Bonilla, Bonnand,
  Booker, Boom, Bork, Boschi, Bose, Bossilkov, Boudart, Bouffanais, Bozzi,
  Bradaschia, Brady, Bramley, Branchesi, Brau, Breschi, Briant, Briggs,
  Brighenti, Brillet, Brinkmann, Brockill, Brooks, Brooks, Brown, Brunett,
  Bruno, Bruntz, Buikema, Bulik, Bulten, Buonanno, Buscicchio, Buskulic, Byer,
  Cabero, Cadonati, Caesar, Cagnoli, Cahillane, {Calder{\'{o}}n Bustillo},
  Callaghan, Callister, Calloni, Camp, Canepa, Cannon, Cao, Cao, Carapella,
  Carbognani, Carney, Carpinelli, Carullo, Carver, {Casanueva Diaz}, Casentini,
  Caudill, Cavagli{\`{a}}, Cavalier, Cavalieri, Cella, Cerd{\'{a}}-Dur{\'{a}}n,
  Cesarini, Chaibi, Chakravarti, Chan, Chan, Chandra, Chanial, Chao, Charlton,
  Chase, Chassande-Mottin, Chatterjee, Chattopadhyay, Chaturvedi,
  Chatziioannou, Chen, Chen, Chen, Chen, Cheng, Cheong, Chia, Chiadini,
  Chierici, Chincarini, Chiummo, Cho, Cho, Cho, Choate, Christensen, Chu, Chua,
  Chung, Chung, Ciani, Ciecielag, Cie{\'{s}}lar, Cifaldi, Ciobanu, Ciolfi,
  Cipriano, Cirone, Clara, Clark, Clark, Clarke, Clearwater, Clesse, Cleva,
  Coccia, Cohadon, Cohen, Colleoni, Collette, Collins, Colpi, Constancio,
  Conti, Cooper, Corban, Corbitt, Cordero-Carri{\'{o}}n, Corezzi, Corley,
  Cornish, Corre, Corsi, Cortese, Costa, Cotesta, Coughlin, Coughlin, Coulon,
  Countryman, Couvares, Covas, Coward, Cowart, Coyne, Coyne, Creighton,
  Creighton, Croquette, Crowder, Cudell, Cullen, Cumming, Cummings, Cunningham,
  Cuoco, Curylo, {Dal Canton}, D{\'{a}}lya, Dana, DaneshgaranBajastani,
  D'Angelo, Danilishin, D'Antonio, Danzmann, Darsow-Fromm, Dasgupta, Datrier,
  Dattilo, Dave, Davier, Davies, Davis, Daw, Dean, DeBra, Deenadayalan,
  Degallaix, {De Laurentis}, Del{\'{e}}glise, {Del Favero}, {De Lillo}, {De
  Lillo}, {Del Pozzo}, DeMarchi, {De Matteis}, D'Emilio, Demos, Denker, Dent,
  Depasse, {De Pietri}, {De Rosa}, {De Rossi}, DeSalvo, de~Varona, Dhurandhar,
  D{\'{i}}az, Diaz-Ortiz, Didio, Dietrich, {Di Fiore}, DiFronzo, {Di Giorgio},
  {Di Giovanni}, {Di Giovanni}, {Di Girolamo}, {Di Lieto}, Ding, {Di Pace}, {Di
  Palma}, {Di Renzo}, Divakarla, Dmitriev, Doctor, D'Onofrio, Donovan, Dooley,
  Doravari, Dorrington, Downes, Drago, Driggers, Du, Ducoin, Dupej, Durante,
  D'Urso, Duverne, Dwyer, Easter, Eddolls, Edelman, Edo, Edy, Effler, Eichholz,
  Eikenberry, Eisenmann, Eisenstein, Ejlli, Errico, Essick, Estell{\'{e}}s,
  Estevez, Etienne, Etzel, Evans, Evans, Ewing, Fafone, Fair, Fairhurst, Fan,
  Farah, Farinon, Farr, Farr, Fauchon-Jones, Favata, Fays, Fazio, Feicht,
  Fejer, Feng, Fenyvesi, Ferguson, Fernandez-Galiana, Ferrante, Ferreira,
  Fidecaro, Figura, Fiori, Fiorucci, Fishbach, Fisher, Fishner, Fittipaldi,
  Fitz-Axen, Fiumara, Flaminio, Floden, Flynn, Fong, Font, Forsyth, Fournier,
  Frasca, Frasconi, Frei, Freise, Frey, Frey, Fritschel, Frolov, Fronz{\'{e}},
  Fulda, Fyffe, Gabbard, Gadre, Gaebel, Gair, Gais, Galaudage, Gamba,
  Ganapathy, Ganguly, Gaonkar, Garaventa, Garc{\'{i}}a-Quir{\'{o}}s, Garufi,
  Gateley, Gaudio, Gayathri, Gemme, Gennai, George, George, Gergely, Ghonge,
  Ghosh, Ghosh, Ghosh, Giacomazzo, Giacoppo, Giaime, Giardina, Gibson, Gier,
  Gill, Giri, Glanzer, Gleckl, Godwin, Goetz, Goetz, Gohlke, Goncharov,
  Gonz{\'{a}}lez, Gopakumar, Gossan, Gosselin, Gouaty, Grace, Grado, Granata,
  Granata, Grant, Gras, Grassia, Gray, Gray, Greco, Green, Green, Gretarsson,
  Griggs, Grignani, Grimaldi, Grimes, Grimm, Grote, Grunewald, Gruning,
  Guerrero, Guidi, Guimaraes, Guix{\'{e}}, Gulati, Guo, Gupta, Gupta, Gupta,
  Gustafson, Gustafson, Guzman, Haegel, Halim, Hall, Hamilton, Hammond, Haney,
  Hanke, Hanks, Hanna, Hannuksela, Hannuksela, Hansen, Hansen, Hanson, Harder,
  Hardwick, Haris, Harms, Harry, Harry, Hartwig, Hasskew, Haster, Haughian,
  Hayes, Healy, Heidmann, Heintze, Heinze, Heinzel, Heitmann, Hellman, Hello,
  Helmling-Cornell, Hemming, Hendry, Heng, Hennes, Hennig, Hennig, {Hernandez
  Vivanco}, Heurs, Hild, Hill, Hines, Hochheim, Hofgard, Hofman, Hohmann,
  Holgado, Holland, Hollows, Holmes, Holt, Holz, Hopkins, Horst, Hough, Howell,
  Hoy, Hoyland, Huang, H{\"{u}}bner, Huddart, Huerta, Hughey, Hui, Husa,
  Huttner, Hutzler, Huxford, Huynh-Dinh, Idzkowski, Iess, Imperato, Inchauspe,
  Ingram, Intini, Isi, Iyer, JaberianHamedan, Jacqmin, Jadhav, Jadhav, James,
  Jani, Janssens, Janthalur, Jaranowski, Jariwala, Jaume, Jenkins, Jeunon,
  Jiang, Johns, Jones, Jones, Jones, Jones, Jones, Jonker, Ju, Junker,
  Kalaghatgi, Kalogera, Kamai, Kandhasamy, Kang, Kanner, Kapadia, Kapasi,
  Karathanasis, Karki, Kashyap, Kasprzack, Kastaun, Katsanevas, Katsavounidis,
  Katzman, Kawabe, K{\'{e}}f{\'{e}}lian, Keitel, Key, Khadka, Khalili, Khan,
  Khan, Khazanov, Khetan, Khursheed, Kijbunchoo, Kim, Kim, Kim, Kim, Kim, Kim,
  Kimball, King, Kinley-Hanlon, Kirchhoff, Kissel, Kleybolte, Klimenko,
  Knowles, Knyazev, Koch, Koehlenbeck, Koekoek, Koley, Kolstein, Komori,
  Kondrashov, Kontos, Koper, Korobko, Korth, Kovalam, Kozak, Kr{\"{a}}mer,
  Kringel, Krishnendu, Kr{\'{o}}lak, Kuehn, Kumar, Kumar, Kumar, Kumar, Kuns,
  Kwang, Lackey, Laghi, Lalande, Lam, Lamberts, Landry, Lane, Lang, Lange,
  Lantz, Lanza, {La Rosa}, Lartaux-Vollard, Lasky, Laxen, Lazzarini, Lazzaro,
  Leaci, Leavey, Lecoeuche, Lee, Lee, Lee, Lee, Lehmann, Leon, Leroy, Letendre,
  Levin, Li, Li, Li, Li, Li, Linde, Linker, Linley, Littenberg, Liu, Liu,
  Llorens-Monteagudo, Lo, Lockwood, London, Longo, Lorenzini, Loriette,
  Lormand, Losurdo, Lough, Lousto, Lovelace, L{\"{u}}ck, Lumaca, Lundgren, Ma,
  Macas, MacInnis, Macleod, MacMillan, Macquet, {Maga{\~{n}}a Hernandez},
  Maga{\~{n}}a-Sandoval, Magazz{\`{u}}, Magee, Majorana, Maksimovic, Maliakal,
  Malik, Man, Mandic, Mangano, Mansell, Manske, Mantovani, Mapelli, Marchesoni,
  Marion, M{\'{a}}rka, M{\'{a}}rka, Markakis, Markosyan, Markowitz, Maros,
  Marquina, Marsat, Martelli, Martin, Martin, Martinez, Martinez, Martynov,
  Masalehdan, Mason, Massera, Masserot, Massinger, Masso-Reid, Mastrogiovanni,
  Matas, Mateu-Lucena, Matichard, Matiushechkina, Mavalvala, Maynard, McCann,
  McCarthy, McClelland, McCormick, McCuller, McGuire, McIsaac, McIver, McManus,
  McRae, McWilliams, Meacher, Meadors, Mehmet, Mehta, Melatos, Melchor,
  Mendell, Menendez-Vazquez, Mercer, Mereni, Merfeld, Merilh, Merritt,
  Merzougui, Meshkov, Messenger, Messick, Metzdorff, Meyers, Meylahn, Mhaske,
  Miani, Miao, Michaloliakos, Michel, Middleton, Milano, Miller, Miller,
  Millhouse, Mills, Milotti, Milovich-Goff, Minazzoli, Minenkov, Mir, Mishkin,
  Mishra, Mistry, Mitra, Mitrofanov, Mitselmakher, Mittleman, Mo, Mogushi,
  Mohapatra, Mohite, Molina, Molina-Ruiz, Mondin, Montani, Moore, Moraru,
  Morawski, Moreno, Morisaki, Mours, Mow-Lowry, Mozzon, Muciaccia, Mukherjee,
  Mukherjee, Mukherjee, Mukherjee, Mukund, Mullavey, Munch, Mu{\~{n}}iz,
  Murray, Nadji, Nagar, Nardecchia, Naticchioni, Nayak, Neil, Neilson,
  Nelemans, Nelson, Nery, Neunzert, Ng, Ng, Nguyen, Nguyen, Nguyen, Nichols,
  Nissanke, Nocera, Noh, North, Nothard, Nuttall, Oberling, O'Brien, O'Dell,
  Oganesyan, Ogin, Oh, Oh, Ohme, Ohta, Okada, Olivetto, Oppermann, Oram,
  O'Reilly, Ormiston, Ormsby, Ortega, O'Shaughnessy, Ossokine, Osthelder,
  Ottaway, Overmier, Owen, Pace, Pagano, Page, Pagliaroli, Pai, Pai, Palamos,
  Palashov, Palomba, Pan, Panda, Pang, Pankow, Pannarale, Pant, Paoletti,
  Paoli, Paolone, Parker, Pascucci, Pasqualetti, Passaquieti, Passuello, Patel,
  Patricelli, Payne, Pechsiri, Pedraza, Pegoraro, Pele, Penn, Perego, Perez,
  P{\'{e}}rigois, Perreca, Perri{\`{e}}s, Petermann, Petterson, Pfeiffer, Pham,
  Phukon, Piccinni, Pichot, Piendibene, Piergiovanni, Pierini, Pierro, Pillant,
  Pilo, Pinard, Pinto, Piotrzkowski, Pirello, Pitkin, Placidi, Plastino,
  Pluchar, Poggiani, Polini, Pong, Ponrathnam, Popolizio, Porter, Poverman,
  Powell, Pracchia, Prajapati, Prasai, Prasanna, Pratten, Prestegard, Principe,
  Prodi, Prokhorov, Prosposito, Puecher, Punturo, Puosi, Puppo, P{\"{u}}rrer,
  Qi, Quetschke, Quinonez, Quitzow-James, Raab, Raaijmakers, Radkins,
  Radulesco, Raffai, Rafferty, Rail, Raja, Rajan, Rajbhandari, Rakhmanov,
  Ramirez, Ramirez, Ramos-Buades, Rana, Rao, Rapagnani, Rapol, Ratto, Raymond,
  Razzano, Read, Regimbau, Rei, Reid, Reitze, Rettegno, Ricci, Richardson,
  Richardson, Richardson, Ricker, Riemenschneider, Riles, Rizzo, Robertson,
  Robinet, Rocchi, Rocha, Rodriguez, Rodriguez-Soto, Rolland, Rollins, Roma,
  Romanelli, Romano, Romel, Romero, Romero-Shaw, Romie, Ronchini, Rose, Rose,
  Rose, Rosell, Rosi{\'{n}}ska, Rosofsky, Ross, Rowan, Rowlinson, Roy, Roy,
  Ruggi, Ryan, Sachdev, Sadecki, Sakellariadou, Salafia, Salconi, Saleem,
  Samajdar, Sanchez, Sanchez, Sanchez, Sanchis-Gual, Sanders, Santiago, Santos,
  Saravanan, Sarin, Sassolas, Sathyaprakash, Sauter, Savage, Savant, Sawant,
  Sayah, Schaetzl, Schale, Scheel, Scheuer, Schindler-Tyka, Schmidt, Schnabel,
  Schofield, Sch{\"{o}}nbeck, Schreiber, Schulte, Schutz, Schwarm, Schwartz,
  Scott, Scott, Seglar-Arroyo, Seidel, Sellers, Sengupta, Sennett, Sentenac,
  Sequino, Sergeev, Setyawati, Shaffer, Shahriar, Sharifi, Sharma, Sharma,
  Shawhan, Shen, Shikauchi, Shink, Shoemaker, Shoemaker, Shukla, ShyamSundar,
  Sieniawska, Sigg, Singer, Singh, Singh, Singha, Singhal, Sintes, Sipala,
  Skliris, Slagmolen, Slaven-Blair, Smetana, Smith, Smith, Somala, Son, Soni,
  Sorazu, Sordini, Sorrentino, Sorrentino, Soulard, Souradeep, Sowell, Spencer,
  Spera, Srivastava, Srivastava, Staats, Stachie, Steer, Steinke, Steinlechner,
  Steinlechner, Steinmeyer, Stevenson, Stolle-McAllister, Stops, Stover,
  Strain, Stratta, Strunk, Sturani, Stuver, S{\"{u}}dbeck, Sudhagar, Sudhir,
  Suh, Summerscales, Sun, Sun, Sunil, Sur, Suresh, Sutton, Swinkels,
  Szczepa{\'{n}}czyk, Tacca, Tait, Talbot, Tanasijczuk, Tanner, Tao, Tapia,
  {Tapia San Martin}, Tasson, Taylor, Tenorio, Terkowski, Thirugnanasambandam,
  Thomas, Thomas, Thomas, Thompson, Thondapu, Thorne, Thrane, Tiwari, Tiwari,
  Tiwari, Toland, Tolley, Tonelli, Tornasi, Torres-Forn{\'{e}}, Torrie, {Tosta
  e Melo}, T{\"{o}}yr{\"{a}}, Tran, Trapananti, Travasso, Traylor, Tringali,
  Tripathee, Trovato, Trudeau, Tsai, Tsang, Tse, Tso, Tsukada, Tsuna, Tsutsui,
  Turconi, Ubhi, Udall, Ueno, Ugolini, Unnikrishnan, Urban, Usman, Utina,
  Vahlbruch, Vajente, Vajpeyi, Valdes, Valentini, Valsan, van Bakel, van
  Beuzekom, van~den Brand, {Van Den Broeck}, Vander-Hyde, van~der Schaaf, van
  Heijningen, Vardaro, Vargas, Varma, Vass, Vas{\'{u}}th, Vecchio, Vedovato,
  Veitch, Veitch, Venkateswara, Venneberg, Venugopalan, Verkindt, Verma, Veske,
  Vetrano, Vicer{\'{e}}, Viets, Villa-Ortega, Vinet, Vitale, Vo, Vocca,
  Vorvick, Vyatchanin, Wade, Wade, Wade, Walet, Walker, Wallace, Wallace,
  Walsh, Wang, Wang, Wang, Wang, Ward, Warner, Was, Washington, Watchi, Weaver,
  Wei, Weinert, Weinstein, Weiss, Wellmann, Wen, We{\ss}els, Westhouse, Wette,
  Whelan, White, White, Whiting, Whittle, Wilken, Williams, Williams,
  Williamson, Willis, Willke, Wilson, Wimmer, Winkler, Wipf, Woan, Woehler,
  Wofford, Wong, Wrangel, Wright, Wu, Wysocki, Xiao, Yamamoto, Yang, Yang,
  Yang, Yap, Yeeles, Yoon, Yu, Yu, Yuen, Zadro{\.{z}}ny, Zanolin, Zelenova,
  Zendri, Zevin, Zhang, Zhang, Zhang, Zhang, Zhao, Zhao, Zhou, Zhou, Zhu,
  Zimmerman, Zucker, \& Zweizig}]{GWTC2_pops}
---. 2021{\natexlab{a}}, The Astrophysical Journal Letters, 913, L7,
  \dodoi{10.3847/2041-8213/abe949}

\bibitem[{Abbott {et~al.}(2021{\natexlab{b}})Abbott, Abbott, Abraham, Acernese,
  Ackley, Adams, Adams, Adhikari, Adya, Affeldt, Agathos, Agatsuma, Aggarwal,
  Aguiar, Aiello, Ain, Ajith, Akcay, Allen, Allocca, Altin, Amato, Anand,
  Ananyeva, Anderson, Anderson, Angelova, Ansoldi, Antelis, Antier, Appert,
  Arai, Araya, Areeda, Ar{\`{e}}ne, Arnaud, Aronson, Arun, Asali, Ascenzi,
  Ashton, Aston, Astone, Aubin, Aufmuth, Aultoneal, Austin, Avendano, Babak,
  Badaracco, Bader, Bae, Baer, Bagnasco, Baird, Ball, Ballardin, Ballmer, Bals,
  Balsamo, Baltus, Banagiri, Bankar, Bankar, Barayoga, Barbieri, Barish,
  Barker, Barneo, Barnum, Barone, Barr, Barsotti, Barsuglia, Barta, Bartlett,
  Bartos, Bassiri, Basti, Bawaj, Bayley, Bazzan, Becher, B{\'{e}}csy,
  Bedakihale, Bejger, Belahcene, Beniwal, Benjamin, Bennett, Bentley, Bergamin,
  Berger, Bergmann, Bernuzzi, Berry, Bersanetti, Bertolini, Betzwieser,
  Bhandare, Bhandari, Bhattacharjee, Bidler, Bilenko, Billingsley, Birney,
  Birnholtz, Biscans, Bischi, Biscoveanu, Bisht, Bitossi, Bizouard, Blackburn,
  Blackman, Blair, Blair, Blair, Blanch, Bobba, Bode, Boer, Boetzel, Bogaert,
  Boldrini, Bondu, Bonilla, Bonnand, Booker, Boom, Bork, Boschi, Bose,
  Bossilkov, Boudart, Bouffanais, Bozzi, Bradaschia, Brady, Bramley, Branchesi,
  Brau, Breschi, Briant, Briggs, Brighenti, Brillet, Brinkmann, Brockill,
  Brooks, Brooks, Brown, Brunett, Bruno, Bruntz, Buikema, Bulik, Bulten,
  Buonanno, Buscicchio, Buskulic, Byer, Cabero, Cadonati, Caesar, Cagnoli,
  Cahillane, {Calder{\'{o}}n Bustillo}, Callaghan, Callister, Calloni, Camp,
  Canepa, Cannon, Cao, Cao, Carapella, Carbognani, Carney, Carpinelli, Carullo,
  Carver, {Casanueva Diaz}, Casentini, Caudill, Cavagli{\`{a}}, Cavalier,
  Cavalieri, Cella, Cerd{\'{a}}-Dur{\'{a}}n, Cesarini, Chaibi, Chakravarti,
  Chan, Chan, Chandra, Chanial, Chao, Charlton, Chase, Chassande-Mottin,
  Chatterjee, Chattopadhyay, Chaturvedi, Chatziioannou, Chen, Chen, Chen, Chen,
  Cheng, Cheong, Chia, Chiadini, Chierici, Chincarini, Chiummo, Cho, Cho, Cho,
  Choate, Christensen, Chu, Chua, Chung, Chung, Ciani, Ciecielag,
  Cie{\'{s}}lar, Cifaldi, Ciobanu, Ciolfi, Cipriano, Cirone, Clara, Clark,
  Clark, Clarke, Clearwater, Clesse, Cleva, Coccia, Cohadon, Cohen, Colleoni,
  Collette, Collins, Colpi, Constancio, Conti, Cooper, Corban, Corbitt,
  Cordero-Carri{\'{o}}n, Corezzi, Corley, Cornish, Corre, Corsi, Cortese,
  Costa, Cotesta, Coughlin, Coughlin, Coulon, Countryman, Cousins, Couvares,
  Covas, Coward, Cowart, Coyne, Coyne, Creighton, Creighton, Croquette,
  Crowder, Cudell, Cullen, Cumming, Cummings, Cunningham, Cuoco, Cury{\l}o,
  Canton, D{\'{a}}lya, Dana, Daneshgaranbajastani, D'Angelo, Danila,
  Danilishin, D'Antonio, Danzmann, Darsow-Fromm, Dasgupta, Datrier, Dattilo,
  Dave, Davier, Davies, Davis, Daw, Dean, Debra, Deenadayalan, Degallaix, {De
  Laurentis}, Del{\'{e}}glise, {Del Favero}, {De Lillo}, {De Lillo}, {Del
  Pozzo}, Demarchi, {De Matteis}, D'Emilio, Demos, Denker, Dent, Depasse, {De
  Pietri}, {De Rosa}, {De Rossi}, Desalvo, {De Varona}, Dhurandhar, D{\'{i}}az,
  Diaz-Ortiz, Didio, Dietrich, {Di Fiore}, Difronzo, {Di Giorgio}, {Di
  Giovanni}, {Di Giovanni}, {Di Girolamo}, {Di Lieto}, Ding, {Di Pace}, {Di
  Palma}, {Di Renzo}, Divakarla, Dmitriev, Doctor, D'Onofrio, Donovan, Dooley,
  Doravari, Dorrington, Downes, Drago, Driggers, Du, Ducoin, Dupej, Durante,
  D'Urso, Duverne, Dwyer, Easter, Eddolls, Edelman, Edo, Edy, Effler, Eichholz,
  Eikenberry, Eisenmann, Eisenstein, Ejlli, Errico, Essick, Estell{\'{e}}s,
  Estevez, Etienne, Etzel, Evans, Evans, Ewing, Fafone, Fair, Fairhurst, Fan,
  Farah, Farinon, Farr, Farr, Fauchon-Jones, Favata, Fays, Fazio, Feicht,
  Fejer, Feng, Fenyvesi, Ferguson, Fernandez-Galiana, Ferrante, Ferreira,
  Fidecaro, Figura, Fiori, Fiorucci, Fishbach, Fisher, Fishner, Fittipaldi,
  Fitz-Axen, Fiumara, Flaminio, Floden, Flynn, Fong, Font, Forsyth, Fournier,
  Frasca, Frasconi, Frei, Freise, Frey, Frey, Fritschel, Frolov, Fronz{\'{e}},
  Fulda, Fyffe, Gabbard, Gadre, Gaebel, Gair, Gais, Galaudage, Gamba,
  Ganapathy, Ganguly, Gaonkar, Garaventa, Garc{\'{i}}a-Quir{\'{o}}s, Garufi,
  Gateley, Gaudio, Gayathri, Gemme, Gennai, George, George, George, Gergely,
  Ghonge, Ghosh, Ghosh, Ghosh, Giacomazzo, Giacoppo, Giaime, Giardina, Gibson,
  Gier, Gill, Giri, Glanzer, Gleckl, Godwin, Goetz, Goetz, Gohlke, Goncharov,
  Gonz{\'{a}}lez, Gopakumar, Gossan, Gosselin, Gouaty, Grace, Grado, Granata,
  Granata, Grant, Gras, Grassia, Gray, Gray, Greco, Green, Green, Gretarsson,
  Griggs, Grignani, Grimaldi, Grimes, Grimm, Grote, Grunewald, Gruning,
  Guerrero, Guidi, Guimaraes, Guix{\'{e}}, Gulati, Guo, Gupta, Gupta, Gupta,
  Gustafson, Gustafson, Guzman, Haegel, Halim, Hall, Hamilton, Hammond, Haney,
  Hanke, Hanks, Hanna, Hannam, Hannuksela, Hannuksela, Hansen, Hansen, Hanson,
  Harder, Hardwick, Haris, Harms, Harry, Harry, Hartwig, Hasskew, Haster,
  Haughian, Hayes, Healy, Heidmann, Heintze, Heinze, Heinzel, Heitmann,
  Hellman, Hello, Helmling-Cornell, Hemming, Hendry, Heng, Hennes, Hennig,
  Hennig, {Hernandez Vivanco}, Heurs, Hild, Hill, Hines, Hochheim, Hofgard,
  Hofman, Hohmann, Holgado, Holland, Hollows, Holmes, Holt, Holz, Hopkins,
  Horst, Hough, Howell, Hoy, Hoyland, Huang, H{\"{u}}bner, Huddart, Huerta,
  Hughey, Hui, Husa, Huttner, Hutzler, Huxford, Huynh-Dinh, Idzkowski, Iess,
  Imperato, Inchauspe, Ingram, Intini, Isi, Iyer, Jaberianhamedan, Jacqmin,
  Jadhav, Jadhav, James, Jani, Janssens, Janthalur, Jaranowski, Jariwala,
  Jaume, Jenkins, Jeunon, Jiang, Johns, Johnson-Mcdaniel, Jones, Jones, Jones,
  Jones, Jones, Jonker, Ju, Junker, Kalaghatgi, Kalogera, Kamai, Kandhasamy,
  Kang, Kanner, Kapadia, Kapasi, Karathanasis, Karki, Kashyap, Kasprzack,
  Kastaun, Katsanevas, Katsavounidis, Katzman, Kawabe, K{\'{e}}f{\'{e}}lian,
  Keitel, Key, Khadka, Khalili, Khan, Khan, Khazanov, Khetan, Khursheed,
  Kijbunchoo, Kim, Kim, Kim, Kim, Kim, Kim, Kimball, King, Kinley-Hanlon,
  Kirchhoff, Kissel, Kleybolte, Klimenko, Knowles, Knyazev, Koch, Koehlenbeck,
  Koekoek, Koley, Kolstein, Komori, Kondrashov, Kontos, Koper, Korobko, Korth,
  Kovalam, Kozak, Kr{\"{a}}mer, Kringel, Krishnendu, Kr{\'{o}}lak, Kuehn,
  Kumar, Kumar, Kumar, Kumar, Kuns, Kwang, Lackey, Laghi, Lalande, Lam,
  Lamberts, Landry, Lane, Lang, Lange, Lantz, Lanza, {La Rosa},
  Lartaux-Vollard, Lasky, Laxen, Lazzarini, Lazzaro, Leaci, Leavey, Lecoeuche,
  Lee, Lee, Lee, Lee, Lehmann, Leon, Leroy, Letendre, Levin, Li, Li, Li, Li,
  Li, Linde, Linker, Linley, Littenberg, Liu, Liu, Llorens-Monteagudo, Lo,
  Lockwood, London, Longo, Lorenzini, Loriette, Lormand, Losurdo, Lough,
  Lousto, Lovelace, L{\"{u}}ck, Lumaca, Lundgren, Ma, Macas, Macinnis, Macleod,
  Macmillan, Macquet, {Maga{\~{n}}a Hernandez}, Maga{\~{n}}a-Sandoval,
  Magazz{\`{u}}, Magee, Majorana, Maksimovic, Maliakal, Malik, Man, Mandic,
  Mangano, Mansell, Manske, Mantovani, Mapelli, Marchesoni, Marion,
  M{\'{a}}rka, M{\'{a}}rka, Markakis, Markosyan, Markowitz, Maros, Marquina,
  Marsat, Martelli, Martin, Martin, Martinez, Martinez, Martynov, Masalehdan,
  Mason, Massera, Masserot, Massinger, Masso-Reid, Mastrogiovanni, Matas,
  Mateu-Lucena, Matichard, Matiushechkina, Mavalvala, Maynard, McCann,
  McCarthy, McClelland, McCormick, McCuller, McGuire, McIsaac, McIver, McManus,
  McRae, McWilliams, Meacher, Meadors, Mehmet, Mehta, Melatos, Melchor,
  Mendell, Menendez-Vazquez, Mercer, Mereni, Merfeld, Merilh, Merritt,
  Merzougui, Meshkov, Messenger, Messick, Metzdorff, Meyers, Meylahn, Mhaske,
  Miani, Miao, Michaloliakos, Michel, Middleton, Milano, Miller, Millhouse,
  Mills, Milotti, Milovich-Goff, Minazzoli, Minenkov, Mir, Mishkin, Mishra,
  Mistry, Mitra, Mitrofanov, Mitselmakher, Mittleman, Mo, Mogushi, Mohapatra,
  Mohite, Molina, Molina-Ruiz, Mondin, Montani, Moore, Moraru, Morawski,
  Moreno, Morisaki, Mours, Mow-Lowry, Mozzon, Muciaccia, Mukherjee, Mukherjee,
  Mukherjee, Mukherjee, Mukund, Mullavey, Munch, Mu{\~{n}}iz, Murray, Nadji,
  Nagar, Nardecchia, Naticchioni, Nayak, Neil, Neilson, Nelemans, Nelson, Nery,
  Neunzert, Nitz, Ng, Ng, Nguyen, Nguyen, Nguyen, Nichols, Nissanke, Nocera,
  Noh, North, Nothard, Nuttall, Oberling, O'Brien, O'Dell, Oganesyan, Ogin, Oh,
  Oh, Ohme, Ohta, Okada, Olivetto, Oppermann, Oram, O'Reilly, Ormiston, Ortega,
  O'Shaughnessy, Ossokine, Osthelder, Ottaway, Overmier, Owen, Pace, Pagano,
  Page, Pagliaroli, Pai, Pai, Palamos, Palashov, Palomba, Pan, Panda, Pang,
  Pankow, Pannarale, Pant, Paoletti, Paoli, Paolone, Parker, Pascucci,
  Pasqualetti, Passaquieti, Passuello, Patel, Patricelli, Payne, Pechsiri,
  Pedraza, Pegoraro, Pele, Penn, Perego, Perez, P{\'{e}}rigois, Perreca,
  Perri{\`{e}}s, Petermann, Petterson, Pfeiffer, Pham, Phukon, Piccinni,
  Pichot, Piendibene, Piergiovanni, Pierini, Pierro, Pillant, Pilo, Pinard,
  Pinto, Piotrzkowski, Pirello, Pitkin, Placidi, Plastino, Pluchar, Poggiani,
  Polini, Pong, Ponrathnam, Popolizio, Porter, Poverman, Powell, Pracchia,
  Prajapati, Prasai, Prasanna, Pratten, Prestegard, Principe, Prodi, Prokhorov,
  Prosposito, Prudenzi, Puecher, Punturo, Puosi, Puppo, P{\"{u}}rrer, Qi,
  Quetschke, Quinonez, Quitzow-James, Raab, Raaijmakers, Radkins, Radulesco,
  Raffai, Rafferty, Rail, Raja, Rajan, Rajbhandari, Rakhmanov, Ramirez,
  Ramirez, Ramos-Buades, Rana, Rao, Rapagnani, Rapol, Ratto, Raymond, Razzano,
  Read, Regimbau, Rei, Reid, Reitze, Rettegno, Ricci, Richardson, Richardson,
  Richardson, Ricker, Riemenschneider, Riles, Rizzo, Robertson, Robinet,
  Rocchi, Rocha, Rodriguez, Rodriguez-Soto, Rolland, Rollins, Roma, Romanelli,
  Romano, Romel, Romero, Romero-Shaw, Romie, Ronchini, Rose, Rose, Rose,
  Rosell, Rosi{\'{n}}ska, Rosofsky, Ross, Rowan, Rowlinson, Roy, Roy, Ruggi,
  Ryan, Sachdev, Sadecki, Sadiq, Sakellariadou, Salafia, Salconi, Saleem,
  Samajdar, Sanchez, Sanchez, Sanchez, Sanchis-Gual, Sanders, Sandles,
  Santiago, Santos, Saravanan, Sarin, Sassolas, Sathyaprakash, Sauter, Savage,
  Savant, Sawant, Sayah, Schaetzl, Schale, Scheel, Scheuer, Schindler-Tyka,
  Schmidt, Schnabel, Schofield, Sch{\"{o}}nbeck, Schreiber, Schulte, Schutz,
  Schwarm, Schwartz, Scott, Scott, Seglar-Arroyo, Seidel, Sellers, Sengupta,
  Sennett, Sentenac, Sequino, Sergeev, Setyawati, Shaffer, Shahriar, Sharifi,
  Sharma, Sharma, Shawhan, Shen, Shikauchi, Shink, Shoemaker, Shoemaker,
  Shukla, Shyamsundar, Sieniawska, Sigg, Singer, Singh, Singh, Singha, Singhal,
  Sintes, Sipala, Skliris, Slagmolen, Slaven-Blair, Smetana, Smith, Smith,
  Somala, Son, Soni, Soni, Sorazu, Sordini, Sorrentino, Sorrentino, Soulard,
  Souradeep, Sowell, Spencer, Spera, Srivastava, Srivastava, Staats, Stachie,
  Steer, Steinhoff, Steinke, Steinlechner, Steinlechner, Steinmeyer, Stevenson,
  Stolle-Mcallister, Stops, Stover, Strain, Stratta, Strunk, Sturani, Stuver,
  S{\"{u}}dbeck, Sudhagar, Sudhir, Suh, Summerscales, Sun, Sun, Sunil, Sur,
  Suresh, Sutton, Swinkels, Szczepa{\'{n}}czyk, Tacca, Tait, Talbot,
  Tanasijczuk, Tanner, Tao, Tapia, {Tapia San Martin}, Tasson, Taylor, Tenorio,
  Terkowski, Thirugnanasambandam, Thomas, Thomas, Thomas, Thompson, Thondapu,
  Thorne, Thrane, Tiwari, Tiwari, Tiwari, Toland, Tolley, Tonelli, Tornasi,
  Torres-Forn{\'{e}}, Torrie, {E Melo}, T{\"{o}}yr{\"{a}}, Tran, Trapananti,
  Travasso, Traylor, Tringali, Tripathee, Trovato, Trudeau, Tsai, Tsang, Tse,
  Tso, Tsukada, Tsuna, Tsutsui, Turconi, Ubhi, Udall, Ueno, Ugolini,
  Unnikrishnan, Urban, Usman, Utina, Vahlbruch, Vajente, Vajpeyi, Valdes,
  Valentini, Valsan, {Van Bakel}, {Van Beuzekom}, {Van Den Brand}, {Van Den
  Broeck}, Vander-Hyde, {Van Der Schaaf}, {Van Heijningen}, Vardaro, Vargas,
  Varma, Vass, Vas{\'{u}}th, Vecchio, Vedovato, Veitch, Veitch, Venkateswara,
  Venneberg, Venugopalan, Verkindt, Verma, Veske, Vetrano, Vicer{\'{e}}, Viets,
  Vijaykumar, Villa-Ortega, Vinet, Vitale, Vo, Vocca, Vorvick, Vyatchanin,
  Wade, Wade, Wade, Walet, Walker, Wallace, Wallace, Walsh, Wang, Wang, Wang,
  Wang, Ward, Warner, Was, Washington, Watchi, Weaver, Wei, Weinert, Weinstein,
  Weiss, Wellmann, Wen, We{\ss}els, Westhouse, Wette, Whelan, White, White,
  Whiting, Whittle, Wilken, Williams, Williams, Williamson, Willis, Willke,
  Wilson, Wimmer, Winkler, Wipf, Woan, Woehler, Wofford, Wong, Wrangel, Wright,
  Wu, Wysocki, Xiao, Yamamoto, Yang, Yang, Yang, Yap, Yeeles, Yoon, Yu, Yu,
  Yuen, Zadro{\.{z}}ny, Zanolin, Zelenova, Zendri, Zevin, Zhang, Zhang, Zhang,
  Zhang, Zhao, Zhao, Zheng, Zhou, Zhou, Zhu, Zimmerman, Zlochower, Zucker, \&
  Zweizig}]{GWTC2.1}
---. 2021{\natexlab{b}}, arXiv e-prints.
\newblock \doarXiv{2108.01045}

\bibitem[{Abbott {et~al.}(2021{\natexlab{c}})Abbott, Abbott, Abraham, Acernese,
  Ackley, Adams, Adams, Adhikari, Adya, Affeldt, Agathos, Agatsuma, Aggarwal,
  Aguiar, Aiello, Ain, Ajith, Akcay, Allen, Allocca, Altin, Amato, Anand,
  Ananyeva, Anderson, Anderson, Angelova, Ansoldi, Antelis, Antier, Appert,
  Arai, Araya, Areeda, Ar{\`{e}}ne, Arnaud, Aronson, Arun, Asali, Ascenzi,
  Ashton, Aston, Astone, Aubin, Aufmuth, Aultoneal, Austin, Avendano, Babak,
  Badaracco, Bader, Bae, Baer, Bagnasco, Baird, Ball, Ballardin, Ballmer, Bals,
  Balsamo, Baltus, Banagiri, Bankar, Bankar, Barayoga, Barbieri, Barish,
  Barker, Barneo, Barnum, Barone, Barr, Barsotti, Barsuglia, Barta, Bartlett,
  Bartos, Bassiri, Basti, Bawaj, Bayley, Bazzan, Becher, B{\'{e}}csy,
  Bedakihale, Bejger, Belahcene, Beniwal, Benjamin, Bennett, Bentley, Bergamin,
  Berger, Bergmann, Bernuzzi, Berry, Bersanetti, Bertolini, Betzwieser,
  Bhandare, Bhandari, Bhattacharjee, Bidler, Bilenko, Billingsley, Birney,
  Birnholtz, Biscans, Bischi, Biscoveanu, Bisht, Bitossi, Bizouard, Blackburn,
  Blackman, Blair, Blair, Blair, Blanch, Bobba, Bode, Boer, Boetzel, Bogaert,
  Boldrini, Bondu, Bonilla, Bonnand, Booker, Boom, Bork, Boschi, Bose,
  Bossilkov, Boudart, Bouffanais, Bozzi, Bradaschia, Brady, Bramley, Branchesi,
  Brau, Breschi, Briant, Briggs, Brighenti, Brillet, Brinkmann, Brockill,
  Brooks, Brooks, Brown, Brunett, Bruno, Bruntz, Buikema, Bulik, Bulten,
  Buonanno, Buscicchio, Buskulic, Byer, Cabero, Cadonati, Caesar, Cagnoli,
  Cahillane, {Calder{\'{o}}n Bustillo}, Callaghan, Callister, Calloni, Camp,
  Canepa, Cannon, Cao, Cao, Carapella, Carbognani, Carney, Carpinelli, Carullo,
  Carver, {Casanueva Diaz}, Casentini, Caudill, Cavagli{\`{a}}, Cavalier,
  Cavalieri, Cella, Cerd{\'{a}}-Dur{\'{a}}n, Cesarini, Chaibi, Chakravarti,
  Chan, Chan, Chandra, Chanial, Chao, Charlton, Chase, Chassande-Mottin,
  Chatterjee, Chattopadhyay, Chaturvedi, Chatziioannou, Chen, Chen, Chen, Chen,
  Cheng, Cheong, Chia, Chiadini, Chierici, Chincarini, Chiummo, Cho, Cho, Cho,
  Choate, Christensen, Chu, Chua, Chung, Chung, Ciani, Ciecielag,
  Cie{\'{s}}lar, Cifaldi, Ciobanu, Ciolfi, Cipriano, Cirone, Clara, Clark,
  Clark, Clarke, Clearwater, Clesse, Cleva, Coccia, Cohadon, Cohen, Colleoni,
  Collette, Collins, Colpi, Constancio, Conti, Cooper, Corban, Corbitt,
  Cordero-Carri{\'{o}}n, Corezzi, Corley, Cornish, Corre, Corsi, Cortese,
  Costa, Cotesta, Coughlin, Coughlin, Coulon, Countryman, Cousins, Couvares,
  Covas, Coward, Cowart, Coyne, Coyne, Creighton, Creighton, Croquette,
  Crowder, Cudell, Cullen, Cumming, Cummings, Cunningham, Cuoco, Cury{\l}o,
  Canton, D{\'{a}}lya, Dana, Daneshgaranbajastani, D'Angelo, Danila,
  Danilishin, D'Antonio, Danzmann, Darsow-Fromm, Dasgupta, Datrier, Dattilo,
  Dave, Davier, Davies, Davis, Daw, Dean, Debra, Deenadayalan, Degallaix, {De
  Laurentis}, Del{\'{e}}glise, {Del Favero}, {De Lillo}, {De Lillo}, {Del
  Pozzo}, Demarchi, {De Matteis}, D'Emilio, Demos, Denker, Dent, Depasse, {De
  Pietri}, {De Rosa}, {De Rossi}, Desalvo, {De Varona}, Dhurandhar, D{\'{i}}az,
  Diaz-Ortiz, Didio, Dietrich, {Di Fiore}, Difronzo, {Di Giorgio}, {Di
  Giovanni}, {Di Giovanni}, {Di Girolamo}, {Di Lieto}, Ding, {Di Pace}, {Di
  Palma}, {Di Renzo}, Divakarla, Dmitriev, Doctor, D'Onofrio, Donovan, Dooley,
  Doravari, Dorrington, Downes, Drago, Driggers, Du, Ducoin, Dupej, Durante,
  D'Urso, Duverne, Dwyer, Easter, Eddolls, Edelman, Edo, Edy, Effler, Eichholz,
  Eikenberry, Eisenmann, Eisenstein, Ejlli, Errico, Essick, Estell{\'{e}}s,
  Estevez, Etienne, Etzel, Evans, Evans, Ewing, Fafone, Fair, Fairhurst, Fan,
  Farah, Farinon, Farr, Farr, Fauchon-Jones, Favata, Fays, Fazio, Feicht,
  Fejer, Feng, Fenyvesi, Ferguson, Fernandez-Galiana, Ferrante, Ferreira,
  Fidecaro, Figura, Fiori, Fiorucci, Fishbach, Fisher, Fishner, Fittipaldi,
  Fitz-Axen, Fiumara, Flaminio, Floden, Flynn, Fong, Font, Forsyth, Fournier,
  Frasca, Frasconi, Frei, Freise, Frey, Frey, Fritschel, Frolov, Fronz{\'{e}},
  Fulda, Fyffe, Gabbard, Gadre, Gaebel, Gair, Gais, Galaudage, Gamba,
  Ganapathy, Ganguly, Gaonkar, Garaventa, Garc{\'{i}}a-Quir{\'{o}}s, Garufi,
  Gateley, Gaudio, Gayathri, Gemme, Gennai, George, George, George, Gergely,
  Ghonge, Ghosh, Ghosh, Ghosh, Giacomazzo, Giacoppo, Giaime, Giardina, Gibson,
  Gier, Gill, Giri, Glanzer, Gleckl, Godwin, Goetz, Goetz, Gohlke, Goncharov,
  Gonz{\'{a}}lez, Gopakumar, Gossan, Gosselin, Gouaty, Grace, Grado, Granata,
  Granata, Grant, Gras, Grassia, Gray, Gray, Greco, Green, Green, Gretarsson,
  Griggs, Grignani, Grimaldi, Grimes, Grimm, Grote, Grunewald, Gruning,
  Guerrero, Guidi, Guimaraes, Guix{\'{e}}, Gulati, Guo, Gupta, Gupta, Gupta,
  Gustafson, Gustafson, Guzman, Haegel, Halim, Hall, Hamilton, Hammond, Haney,
  Hanke, Hanks, Hanna, Hannam, Hannuksela, Hannuksela, Hansen, Hansen, Hanson,
  Harder, Hardwick, Haris, Harms, Harry, Harry, Hartwig, Hasskew, Haster,
  Haughian, Hayes, Healy, Heidmann, Heintze, Heinze, Heinzel, Heitmann,
  Hellman, Hello, Helmling-Cornell, Hemming, Hendry, Heng, Hennes, Hennig,
  Hennig, {Hernandez Vivanco}, Heurs, Hild, Hill, Hines, Hochheim, Hofgard,
  Hofman, Hohmann, Holgado, Holland, Hollows, Holmes, Holt, Holz, Hopkins,
  Horst, Hough, Howell, Hoy, Hoyland, Huang, H{\"{u}}bner, Huddart, Huerta,
  Hughey, Hui, Husa, Huttner, Hutzler, Huxford, Huynh-Dinh, Idzkowski, Iess,
  Imperato, Inchauspe, Ingram, Intini, Isi, Iyer, Jaberianhamedan, Jacqmin,
  Jadhav, Jadhav, James, Jani, Janssens, Janthalur, Jaranowski, Jariwala,
  Jaume, Jenkins, Jeunon, Jiang, Johns, Johnson-Mcdaniel, Jones, Jones, Jones,
  Jones, Jones, Jonker, Ju, Junker, Kalaghatgi, Kalogera, Kamai, Kandhasamy,
  Kang, Kanner, Kapadia, Kapasi, Karathanasis, Karki, Kashyap, Kasprzack,
  Kastaun, Katsanevas, Katsavounidis, Katzman, Kawabe, K{\'{e}}f{\'{e}}lian,
  Keitel, Key, Khadka, Khalili, Khan, Khan, Khazanov, Khetan, Khursheed,
  Kijbunchoo, Kim, Kim, Kim, Kim, Kim, Kim, Kimball, King, Kinley-Hanlon,
  Kirchhoff, Kissel, Kleybolte, Klimenko, Knowles, Knyazev, Koch, Koehlenbeck,
  Koekoek, Koley, Kolstein, Komori, Kondrashov, Kontos, Koper, Korobko, Korth,
  Kovalam, Kozak, Kr{\"{a}}mer, Kringel, Krishnendu, Kr{\'{o}}lak, Kuehn,
  Kumar, Kumar, Kumar, Kumar, Kuns, Kwang, Lackey, Laghi, Lalande, Lam,
  Lamberts, Landry, Lane, Lang, Lange, Lantz, Lanza, {La Rosa},
  Lartaux-Vollard, Lasky, Laxen, Lazzarini, Lazzaro, Leaci, Leavey, Lecoeuche,
  Lee, Lee, Lee, Lee, Lehmann, Leon, Leroy, Letendre, Levin, Li, Li, Li, Li,
  Li, Linde, Linker, Linley, Littenberg, Liu, Liu, Llorens-Monteagudo, Lo,
  Lockwood, London, Longo, Lorenzini, Loriette, Lormand, Losurdo, Lough,
  Lousto, Lovelace, L{\"{u}}ck, Lumaca, Lundgren, Ma, Macas, Macinnis, Macleod,
  Macmillan, Macquet, {Maga{\~{n}}a Hernandez}, Maga{\~{n}}a-Sandoval,
  Magazz{\`{u}}, Magee, Majorana, Maksimovic, Maliakal, Malik, Man, Mandic,
  Mangano, Mansell, Manske, Mantovani, Mapelli, Marchesoni, Marion,
  M{\'{a}}rka, M{\'{a}}rka, Markakis, Markosyan, Markowitz, Maros, Marquina,
  Marsat, Martelli, Martin, Martin, Martinez, Martinez, Martynov, Masalehdan,
  Mason, Massera, Masserot, Massinger, Masso-Reid, Mastrogiovanni, Matas,
  Mateu-Lucena, Matichard, Matiushechkina, Mavalvala, Maynard, McCann,
  McCarthy, McClelland, McCormick, McCuller, McGuire, McIsaac, McIver, McManus,
  McRae, McWilliams, Meacher, Meadors, Mehmet, Mehta, Melatos, Melchor,
  Mendell, Menendez-Vazquez, Mercer, Mereni, Merfeld, Merilh, Merritt,
  Merzougui, Meshkov, Messenger, Messick, Metzdorff, Meyers, Meylahn, Mhaske,
  Miani, Miao, Michaloliakos, Michel, Middleton, Milano, Miller, Millhouse,
  Mills, Milotti, Milovich-Goff, Minazzoli, Minenkov, Mir, Mishkin, Mishra,
  Mistry, Mitra, Mitrofanov, Mitselmakher, Mittleman, Mo, Mogushi, Mohapatra,
  Mohite, Molina, Molina-Ruiz, Mondin, Montani, Moore, Moraru, Morawski,
  Moreno, Morisaki, Mours, Mow-Lowry, Mozzon, Muciaccia, Mukherjee, Mukherjee,
  Mukherjee, Mukherjee, Mukund, Mullavey, Munch, Mu{\~{n}}iz, Murray, Nadji,
  Nagar, Nardecchia, Naticchioni, Nayak, Neil, Neilson, Nelemans, Nelson, Nery,
  Neunzert, Nitz, Ng, Ng, Nguyen, Nguyen, Nguyen, Nichols, Nissanke, Nocera,
  Noh, North, Nothard, Nuttall, Oberling, O'Brien, O'Dell, Oganesyan, Ogin, Oh,
  Oh, Ohme, Ohta, Okada, Olivetto, Oppermann, Oram, O'Reilly, Ormiston, Ortega,
  O'Shaughnessy, Ossokine, Osthelder, Ottaway, Overmier, Owen, Pace, Pagano,
  Page, Pagliaroli, Pai, Pai, Palamos, Palashov, Palomba, Pan, Panda, Pang,
  Pankow, Pannarale, Pant, Paoletti, Paoli, Paolone, Parker, Pascucci,
  Pasqualetti, Passaquieti, Passuello, Patel, Patricelli, Payne, Pechsiri,
  Pedraza, Pegoraro, Pele, Penn, Perego, Perez, P{\'{e}}rigois, Perreca,
  Perri{\`{e}}s, Petermann, Petterson, Pfeiffer, Pham, Phukon, Piccinni,
  Pichot, Piendibene, Piergiovanni, Pierini, Pierro, Pillant, Pilo, Pinard,
  Pinto, Piotrzkowski, Pirello, Pitkin, Placidi, Plastino, Pluchar, Poggiani,
  Polini, Pong, Ponrathnam, Popolizio, Porter, Poverman, Powell, Pracchia,
  Prajapati, Prasai, Prasanna, Pratten, Prestegard, Principe, Prodi, Prokhorov,
  Prosposito, Prudenzi, Puecher, Punturo, Puosi, Puppo, P{\"{u}}rrer, Qi,
  Quetschke, Quinonez, Quitzow-James, Raab, Raaijmakers, Radkins, Radulesco,
  Raffai, Rafferty, Rail, Raja, Rajan, Rajbhandari, Rakhmanov, Ramirez,
  Ramirez, Ramos-Buades, Rana, Rao, Rapagnani, Rapol, Ratto, Raymond, Razzano,
  Read, Regimbau, Rei, Reid, Reitze, Rettegno, Ricci, Richardson, Richardson,
  Richardson, Ricker, Riemenschneider, Riles, Rizzo, Robertson, Robinet,
  Rocchi, Rocha, Rodriguez, Rodriguez-Soto, Rolland, Rollins, Roma, Romanelli,
  Romano, Romel, Romero, Romero-Shaw, Romie, Ronchini, Rose, Rose, Rose,
  Rosell, Rosi{\'{n}}ska, Rosofsky, Ross, Rowan, Rowlinson, Roy, Roy, Ruggi,
  Ryan, Sachdev, Sadecki, Sadiq, Sakellariadou, Salafia, Salconi, Saleem,
  Samajdar, Sanchez, Sanchez, Sanchez, Sanchis-Gual, Sanders, Sandles,
  Santiago, Santos, Saravanan, Sarin, Sassolas, Sathyaprakash, Sauter, Savage,
  Savant, Sawant, Sayah, Schaetzl, Schale, Scheel, Scheuer, Schindler-Tyka,
  Schmidt, Schnabel, Schofield, Sch{\"{o}}nbeck, Schreiber, Schulte, Schutz,
  Schwarm, Schwartz, Scott, Scott, Seglar-Arroyo, Seidel, Sellers, Sengupta,
  Sennett, Sentenac, Sequino, Sergeev, Setyawati, Shaffer, Shahriar, Sharifi,
  Sharma, Sharma, Shawhan, Shen, Shikauchi, Shink, Shoemaker, Shoemaker,
  Shukla, Shyamsundar, Sieniawska, Sigg, Singer, Singh, Singh, Singha, Singhal,
  Sintes, Sipala, Skliris, Slagmolen, Slaven-Blair, Smetana, Smith, Smith,
  Somala, Son, Soni, Soni, Sorazu, Sordini, Sorrentino, Sorrentino, Soulard,
  Souradeep, Sowell, Spencer, Spera, Srivastava, Srivastava, Staats, Stachie,
  Steer, Steinhoff, Steinke, Steinlechner, Steinlechner, Steinmeyer, Stevenson,
  Stolle-Mcallister, Stops, Stover, Strain, Stratta, Strunk, Sturani, Stuver,
  S{\"{u}}dbeck, Sudhagar, Sudhir, Suh, Summerscales, Sun, Sun, Sunil, Sur,
  Suresh, Sutton, Swinkels, Szczepa{\'{n}}czyk, Tacca, Tait, Talbot,
  Tanasijczuk, Tanner, Tao, Tapia, {Tapia San Martin}, Tasson, Taylor, Tenorio,
  Terkowski, Thirugnanasambandam, Thomas, Thomas, Thomas, Thompson, Thondapu,
  Thorne, Thrane, Tiwari, Tiwari, Tiwari, Toland, Tolley, Tonelli, Tornasi,
  Torres-Forn{\'{e}}, Torrie, {E Melo}, T{\"{o}}yr{\"{a}}, Tran, Trapananti,
  Travasso, Traylor, Tringali, Tripathee, Trovato, Trudeau, Tsai, Tsang, Tse,
  Tso, Tsukada, Tsuna, Tsutsui, Turconi, Ubhi, Udall, Ueno, Ugolini,
  Unnikrishnan, Urban, Usman, Utina, Vahlbruch, Vajente, Vajpeyi, Valdes,
  Valentini, Valsan, {Van Bakel}, {Van Beuzekom}, {Van Den Brand}, {Van Den
  Broeck}, Vander-Hyde, {Van Der Schaaf}, {Van Heijningen}, Vardaro, Vargas,
  Varma, Vass, Vas{\'{u}}th, Vecchio, Vedovato, Veitch, Veitch, Venkateswara,
  Venneberg, Venugopalan, Verkindt, Verma, Veske, Vetrano, Vicer{\'{e}}, Viets,
  Vijaykumar, Villa-Ortega, Vinet, Vitale, Vo, Vocca, Vorvick, Vyatchanin,
  Wade, Wade, Wade, Walet, Walker, Wallace, Wallace, Walsh, Wang, Wang, Wang,
  Wang, Ward, Warner, Was, Washington, Watchi, Weaver, Wei, Weinert, Weinstein,
  Weiss, Wellmann, Wen, We{\ss}els, Westhouse, Wette, Whelan, White, White,
  Whiting, Whittle, Wilken, Williams, Williams, Williamson, Willis, Willke,
  Wilson, Wimmer, Winkler, Wipf, Woan, Woehler, Wofford, Wong, Wrangel, Wright,
  Wu, Wysocki, Xiao, Yamamoto, Yang, Yang, Yang, Yap, Yeeles, Yoon, Yu, Yu,
  Yuen, Zadro{\.{z}}ny, Zanolin, Zelenova, Zendri, Zevin, Zhang, Zhang, Zhang,
  Zhang, Zhao, Zhao, Zheng, Zhou, Zhou, Zhu, Zimmerman, Zlochower, Zucker, \&
  Zweizig}]{GWTC3}
---. 2021{\natexlab{c}}, arXiv e-prints.
\newblock \doarXiv{2111.03606}

\bibitem[{Abbott {et~al.}(2021{\natexlab{d}})Abbott, Abbott, Abraham, Acernese,
  Ackley, Adams, Adams, Adhikari, Adya, Affeldt, Agathos, Agatsuma, Aggarwal,
  Aguiar, Aiello, Ain, Ajith, Akcay, Allen, Allocca, Altin, Amato, Anand,
  Ananyeva, Anderson, Anderson, Angelova, Ansoldi, Antelis, Antier, Appert,
  Arai, Araya, Areeda, Ar{\`{e}}ne, Arnaud, Aronson, Arun, Asali, Ascenzi,
  Ashton, Aston, Astone, Aubin, Aufmuth, Aultoneal, Austin, Avendano, Babak,
  Badaracco, Bader, Bae, Baer, Bagnasco, Baird, Ball, Ballardin, Ballmer, Bals,
  Balsamo, Baltus, Banagiri, Bankar, Bankar, Barayoga, Barbieri, Barish,
  Barker, Barneo, Barnum, Barone, Barr, Barsotti, Barsuglia, Barta, Bartlett,
  Bartos, Bassiri, Basti, Bawaj, Bayley, Bazzan, Becher, B{\'{e}}csy,
  Bedakihale, Bejger, Belahcene, Beniwal, Benjamin, Bennett, Bentley, Bergamin,
  Berger, Bergmann, Bernuzzi, Berry, Bersanetti, Bertolini, Betzwieser,
  Bhandare, Bhandari, Bhattacharjee, Bidler, Bilenko, Billingsley, Birney,
  Birnholtz, Biscans, Bischi, Biscoveanu, Bisht, Bitossi, Bizouard, Blackburn,
  Blackman, Blair, Blair, Blair, Blanch, Bobba, Bode, Boer, Boetzel, Bogaert,
  Boldrini, Bondu, Bonilla, Bonnand, Booker, Boom, Bork, Boschi, Bose,
  Bossilkov, Boudart, Bouffanais, Bozzi, Bradaschia, Brady, Bramley, Branchesi,
  Brau, Breschi, Briant, Briggs, Brighenti, Brillet, Brinkmann, Brockill,
  Brooks, Brooks, Brown, Brunett, Bruno, Bruntz, Buikema, Bulik, Bulten,
  Buonanno, Buscicchio, Buskulic, Byer, Cabero, Cadonati, Caesar, Cagnoli,
  Cahillane, {Calder{\'{o}}n Bustillo}, Callaghan, Callister, Calloni, Camp,
  Canepa, Cannon, Cao, Cao, Carapella, Carbognani, Carney, Carpinelli, Carullo,
  Carver, {Casanueva Diaz}, Casentini, Caudill, Cavagli{\`{a}}, Cavalier,
  Cavalieri, Cella, Cerd{\'{a}}-Dur{\'{a}}n, Cesarini, Chaibi, Chakravarti,
  Chan, Chan, Chandra, Chanial, Chao, Charlton, Chase, Chassande-Mottin,
  Chatterjee, Chattopadhyay, Chaturvedi, Chatziioannou, Chen, Chen, Chen, Chen,
  Cheng, Cheong, Chia, Chiadini, Chierici, Chincarini, Chiummo, Cho, Cho, Cho,
  Choate, Christensen, Chu, Chua, Chung, Chung, Ciani, Ciecielag,
  Cie{\'{s}}lar, Cifaldi, Ciobanu, Ciolfi, Cipriano, Cirone, Clara, Clark,
  Clark, Clarke, Clearwater, Clesse, Cleva, Coccia, Cohadon, Cohen, Colleoni,
  Collette, Collins, Colpi, Constancio, Conti, Cooper, Corban, Corbitt,
  Cordero-Carri{\'{o}}n, Corezzi, Corley, Cornish, Corre, Corsi, Cortese,
  Costa, Cotesta, Coughlin, Coughlin, Coulon, Countryman, Cousins, Couvares,
  Covas, Coward, Cowart, Coyne, Coyne, Creighton, Creighton, Croquette,
  Crowder, Cudell, Cullen, Cumming, Cummings, Cunningham, Cuoco, Cury{\l}o,
  Canton, D{\'{a}}lya, Dana, Daneshgaranbajastani, D'Angelo, Danila,
  Danilishin, D'Antonio, Danzmann, Darsow-Fromm, Dasgupta, Datrier, Dattilo,
  Dave, Davier, Davies, Davis, Daw, Dean, Debra, Deenadayalan, Degallaix, {De
  Laurentis}, Del{\'{e}}glise, {Del Favero}, {De Lillo}, {De Lillo}, {Del
  Pozzo}, Demarchi, {De Matteis}, D'Emilio, Demos, Denker, Dent, Depasse, {De
  Pietri}, {De Rosa}, {De Rossi}, Desalvo, {De Varona}, Dhurandhar, D{\'{i}}az,
  Diaz-Ortiz, Didio, Dietrich, {Di Fiore}, Difronzo, {Di Giorgio}, {Di
  Giovanni}, {Di Giovanni}, {Di Girolamo}, {Di Lieto}, Ding, {Di Pace}, {Di
  Palma}, {Di Renzo}, Divakarla, Dmitriev, Doctor, D'Onofrio, Donovan, Dooley,
  Doravari, Dorrington, Downes, Drago, Driggers, Du, Ducoin, Dupej, Durante,
  D'Urso, Duverne, Dwyer, Easter, Eddolls, Edelman, Edo, Edy, Effler, Eichholz,
  Eikenberry, Eisenmann, Eisenstein, Ejlli, Errico, Essick, Estell{\'{e}}s,
  Estevez, Etienne, Etzel, Evans, Evans, Ewing, Fafone, Fair, Fairhurst, Fan,
  Farah, Farinon, Farr, Farr, Fauchon-Jones, Favata, Fays, Fazio, Feicht,
  Fejer, Feng, Fenyvesi, Ferguson, Fernandez-Galiana, Ferrante, Ferreira,
  Fidecaro, Figura, Fiori, Fiorucci, Fishbach, Fisher, Fishner, Fittipaldi,
  Fitz-Axen, Fiumara, Flaminio, Floden, Flynn, Fong, Font, Forsyth, Fournier,
  Frasca, Frasconi, Frei, Freise, Frey, Frey, Fritschel, Frolov, Fronz{\'{e}},
  Fulda, Fyffe, Gabbard, Gadre, Gaebel, Gair, Gais, Galaudage, Gamba,
  Ganapathy, Ganguly, Gaonkar, Garaventa, Garc{\'{i}}a-Quir{\'{o}}s, Garufi,
  Gateley, Gaudio, Gayathri, Gemme, Gennai, George, George, George, Gergely,
  Ghonge, Ghosh, Ghosh, Ghosh, Giacomazzo, Giacoppo, Giaime, Giardina, Gibson,
  Gier, Gill, Giri, Glanzer, Gleckl, Godwin, Goetz, Goetz, Gohlke, Goncharov,
  Gonz{\'{a}}lez, Gopakumar, Gossan, Gosselin, Gouaty, Grace, Grado, Granata,
  Granata, Grant, Gras, Grassia, Gray, Gray, Greco, Green, Green, Gretarsson,
  Griggs, Grignani, Grimaldi, Grimes, Grimm, Grote, Grunewald, Gruning,
  Guerrero, Guidi, Guimaraes, Guix{\'{e}}, Gulati, Guo, Gupta, Gupta, Gupta,
  Gustafson, Gustafson, Guzman, Haegel, Halim, Hall, Hamilton, Hammond, Haney,
  Hanke, Hanks, Hanna, Hannam, Hannuksela, Hannuksela, Hansen, Hansen, Hanson,
  Harder, Hardwick, Haris, Harms, Harry, Harry, Hartwig, Hasskew, Haster,
  Haughian, Hayes, Healy, Heidmann, Heintze, Heinze, Heinzel, Heitmann,
  Hellman, Hello, Helmling-Cornell, Hemming, Hendry, Heng, Hennes, Hennig,
  Hennig, {Hernandez Vivanco}, Heurs, Hild, Hill, Hines, Hochheim, Hofgard,
  Hofman, Hohmann, Holgado, Holland, Hollows, Holmes, Holt, Holz, Hopkins,
  Horst, Hough, Howell, Hoy, Hoyland, Huang, H{\"{u}}bner, Huddart, Huerta,
  Hughey, Hui, Husa, Huttner, Hutzler, Huxford, Huynh-Dinh, Idzkowski, Iess,
  Imperato, Inchauspe, Ingram, Intini, Isi, Iyer, Jaberianhamedan, Jacqmin,
  Jadhav, Jadhav, James, Jani, Janssens, Janthalur, Jaranowski, Jariwala,
  Jaume, Jenkins, Jeunon, Jiang, Johns, Johnson-Mcdaniel, Jones, Jones, Jones,
  Jones, Jones, Jonker, Ju, Junker, Kalaghatgi, Kalogera, Kamai, Kandhasamy,
  Kang, Kanner, Kapadia, Kapasi, Karathanasis, Karki, Kashyap, Kasprzack,
  Kastaun, Katsanevas, Katsavounidis, Katzman, Kawabe, K{\'{e}}f{\'{e}}lian,
  Keitel, Key, Khadka, Khalili, Khan, Khan, Khazanov, Khetan, Khursheed,
  Kijbunchoo, Kim, Kim, Kim, Kim, Kim, Kim, Kimball, King, Kinley-Hanlon,
  Kirchhoff, Kissel, Kleybolte, Klimenko, Knowles, Knyazev, Koch, Koehlenbeck,
  Koekoek, Koley, Kolstein, Komori, Kondrashov, Kontos, Koper, Korobko, Korth,
  Kovalam, Kozak, Kr{\"{a}}mer, Kringel, Krishnendu, Kr{\'{o}}lak, Kuehn,
  Kumar, Kumar, Kumar, Kumar, Kuns, Kwang, Lackey, Laghi, Lalande, Lam,
  Lamberts, Landry, Lane, Lang, Lange, Lantz, Lanza, {La Rosa},
  Lartaux-Vollard, Lasky, Laxen, Lazzarini, Lazzaro, Leaci, Leavey, Lecoeuche,
  Lee, Lee, Lee, Lee, Lehmann, Leon, Leroy, Letendre, Levin, Li, Li, Li, Li,
  Li, Linde, Linker, Linley, Littenberg, Liu, Liu, Llorens-Monteagudo, Lo,
  Lockwood, London, Longo, Lorenzini, Loriette, Lormand, Losurdo, Lough,
  Lousto, Lovelace, L{\"{u}}ck, Lumaca, Lundgren, Ma, Macas, Macinnis, Macleod,
  Macmillan, Macquet, {Maga{\~{n}}a Hernandez}, Maga{\~{n}}a-Sandoval,
  Magazz{\`{u}}, Magee, Majorana, Maksimovic, Maliakal, Malik, Man, Mandic,
  Mangano, Mansell, Manske, Mantovani, Mapelli, Marchesoni, Marion,
  M{\'{a}}rka, M{\'{a}}rka, Markakis, Markosyan, Markowitz, Maros, Marquina,
  Marsat, Martelli, Martin, Martin, Martinez, Martinez, Martynov, Masalehdan,
  Mason, Massera, Masserot, Massinger, Masso-Reid, Mastrogiovanni, Matas,
  Mateu-Lucena, Matichard, Matiushechkina, Mavalvala, Maynard, McCann,
  McCarthy, McClelland, McCormick, McCuller, McGuire, McIsaac, McIver, McManus,
  McRae, McWilliams, Meacher, Meadors, Mehmet, Mehta, Melatos, Melchor,
  Mendell, Menendez-Vazquez, Mercer, Mereni, Merfeld, Merilh, Merritt,
  Merzougui, Meshkov, Messenger, Messick, Metzdorff, Meyers, Meylahn, Mhaske,
  Miani, Miao, Michaloliakos, Michel, Middleton, Milano, Miller, Millhouse,
  Mills, Milotti, Milovich-Goff, Minazzoli, Minenkov, Mir, Mishkin, Mishra,
  Mistry, Mitra, Mitrofanov, Mitselmakher, Mittleman, Mo, Mogushi, Mohapatra,
  Mohite, Molina, Molina-Ruiz, Mondin, Montani, Moore, Moraru, Morawski,
  Moreno, Morisaki, Mours, Mow-Lowry, Mozzon, Muciaccia, Mukherjee, Mukherjee,
  Mukherjee, Mukherjee, Mukund, Mullavey, Munch, Mu{\~{n}}iz, Murray, Nadji,
  Nagar, Nardecchia, Naticchioni, Nayak, Neil, Neilson, Nelemans, Nelson, Nery,
  Neunzert, Nitz, Ng, Ng, Nguyen, Nguyen, Nguyen, Nichols, Nissanke, Nocera,
  Noh, North, Nothard, Nuttall, Oberling, O'Brien, O'Dell, Oganesyan, Ogin, Oh,
  Oh, Ohme, Ohta, Okada, Olivetto, Oppermann, Oram, O'Reilly, Ormiston, Ortega,
  O'Shaughnessy, Ossokine, Osthelder, Ottaway, Overmier, Owen, Pace, Pagano,
  Page, Pagliaroli, Pai, Pai, Palamos, Palashov, Palomba, Pan, Panda, Pang,
  Pankow, Pannarale, Pant, Paoletti, Paoli, Paolone, Parker, Pascucci,
  Pasqualetti, Passaquieti, Passuello, Patel, Patricelli, Payne, Pechsiri,
  Pedraza, Pegoraro, Pele, Penn, Perego, Perez, P{\'{e}}rigois, Perreca,
  Perri{\`{e}}s, Petermann, Petterson, Pfeiffer, Pham, Phukon, Piccinni,
  Pichot, Piendibene, Piergiovanni, Pierini, Pierro, Pillant, Pilo, Pinard,
  Pinto, Piotrzkowski, Pirello, Pitkin, Placidi, Plastino, Pluchar, Poggiani,
  Polini, Pong, Ponrathnam, Popolizio, Porter, Poverman, Powell, Pracchia,
  Prajapati, Prasai, Prasanna, Pratten, Prestegard, Principe, Prodi, Prokhorov,
  Prosposito, Prudenzi, Puecher, Punturo, Puosi, Puppo, P{\"{u}}rrer, Qi,
  Quetschke, Quinonez, Quitzow-James, Raab, Raaijmakers, Radkins, Radulesco,
  Raffai, Rafferty, Rail, Raja, Rajan, Rajbhandari, Rakhmanov, Ramirez,
  Ramirez, Ramos-Buades, Rana, Rao, Rapagnani, Rapol, Ratto, Raymond, Razzano,
  Read, Regimbau, Rei, Reid, Reitze, Rettegno, Ricci, Richardson, Richardson,
  Richardson, Ricker, Riemenschneider, Riles, Rizzo, Robertson, Robinet,
  Rocchi, Rocha, Rodriguez, Rodriguez-Soto, Rolland, Rollins, Roma, Romanelli,
  Romano, Romel, Romero, Romero-Shaw, Romie, Ronchini, Rose, Rose, Rose,
  Rosell, Rosi{\'{n}}ska, Rosofsky, Ross, Rowan, Rowlinson, Roy, Roy, Ruggi,
  Ryan, Sachdev, Sadecki, Sadiq, Sakellariadou, Salafia, Salconi, Saleem,
  Samajdar, Sanchez, Sanchez, Sanchez, Sanchis-Gual, Sanders, Sandles,
  Santiago, Santos, Saravanan, Sarin, Sassolas, Sathyaprakash, Sauter, Savage,
  Savant, Sawant, Sayah, Schaetzl, Schale, Scheel, Scheuer, Schindler-Tyka,
  Schmidt, Schnabel, Schofield, Sch{\"{o}}nbeck, Schreiber, Schulte, Schutz,
  Schwarm, Schwartz, Scott, Scott, Seglar-Arroyo, Seidel, Sellers, Sengupta,
  Sennett, Sentenac, Sequino, Sergeev, Setyawati, Shaffer, Shahriar, Sharifi,
  Sharma, Sharma, Shawhan, Shen, Shikauchi, Shink, Shoemaker, Shoemaker,
  Shukla, Shyamsundar, Sieniawska, Sigg, Singer, Singh, Singh, Singha, Singhal,
  Sintes, Sipala, Skliris, Slagmolen, Slaven-Blair, Smetana, Smith, Smith,
  Somala, Son, Soni, Soni, Sorazu, Sordini, Sorrentino, Sorrentino, Soulard,
  Souradeep, Sowell, Spencer, Spera, Srivastava, Srivastava, Staats, Stachie,
  Steer, Steinhoff, Steinke, Steinlechner, Steinlechner, Steinmeyer, Stevenson,
  Stolle-Mcallister, Stops, Stover, Strain, Stratta, Strunk, Sturani, Stuver,
  S{\"{u}}dbeck, Sudhagar, Sudhir, Suh, Summerscales, Sun, Sun, Sunil, Sur,
  Suresh, Sutton, Swinkels, Szczepa{\'{n}}czyk, Tacca, Tait, Talbot,
  Tanasijczuk, Tanner, Tao, Tapia, {Tapia San Martin}, Tasson, Taylor, Tenorio,
  Terkowski, Thirugnanasambandam, Thomas, Thomas, Thomas, Thompson, Thondapu,
  Thorne, Thrane, Tiwari, Tiwari, Tiwari, Toland, Tolley, Tonelli, Tornasi,
  Torres-Forn{\'{e}}, Torrie, {E Melo}, T{\"{o}}yr{\"{a}}, Tran, Trapananti,
  Travasso, Traylor, Tringali, Tripathee, Trovato, Trudeau, Tsai, Tsang, Tse,
  Tso, Tsukada, Tsuna, Tsutsui, Turconi, Ubhi, Udall, Ueno, Ugolini,
  Unnikrishnan, Urban, Usman, Utina, Vahlbruch, Vajente, Vajpeyi, Valdes,
  Valentini, Valsan, {Van Bakel}, {Van Beuzekom}, {Van Den Brand}, {Van Den
  Broeck}, Vander-Hyde, {Van Der Schaaf}, {Van Heijningen}, Vardaro, Vargas,
  Varma, Vass, Vas{\'{u}}th, Vecchio, Vedovato, Veitch, Veitch, Venkateswara,
  Venneberg, Venugopalan, Verkindt, Verma, Veske, Vetrano, Vicer{\'{e}}, Viets,
  Vijaykumar, Villa-Ortega, Vinet, Vitale, Vo, Vocca, Vorvick, Vyatchanin,
  Wade, Wade, Wade, Walet, Walker, Wallace, Wallace, Walsh, Wang, Wang, Wang,
  Wang, Ward, Warner, Was, Washington, Watchi, Weaver, Wei, Weinert, Weinstein,
  Weiss, Wellmann, Wen, We{\ss}els, Westhouse, Wette, Whelan, White, White,
  Whiting, Whittle, Wilken, Williams, Williams, Williamson, Willis, Willke,
  Wilson, Wimmer, Winkler, Wipf, Woan, Woehler, Wofford, Wong, Wrangel, Wright,
  Wu, Wysocki, Xiao, Yamamoto, Yang, Yang, Yang, Yap, Yeeles, Yoon, Yu, Yu,
  Yuen, Zadro{\.{z}}ny, Zanolin, Zelenova, Zendri, Zevin, Zhang, Zhang, Zhang,
  Zhang, Zhao, Zhao, Zheng, Zhou, Zhou, Zhu, Zimmerman, Zlochower, Zucker, \&
  Zweizig}]{GWTC3_pops}
---. 2021{\natexlab{d}}, arXiv e-prints.
\newblock \doarXiv{2111.03634}

\bibitem[{Abbott {et~al.}(2021{\natexlab{e}})Abbott, Abbott, Abraham, Acernese,
  Ackley, Adams, Adams, Adhikari, Adya, Affeldt, Agathos, Agatsuma, Aggarwal,
  Aguiar, Aiello, Ain, Ajith, Akcay, Allen, Allocca, Altin, Amato, Anand,
  Ananyeva, Anderson, Anderson, Angelova, Ansoldi, Antelis, Antier, Appert,
  Arai, Araya, Areeda, Ar{\`{e}}ne, Arnaud, Aronson, Arun, Asali, Ascenzi,
  Ashton, Aston, Astone, Aubin, Aufmuth, Aultoneal, Austin, Avendano, Babak,
  Badaracco, Bader, Bae, Baer, Bagnasco, Baird, Ball, Ballardin, Ballmer, Bals,
  Balsamo, Baltus, Banagiri, Bankar, Bankar, Barayoga, Barbieri, Barish,
  Barker, Barneo, Barnum, Barone, Barr, Barsotti, Barsuglia, Barta, Bartlett,
  Bartos, Bassiri, Basti, Bawaj, Bayley, Bazzan, Becher, B{\'{e}}csy,
  Bedakihale, Bejger, Belahcene, Beniwal, Benjamin, Bennett, Bentley, Bergamin,
  Berger, Bergmann, Bernuzzi, Berry, Bersanetti, Bertolini, Betzwieser,
  Bhandare, Bhandari, Bhattacharjee, Bidler, Bilenko, Billingsley, Birney,
  Birnholtz, Biscans, Bischi, Biscoveanu, Bisht, Bitossi, Bizouard, Blackburn,
  Blackman, Blair, Blair, Blair, Blanch, Bobba, Bode, Boer, Boetzel, Bogaert,
  Boldrini, Bondu, Bonilla, Bonnand, Booker, Boom, Bork, Boschi, Bose,
  Bossilkov, Boudart, Bouffanais, Bozzi, Bradaschia, Brady, Bramley, Branchesi,
  Brau, Breschi, Briant, Briggs, Brighenti, Brillet, Brinkmann, Brockill,
  Brooks, Brooks, Brown, Brunett, Bruno, Bruntz, Buikema, Bulik, Bulten,
  Buonanno, Buscicchio, Buskulic, Byer, Cabero, Cadonati, Caesar, Cagnoli,
  Cahillane, {Calder{\'{o}}n Bustillo}, Callaghan, Callister, Calloni, Camp,
  Canepa, Cannon, Cao, Cao, Carapella, Carbognani, Carney, Carpinelli, Carullo,
  Carver, {Casanueva Diaz}, Casentini, Caudill, Cavagli{\`{a}}, Cavalier,
  Cavalieri, Cella, Cerd{\'{a}}-Dur{\'{a}}n, Cesarini, Chaibi, Chakravarti,
  Chan, Chan, Chandra, Chanial, Chao, Charlton, Chase, Chassande-Mottin,
  Chatterjee, Chattopadhyay, Chaturvedi, Chatziioannou, Chen, Chen, Chen, Chen,
  Cheng, Cheong, Chia, Chiadini, Chierici, Chincarini, Chiummo, Cho, Cho, Cho,
  Choate, Christensen, Chu, Chua, Chung, Chung, Ciani, Ciecielag,
  Cie{\'{s}}lar, Cifaldi, Ciobanu, Ciolfi, Cipriano, Cirone, Clara, Clark,
  Clark, Clarke, Clearwater, Clesse, Cleva, Coccia, Cohadon, Cohen, Colleoni,
  Collette, Collins, Colpi, Constancio, Conti, Cooper, Corban, Corbitt,
  Cordero-Carri{\'{o}}n, Corezzi, Corley, Cornish, Corre, Corsi, Cortese,
  Costa, Cotesta, Coughlin, Coughlin, Coulon, Countryman, Cousins, Couvares,
  Covas, Coward, Cowart, Coyne, Coyne, Creighton, Creighton, Croquette,
  Crowder, Cudell, Cullen, Cumming, Cummings, Cunningham, Cuoco, Cury{\l}o,
  Canton, D{\'{a}}lya, Dana, Daneshgaranbajastani, D'Angelo, Danila,
  Danilishin, D'Antonio, Danzmann, Darsow-Fromm, Dasgupta, Datrier, Dattilo,
  Dave, Davier, Davies, Davis, Daw, Dean, Debra, Deenadayalan, Degallaix, {De
  Laurentis}, Del{\'{e}}glise, {Del Favero}, {De Lillo}, {De Lillo}, {Del
  Pozzo}, Demarchi, {De Matteis}, D'Emilio, Demos, Denker, Dent, Depasse, {De
  Pietri}, {De Rosa}, {De Rossi}, Desalvo, {De Varona}, Dhurandhar, D{\'{i}}az,
  Diaz-Ortiz, Didio, Dietrich, {Di Fiore}, Difronzo, {Di Giorgio}, {Di
  Giovanni}, {Di Giovanni}, {Di Girolamo}, {Di Lieto}, Ding, {Di Pace}, {Di
  Palma}, {Di Renzo}, Divakarla, Dmitriev, Doctor, D'Onofrio, Donovan, Dooley,
  Doravari, Dorrington, Downes, Drago, Driggers, Du, Ducoin, Dupej, Durante,
  D'Urso, Duverne, Dwyer, Easter, Eddolls, Edelman, Edo, Edy, Effler, Eichholz,
  Eikenberry, Eisenmann, Eisenstein, Ejlli, Errico, Essick, Estell{\'{e}}s,
  Estevez, Etienne, Etzel, Evans, Evans, Ewing, Fafone, Fair, Fairhurst, Fan,
  Farah, Farinon, Farr, Farr, Fauchon-Jones, Favata, Fays, Fazio, Feicht,
  Fejer, Feng, Fenyvesi, Ferguson, Fernandez-Galiana, Ferrante, Ferreira,
  Fidecaro, Figura, Fiori, Fiorucci, Fishbach, Fisher, Fishner, Fittipaldi,
  Fitz-Axen, Fiumara, Flaminio, Floden, Flynn, Fong, Font, Forsyth, Fournier,
  Frasca, Frasconi, Frei, Freise, Frey, Frey, Fritschel, Frolov, Fronz{\'{e}},
  Fulda, Fyffe, Gabbard, Gadre, Gaebel, Gair, Gais, Galaudage, Gamba,
  Ganapathy, Ganguly, Gaonkar, Garaventa, Garc{\'{i}}a-Quir{\'{o}}s, Garufi,
  Gateley, Gaudio, Gayathri, Gemme, Gennai, George, George, George, Gergely,
  Ghonge, Ghosh, Ghosh, Ghosh, Giacomazzo, Giacoppo, Giaime, Giardina, Gibson,
  Gier, Gill, Giri, Glanzer, Gleckl, Godwin, Goetz, Goetz, Gohlke, Goncharov,
  Gonz{\'{a}}lez, Gopakumar, Gossan, Gosselin, Gouaty, Grace, Grado, Granata,
  Granata, Grant, Gras, Grassia, Gray, Gray, Greco, Green, Green, Gretarsson,
  Griggs, Grignani, Grimaldi, Grimes, Grimm, Grote, Grunewald, Gruning,
  Guerrero, Guidi, Guimaraes, Guix{\'{e}}, Gulati, Guo, Gupta, Gupta, Gupta,
  Gustafson, Gustafson, Guzman, Haegel, Halim, Hall, Hamilton, Hammond, Haney,
  Hanke, Hanks, Hanna, Hannam, Hannuksela, Hannuksela, Hansen, Hansen, Hanson,
  Harder, Hardwick, Haris, Harms, Harry, Harry, Hartwig, Hasskew, Haster,
  Haughian, Hayes, Healy, Heidmann, Heintze, Heinze, Heinzel, Heitmann,
  Hellman, Hello, Helmling-Cornell, Hemming, Hendry, Heng, Hennes, Hennig,
  Hennig, {Hernandez Vivanco}, Heurs, Hild, Hill, Hines, Hochheim, Hofgard,
  Hofman, Hohmann, Holgado, Holland, Hollows, Holmes, Holt, Holz, Hopkins,
  Horst, Hough, Howell, Hoy, Hoyland, Huang, H{\"{u}}bner, Huddart, Huerta,
  Hughey, Hui, Husa, Huttner, Hutzler, Huxford, Huynh-Dinh, Idzkowski, Iess,
  Imperato, Inchauspe, Ingram, Intini, Isi, Iyer, Jaberianhamedan, Jacqmin,
  Jadhav, Jadhav, James, Jani, Janssens, Janthalur, Jaranowski, Jariwala,
  Jaume, Jenkins, Jeunon, Jiang, Johns, Johnson-Mcdaniel, Jones, Jones, Jones,
  Jones, Jones, Jonker, Ju, Junker, Kalaghatgi, Kalogera, Kamai, Kandhasamy,
  Kang, Kanner, Kapadia, Kapasi, Karathanasis, Karki, Kashyap, Kasprzack,
  Kastaun, Katsanevas, Katsavounidis, Katzman, Kawabe, K{\'{e}}f{\'{e}}lian,
  Keitel, Key, Khadka, Khalili, Khan, Khan, Khazanov, Khetan, Khursheed,
  Kijbunchoo, Kim, Kim, Kim, Kim, Kim, Kim, Kimball, King, Kinley-Hanlon,
  Kirchhoff, Kissel, Kleybolte, Klimenko, Knowles, Knyazev, Koch, Koehlenbeck,
  Koekoek, Koley, Kolstein, Komori, Kondrashov, Kontos, Koper, Korobko, Korth,
  Kovalam, Kozak, Kr{\"{a}}mer, Kringel, Krishnendu, Kr{\'{o}}lak, Kuehn,
  Kumar, Kumar, Kumar, Kumar, Kuns, Kwang, Lackey, Laghi, Lalande, Lam,
  Lamberts, Landry, Lane, Lang, Lange, Lantz, Lanza, {La Rosa},
  Lartaux-Vollard, Lasky, Laxen, Lazzarini, Lazzaro, Leaci, Leavey, Lecoeuche,
  Lee, Lee, Lee, Lee, Lehmann, Leon, Leroy, Letendre, Levin, Li, Li, Li, Li,
  Li, Linde, Linker, Linley, Littenberg, Liu, Liu, Llorens-Monteagudo, Lo,
  Lockwood, London, Longo, Lorenzini, Loriette, Lormand, Losurdo, Lough,
  Lousto, Lovelace, L{\"{u}}ck, Lumaca, Lundgren, Ma, Macas, Macinnis, Macleod,
  Macmillan, Macquet, {Maga{\~{n}}a Hernandez}, Maga{\~{n}}a-Sandoval,
  Magazz{\`{u}}, Magee, Majorana, Maksimovic, Maliakal, Malik, Man, Mandic,
  Mangano, Mansell, Manske, Mantovani, Mapelli, Marchesoni, Marion,
  M{\'{a}}rka, M{\'{a}}rka, Markakis, Markosyan, Markowitz, Maros, Marquina,
  Marsat, Martelli, Martin, Martin, Martinez, Martinez, Martynov, Masalehdan,
  Mason, Massera, Masserot, Massinger, Masso-Reid, Mastrogiovanni, Matas,
  Mateu-Lucena, Matichard, Matiushechkina, Mavalvala, Maynard, McCann,
  McCarthy, McClelland, McCormick, McCuller, McGuire, McIsaac, McIver, McManus,
  McRae, McWilliams, Meacher, Meadors, Mehmet, Mehta, Melatos, Melchor,
  Mendell, Menendez-Vazquez, Mercer, Mereni, Merfeld, Merilh, Merritt,
  Merzougui, Meshkov, Messenger, Messick, Metzdorff, Meyers, Meylahn, Mhaske,
  Miani, Miao, Michaloliakos, Michel, Middleton, Milano, Miller, Millhouse,
  Mills, Milotti, Milovich-Goff, Minazzoli, Minenkov, Mir, Mishkin, Mishra,
  Mistry, Mitra, Mitrofanov, Mitselmakher, Mittleman, Mo, Mogushi, Mohapatra,
  Mohite, Molina, Molina-Ruiz, Mondin, Montani, Moore, Moraru, Morawski,
  Moreno, Morisaki, Mours, Mow-Lowry, Mozzon, Muciaccia, Mukherjee, Mukherjee,
  Mukherjee, Mukherjee, Mukund, Mullavey, Munch, Mu{\~{n}}iz, Murray, Nadji,
  Nagar, Nardecchia, Naticchioni, Nayak, Neil, Neilson, Nelemans, Nelson, Nery,
  Neunzert, Nitz, Ng, Ng, Nguyen, Nguyen, Nguyen, Nichols, Nissanke, Nocera,
  Noh, North, Nothard, Nuttall, Oberling, O'Brien, O'Dell, Oganesyan, Ogin, Oh,
  Oh, Ohme, Ohta, Okada, Olivetto, Oppermann, Oram, O'Reilly, Ormiston, Ortega,
  O'Shaughnessy, Ossokine, Osthelder, Ottaway, Overmier, Owen, Pace, Pagano,
  Page, Pagliaroli, Pai, Pai, Palamos, Palashov, Palomba, Pan, Panda, Pang,
  Pankow, Pannarale, Pant, Paoletti, Paoli, Paolone, Parker, Pascucci,
  Pasqualetti, Passaquieti, Passuello, Patel, Patricelli, Payne, Pechsiri,
  Pedraza, Pegoraro, Pele, Penn, Perego, Perez, P{\'{e}}rigois, Perreca,
  Perri{\`{e}}s, Petermann, Petterson, Pfeiffer, Pham, Phukon, Piccinni,
  Pichot, Piendibene, Piergiovanni, Pierini, Pierro, Pillant, Pilo, Pinard,
  Pinto, Piotrzkowski, Pirello, Pitkin, Placidi, Plastino, Pluchar, Poggiani,
  Polini, Pong, Ponrathnam, Popolizio, Porter, Poverman, Powell, Pracchia,
  Prajapati, Prasai, Prasanna, Pratten, Prestegard, Principe, Prodi, Prokhorov,
  Prosposito, Prudenzi, Puecher, Punturo, Puosi, Puppo, P{\"{u}}rrer, Qi,
  Quetschke, Quinonez, Quitzow-James, Raab, Raaijmakers, Radkins, Radulesco,
  Raffai, Rafferty, Rail, Raja, Rajan, Rajbhandari, Rakhmanov, Ramirez,
  Ramirez, Ramos-Buades, Rana, Rao, Rapagnani, Rapol, Ratto, Raymond, Razzano,
  Read, Regimbau, Rei, Reid, Reitze, Rettegno, Ricci, Richardson, Richardson,
  Richardson, Ricker, Riemenschneider, Riles, Rizzo, Robertson, Robinet,
  Rocchi, Rocha, Rodriguez, Rodriguez-Soto, Rolland, Rollins, Roma, Romanelli,
  Romano, Romel, Romero, Romero-Shaw, Romie, Ronchini, Rose, Rose, Rose,
  Rosell, Rosi{\'{n}}ska, Rosofsky, Ross, Rowan, Rowlinson, Roy, Roy, Ruggi,
  Ryan, Sachdev, Sadecki, Sadiq, Sakellariadou, Salafia, Salconi, Saleem,
  Samajdar, Sanchez, Sanchez, Sanchez, Sanchis-Gual, Sanders, Sandles,
  Santiago, Santos, Saravanan, Sarin, Sassolas, Sathyaprakash, Sauter, Savage,
  Savant, Sawant, Sayah, Schaetzl, Schale, Scheel, Scheuer, Schindler-Tyka,
  Schmidt, Schnabel, Schofield, Sch{\"{o}}nbeck, Schreiber, Schulte, Schutz,
  Schwarm, Schwartz, Scott, Scott, Seglar-Arroyo, Seidel, Sellers, Sengupta,
  Sennett, Sentenac, Sequino, Sergeev, Setyawati, Shaffer, Shahriar, Sharifi,
  Sharma, Sharma, Shawhan, Shen, Shikauchi, Shink, Shoemaker, Shoemaker,
  Shukla, Shyamsundar, Sieniawska, Sigg, Singer, Singh, Singh, Singha, Singhal,
  Sintes, Sipala, Skliris, Slagmolen, Slaven-Blair, Smetana, Smith, Smith,
  Somala, Son, Soni, Soni, Sorazu, Sordini, Sorrentino, Sorrentino, Soulard,
  Souradeep, Sowell, Spencer, Spera, Srivastava, Srivastava, Staats, Stachie,
  Steer, Steinhoff, Steinke, Steinlechner, Steinlechner, Steinmeyer, Stevenson,
  Stolle-Mcallister, Stops, Stover, Strain, Stratta, Strunk, Sturani, Stuver,
  S{\"{u}}dbeck, Sudhagar, Sudhir, Suh, Summerscales, Sun, Sun, Sunil, Sur,
  Suresh, Sutton, Swinkels, Szczepa{\'{n}}czyk, Tacca, Tait, Talbot,
  Tanasijczuk, Tanner, Tao, Tapia, {Tapia San Martin}, Tasson, Taylor, Tenorio,
  Terkowski, Thirugnanasambandam, Thomas, Thomas, Thomas, Thompson, Thondapu,
  Thorne, Thrane, Tiwari, Tiwari, Tiwari, Toland, Tolley, Tonelli, Tornasi,
  Torres-Forn{\'{e}}, Torrie, {E Melo}, T{\"{o}}yr{\"{a}}, Tran, Trapananti,
  Travasso, Traylor, Tringali, Tripathee, Trovato, Trudeau, Tsai, Tsang, Tse,
  Tso, Tsukada, Tsuna, Tsutsui, Turconi, Ubhi, Udall, Ueno, Ugolini,
  Unnikrishnan, Urban, Usman, Utina, Vahlbruch, Vajente, Vajpeyi, Valdes,
  Valentini, Valsan, {Van Bakel}, {Van Beuzekom}, {Van Den Brand}, {Van Den
  Broeck}, Vander-Hyde, {Van Der Schaaf}, {Van Heijningen}, Vardaro, Vargas,
  Varma, Vass, Vas{\'{u}}th, Vecchio, Vedovato, Veitch, Veitch, Venkateswara,
  Venneberg, Venugopalan, Verkindt, Verma, Veske, Vetrano, Vicer{\'{e}}, Viets,
  Vijaykumar, Villa-Ortega, Vinet, Vitale, Vo, Vocca, Vorvick, Vyatchanin,
  Wade, Wade, Wade, Walet, Walker, Wallace, Wallace, Walsh, Wang, Wang, Wang,
  Wang, Ward, Warner, Was, Washington, Watchi, Weaver, Wei, Weinert, Weinstein,
  Weiss, Wellmann, Wen, We{\ss}els, Westhouse, Wette, Whelan, White, White,
  Whiting, Whittle, Wilken, Williams, Williams, Williamson, Willis, Willke,
  Wilson, Wimmer, Winkler, Wipf, Woan, Woehler, Wofford, Wong, Wrangel, Wright,
  Wu, Wysocki, Xiao, Yamamoto, Yang, Yang, Yang, Yap, Yeeles, Yoon, Yu, Yu,
  Yuen, Zadro{\.{z}}ny, Zanolin, Zelenova, Zendri, Zevin, Zhang, Zhang, Zhang,
  Zhang, Zhao, Zhao, Zheng, Zhou, Zhou, Zhu, Zimmerman, Zlochower, Zucker, \&
  Zweizig}]{GWTC2}
---. 2021{\natexlab{e}}, Physical Review X, 11, 21053,
  \dodoi{10.1103/PhysRevX.11.021053}

\bibitem[{Alves {et~al.}(2020)Alves, Forveille, Pentericci, \&
  Shore}]{PlanckCollaboration2018}
Alves, J., Forveille, T., Pentericci, L., \& Shore, S. 2020, Astronomy and
  Astrophysics, 641, 1, \dodoi{10.1051/0004-6361/202039265}

\bibitem[{Bavera {et~al.}(2021{\natexlab{a}})Bavera, Zevin, \&
  Fragos}]{Bavera2021b}
Bavera, S.~S., Zevin, M., \& Fragos, T. 2021{\natexlab{a}}, Research Notes of
  the American Astronomical Society, 5, 127.
\newblock \doarXiv{2105.09077}

\bibitem[{Bavera {et~al.}(2020)Bavera, Fragos, Qin, Zapartas, Neijssel, Mandel,
  Batta, Gaebel, Kimball, \& Stevenson}]{Bavera2020}
Bavera, S.~S., Fragos, T., Qin, Y., {et~al.} 2020, Astronomy {\&} Astrophysics,
  635, A97, \dodoi{10.1051/0004-6361/201936204}

\bibitem[{Bavera {et~al.}(2021{\natexlab{b}})Bavera, Fragos, Zevin, Berry,
  Marchant, Andrews, Coughlin, Dotter, Kovlakas, Misra, Serra-Perez, Qin,
  Rocha, Rom{\'{a}}n-Garza, Tran, \& Zapartas}]{Bavera2021}
Bavera, S.~S., Fragos, T., Zevin, M., {et~al.} 2021{\natexlab{b}}, Astronomy
  and Astrophysics, 647, A153, \dodoi{10.1051/0004-6361/202039804}

\bibitem[{Belczynski {et~al.}(2011)Belczynski, Bulik, \&
  Bailyn}]{Belczynski2011}
Belczynski, K., Bulik, T., \& Bailyn, C. 2011, Astrophysical Journal Letters,
  742, 4, \dodoi{10.1088/2041-8205/742/1/L2}

\bibitem[{Belczynski {et~al.}(2012)Belczynski, Bulik, \&
  Fryer}]{Belczynski2012}
Belczynski, K., Bulik, T., \& Fryer, C.~L. 2012, arXiv e-prints.
\newblock \doarXiv{1208.2422}

\bibitem[{Belczynski {et~al.}(2008)Belczynski, Kalogera, Rasio, Taam, Zezas,
  Bulik, Maccarone, \& Ivanova}]{Belczynski2008}
Belczynski, K., Kalogera, V., Rasio, F.~A., {et~al.} 2008, The Astrophysical
  Journal Supplement Series, 174, 223, \dodoi{10.1086/521026}

\bibitem[{Belczynski {et~al.}(2020)Belczynski, Klencki, Fields, Olejak, Berti,
  Meynet, Fryer, Holz, \& Shaughnessy}]{Belczynski2020}
Belczynski, K., Klencki, J., Fields, C.~E., {et~al.} 2020, Astronomy {\&}
  Astrophysics, 636, A104, \dodoi{10.1051/0004-6361/201936528}

\bibitem[{Belczynski {et~al.}(2022)Belczynski, Romagnolo, Olejak, Klencki,
  Chattopadhyay, Stevenson, {Coleman Miller}, Lasota, \&
  Crowther}]{Belczynski2022a}
Belczynski, K., Romagnolo, A., Olejak, A., {et~al.} 2022, The Astrophysical
  Journal, 925, 69, \dodoi{10.3847/1538-4357/ac375a}

\bibitem[{Biscoveanu {et~al.}(2021)Biscoveanu, Isi, Vitale, \&
  Varma}]{Biscoveanu2021a}
Biscoveanu, S., Isi, M., Vitale, S., \& Varma, V. 2021, Physical Review
  Letters, 126, 171103, \dodoi{10.1103/PhysRevLett.126.171103}

\bibitem[{Bouffanais {et~al.}(2019)Bouffanais, Mapelli, Gerosa, {Di Carlo},
  Giacobbo, Berti, \& Baibhav}]{Bouffanais2019}
Bouffanais, Y., Mapelli, M., Gerosa, D., {et~al.} 2019, The Astrophysical
  Journal, 886, 25, \dodoi{10.3847/1538-4357/ab4a79}

\bibitem[{Bouffanais {et~al.}(2021{\natexlab{a}})Bouffanais, Mapelli,
  Santoliquido, Giacobbo, {Di Carlo}, Rastello, Artale, \&
  Iorio}]{Bouffanais2021}
Bouffanais, Y., Mapelli, M., Santoliquido, F., {et~al.} 2021{\natexlab{a}},
  Monthly Notices of the Royal Astronomical Society, 507, 5224,
  \dodoi{10.1093/mnras/stab2438}

\bibitem[{Bouffanais {et~al.}(2021{\natexlab{b}})Bouffanais, Mapelli,
  Santoliquido, Giacobbo, Iorio, \& Costa}]{Bouffanais2021a}
---. 2021{\natexlab{b}}, Monthly Notices of the Royal Astronomical Society,
  505, 3873, \dodoi{10.1093/mnras/stab1589}

\bibitem[{Breivik {et~al.}(2020)Breivik, Coughlin, Zevin, Rodriguez, Kremer,
  Ye, Andrews, Kurkowski, Digman, Larson, \& Rasio}]{Breivik2020}
Breivik, K., Coughlin, S., Zevin, M., {et~al.} 2020, The Astrophysical Journal,
  898, 71, \dodoi{10.3847/1538-4357/ab9d85}

\bibitem[{Broekgaarden {et~al.}(2022)Broekgaarden, Stevenson, \&
  Thrane}]{Broekgaarden2022}
Broekgaarden, F.~S., Stevenson, S., \& Thrane, E. 2022, arXiv e-prints.
\newblock \doarXiv{2205.01693}

\bibitem[{Broekgaarden {et~al.}(2021)Broekgaarden, Berger, Neijssel,
  Vigna-G{\'{o}}mez, Chattopadhyay, Stevenson, Chruslinska, Justham, {De Mink},
  \& Mandel}]{Broekgaarden2021}
Broekgaarden, F.~S., Berger, E., Neijssel, C.~J., {et~al.} 2021, Monthly
  Notices of the Royal Astronomical Society, 508, 5028,
  \dodoi{10.1093/mnras/stab2716}

\bibitem[{Brown(1995)}]{Brown1995}
Brown, G. 1995, The Astrophysical Journal, 440, 270, \dodoi{10.1086/175268}

\bibitem[{Callister {et~al.}(2021)Callister, Haster, Ng, Vitale, \&
  Farr}]{Callister2021a}
Callister, T.~A., Haster, C.-J., Ng, K. K.~Y., Vitale, S., \& Farr, W.~M. 2021,
  The Astrophysical Journal Letters, 922, L5, \dodoi{10.3847/2041-8213/ac2ccc}

\bibitem[{Claeys {et~al.}(2014)Claeys, Pols, Izzard, Vink, \&
  Verbunt}]{Claeys2014}
Claeys, J. S.~W., Pols, O.~R., Izzard, R.~G., Vink, J., \& Verbunt, F. W.~M.
  2014, Astronomy {\&} Astrophysics, 563, A83,
  \dodoi{10.1051/0004-6361/201322714}

\bibitem[{Cruz-Osorio {et~al.}(2021)Cruz-Osorio, Lora-Clavijo, \&
  Herdeiro}]{Cruz-Osorio2021}
Cruz-Osorio, A., Lora-Clavijo, F.~D., \& Herdeiro, C. 2021, Journal of
  Cosmology and Astroparticle Physics, 2021, 32,
  \dodoi{10.1088/1475-7516/2021/07/032}

\bibitem[{Cruz-Osorio \& Rezzolla(2020)}]{Cruz-Osorio2020}
Cruz-Osorio, A., \& Rezzolla, L. 2020, The Astrophysical Journal, 894, 147,
  \dodoi{10.3847/1538-4357/ab89aa}

\bibitem[{Damour(2001)}]{Damour2001}
Damour, T. 2001, Physical Review D, 64, 124013,
  \dodoi{10.1103/PhysRevD.64.124013}

\bibitem[{{De Mink} \& Mandel(2016)}]{DeMink2016}
{De Mink}, S.~E., \& Mandel, I. 2016, Monthly Notices of the Royal Astronomical
  Society, 460, 3545, \dodoi{10.1093/mnras/stw1219}

\bibitem[{Detmers {et~al.}(2008)Detmers, Langer, Podsiadlowski, \&
  Izzard}]{Detmers2008}
Detmers, R.~G., Langer, N., Podsiadlowski, P., \& Izzard, R.~G. 2008, Astronomy
  and Astrophysics, 484, 831, \dodoi{10.1051/0004-6361:200809371}

\bibitem[{Dewi {et~al.}(2006)Dewi, Podsiadlowski, \& Sena}]{Dewi2006}
Dewi, J.~D., Podsiadlowski, P., \& Sena, A. 2006, Monthly Notices of the Royal
  Astronomical Society, 368, 1742, \dodoi{10.1111/j.1365-2966.2006.10233.x}

\bibitem[{du~Buisson {et~al.}(2020)du~Buisson, Marchant, Podsiadlowski,
  Kobayashi, Abdalla, Taylor, Mandel, de~Mink, Moriya, \&
  Langer}]{duBuisson2020}
du~Buisson, L., Marchant, P., Podsiadlowski, P., {et~al.} 2020, Monthly Notices
  of the Royal Astronomical Society, 499, 5941, \dodoi{10.1093/mnras/staa3225}

\bibitem[{Fishbach \& Holz(2017)}]{Fishbach2017a}
Fishbach, M., \& Holz, D.~E. 2017, The Astrophysical Journal, 851, L25,
  \dodoi{10.3847/2041-8213/aa9bf6}

\bibitem[{Fragos {et~al.}(2022)Fragos, Andrews, Bavera, Berry, Coughlin,
  Dotter, Giri, Kalogera, Katsaggelos, Kovlakas, Lalvani, Misra, Srivastava,
  Qin, Rocha, Roman-Garza, Serra, Stahle, Sun, Teng, Trajcevski, Tran, Xing,
  Zapartas, \& Zevin}]{Fragos2022}
Fragos, T., Andrews, J.~J., Bavera, S.~S., {et~al.} 2022, arXiv e-prints.
\newblock \doarXiv{2202.05892}

\bibitem[{Fryer {et~al.}(2012)Fryer, Belczynski, Wiktorowicz, Dominik,
  Kalogera, \& Holz}]{Fryer2012}
Fryer, C.~L., Belczynski, K., Wiktorowicz, G., {et~al.} 2012, The Astrophysical
  Journal, 749, 14, \dodoi{10.1088/0004-637X/749/1/91}

\bibitem[{Fuller \& Lu(2022)}]{Fuller2022}
Fuller, J., \& Lu, W. 2022, arXiv e-prints.
\newblock \doarXiv{2201.08407}

\bibitem[{Fuller \& Ma(2019)}]{Fuller2019b}
Fuller, J., \& Ma, L. 2019, The Astrophysical Journal Letters, 881, L1,
  \dodoi{10.3847/2041-8213/ab339b}

\bibitem[{Fuller {et~al.}(2019)Fuller, Piro, \& Jermyn}]{Fuller2019a}
Fuller, J., Piro, A.~L., \& Jermyn, A.~S. 2019, Monthly Notices of the Royal
  Astronomical Society, 485, 3661, \dodoi{10.1093/mnras/stz514}

\bibitem[{Gallegos-Garcia {et~al.}(2021)Gallegos-Garcia, Berry, Marchant, \&
  Kalogera}]{Gallegos-Garcia2021}
Gallegos-Garcia, M., Berry, C. P.~L., Marchant, P., \& Kalogera, V. 2021, The
  Astrophysical Journal, 922, 110, \dodoi{10.3847/1538-4357/ac2610}

\bibitem[{Giacobbo \& Mapelli(2018)}]{Giacobbo2018b}
Giacobbo, N., \& Mapelli, M. 2018, Monthly Notices of the Royal Astronomical
  Society, 480, 2011, \dodoi{10.1093/mnras/sty1999}

\bibitem[{Giacobbo {et~al.}(2018)Giacobbo, Mapelli, \& Spera}]{Giacobbo2018a}
Giacobbo, N., Mapelli, M., \& Spera, M. 2018, Monthly Notices of the Royal
  Astronomical Society, 474, 2959, \dodoi{10.1093/mnras/stx2933}

\bibitem[{Grevesse \& Sauval(1998)}]{Grevesse1998}
Grevesse, N., \& Sauval, A.~J. 1998, Space Science Reviews, 85, 161,
  \dodoi{10.1023/A:1005161325181}

\bibitem[{Hamers \& Safarzadeh(2020)}]{Hamers2020}
Hamers, A.~S., \& Safarzadeh, M. 2020, The Astrophysical Journal, 99, 99,
  \dodoi{10.3847/1538-4357/ab9b27}

\bibitem[{Heger {et~al.}(2005)Heger, Woosley, \& Spruit}]{Heger2005}
Heger, A., Woosley, S.~E., \& Spruit, H.~C. 2005, The Astrophysical Journal,
  626, 350, \dodoi{10.1086/429868}

\bibitem[{Hobbs {et~al.}(2005)Hobbs, Lorimer, Lyne, \& Kramer}]{Hobbs2005}
Hobbs, G., Lorimer, D.~R., Lyne, A.~G., \& Kramer, M. 2005, Monthly Notices of
  the Royal Astronomical Society, 360, 974,
  \dodoi{10.1111/j.1365-2966.2005.09087.x}

\bibitem[{Hotokezaka \& Piran(2017)}]{Hotokezaka2017}
Hotokezaka, K., \& Piran, T. 2017, The Astrophysical Journal, 842, 111,
  \dodoi{10.3847/1538-4357/aa6f61}

\bibitem[{Hu {et~al.}(2022)Hu, Zhu, Qin, Zhang, Liang, \& Shao}]{Hu2022}
Hu, R.-C., Zhu, J.-P., Qin, Y., {et~al.} 2022, The Astrophysical Journal, 928,
  163, \dodoi{10.3847/1538-4357/ac573f}

\bibitem[{Hunter(2007)}]{matplotlib}
Hunter, J.~D. 2007, Computing in Science and Engineering, 9, 99,
  \dodoi{10.1109/MCSE.2007.55}

\bibitem[{Hurley {et~al.}(2000)Hurley, Pols, \& Tout}]{Hurley2000}
Hurley, J.~R., Pols, O.~R., \& Tout, C.~A. 2000, Monthly Notices of the Royal
  Astronomical Society, 315, 543, \dodoi{10.1046/j.1365-8711.2000.03426.x}

\bibitem[{Hurley {et~al.}(2002)Hurley, Tout, \& Pols}]{Hurley2002}
Hurley, J.~R., Tout, C.~A., \& Pols, O.~R. 2002, Monthly Notices of the Royal
  Astronomical Society, 329, 897, \dodoi{10.1046/j.1365-8711.2002.05038.x}

\bibitem[{Kroupa(2001)}]{Kroupa2001}
Kroupa, P. 2001, Monthly Notices of the Royal Astronomical Society, 322, 231,
  \dodoi{10.1046/j.1365-8711.2001.04022.x}

\bibitem[{Kruckow {et~al.}(2018)Kruckow, Tauris, Langer, Kramer, \&
  Izzard}]{Kruckow2018}
Kruckow, M.~U., Tauris, T.~M., Langer, N., Kramer, M., \& Izzard, R.~G. 2018,
  Monthly Notices of the Royal Astronomical Society, 481, 1908,
  \dodoi{10.1093/MNRAS/STY2190}

\bibitem[{Kushnir {et~al.}(2017)Kushnir, Zaldarriaga, Kollmeier, \&
  Waldman}]{Kushnir2017}
Kushnir, D., Zaldarriaga, M., Kollmeier, J.~A., \& Waldman, R. 2017, Monthly
  Notices of the Royal Astronomical Society, 467, 2146,
  \dodoi{10.1093/mnras/stx255}

\bibitem[{MacLeod \& Ramirez-Ruiz(2015)}]{MacLeod2015a}
MacLeod, M., \& Ramirez-Ruiz, E. 2015, Astrophysical Journal, 803, 41,
  \dodoi{10.1088/0004-637X/803/1/41}

\bibitem[{Mandel \& {De Mink}(2016)}]{Mandel2016b}
Mandel, I., \& {De Mink}, S.~E. 2016, Monthly Notices of the Royal Astronomical
  Society, 458, 2634, \dodoi{10.1093/mnras/stw379}

\bibitem[{Mandel \& Farmer(2022)}]{Mandel2022}
Mandel, I., \& Farmer, A. 2022, Physics Reports, 955, 1,
  \dodoi{10.1016/j.physrep.2022.01.003}

\bibitem[{Mandel {et~al.}(2019)Mandel, Farr, \& Gair}]{Mandel2019}
Mandel, I., Farr, W.~M., \& Gair, J.~R. 2019, Monthly Notices of the Royal
  Astronomical Society, 486, 1086, \dodoi{10.1093/mnras/stz896}

\bibitem[{Mandel \& Fragos(2020)}]{Mandel2020}
Mandel, I., \& Fragos, T. 2020, The Astrophysical Journal, 895, L28,
  \dodoi{10.3847/2041-8213/ab8e41}

\bibitem[{Marchant {et~al.}(2016)Marchant, Langer, Podsiadlowski, Tauris, \&
  Moriya}]{Marchant2016}
Marchant, P., Langer, N., Podsiadlowski, P., Tauris, T., \& Moriya, T. 2016,
  Astronomy {\&} Astrophysics, 588, A50, \dodoi{10.1051/0004-6361/201628133}

\bibitem[{Marchant {et~al.}(2019)Marchant, Renzo, Farmer, Pappas, Taam,
  de~Mink, \& Kalogera}]{Marchant2019}
Marchant, P., Renzo, M., Farmer, R., {et~al.} 2019, The Astrophysical Journal,
  882, 36, \dodoi{10.3847/1538-4357/ab3426}

\bibitem[{McKernan {et~al.}(2021)McKernan, Ford, Callister, Farr,
  O'Shaughnessy, Smith, Thrane, \& Vajpeyi}]{McKernan2021}
McKernan, B., Ford, K. E.~S., Callister, T., {et~al.} 2021, arXiv e-prints.
\newblock \doarXiv{2107.07551}

\bibitem[{McKinney(2010)}]{pandas}
McKinney, W. 2010, in Proceedings of the 9th Python in Science Conference, ed.
  S.~van~der Walt \& J.~Millman, 51--56, \dodoi{10.25080/Majora-92bf1922-00a}

\bibitem[{Mehta {et~al.}(2022)Mehta, Buonanno, Gair, Miller, Farag, DeBoer,
  Wiescher, \& Timmes}]{Mehta2022}
Mehta, A.~K., Buonanno, A., Gair, J., {et~al.} 2022, The Astrophysical Journal,
  924, 39, \dodoi{10.3847/1538-4357/ac3130}

\bibitem[{Miller {et~al.}(2020)Miller, Callister, \& Farr}]{Miller2020}
Miller, S., Callister, T.~A., \& Farr, W.~M. 2020, The Astrophysical Journal,
  895, 128, \dodoi{10.3847/1538-4357/ab80c0}

\bibitem[{Neijssel {et~al.}(2021)Neijssel, Vinciguerra, Vigna-G{\'{o}}mez,
  Hirai, Miller-Jones, Bahramian, Maccarone, \& Mandel}]{Neijssel2021}
Neijssel, C.~J., Vinciguerra, S., Vigna-G{\'{o}}mez, A., {et~al.} 2021, The
  Astrophysical Journal, 908, 118, \dodoi{10.3847/1538-4357/abde4a}

\bibitem[{Neijssel {et~al.}(2019)Neijssel, Vigna-G{\'{o}}mez, Stevenson,
  Barrett, Gaebel, Broekgaarden, de~Mink, Sz{\'{e}}csi, Vinciguerra, \&
  Mandel}]{Neijssel2019}
Neijssel, C.~J., Vigna-G{\'{o}}mez, A., Stevenson, S., {et~al.} 2019, Monthly
  Notices of the Royal Astronomical Society, 490, 3740,
  \dodoi{10.1093/mnras/stz2840}

\bibitem[{Nelson {et~al.}(2015)Nelson, Pillepich, Genel, Vogelsberger,
  Springel, Torrey, Rodriguez-Gomez, Sijacki, Snyder, Griffen, Marinacci,
  Blecha, Sales, Xu, \& Hernquist}]{Nelson2015}
Nelson, D., Pillepich, A., Genel, S., {et~al.} 2015, Astronomy and Computing,
  13, 12, \dodoi{10.1016/j.ascom.2015.09.003}

\bibitem[{Ng {et~al.}(2018)Ng, Vitale, Zimmerman, Chatziioannou, Gerosa, \&
  Haster}]{Ng2018}
Ng, K.~K., Vitale, S., Zimmerman, A., {et~al.} 2018, Physical Review D, 98,
  83007, \dodoi{10.1103/PhysRevD.98.083007}

\bibitem[{Nitz {et~al.}(2021)Nitz, Capano, Kumar, Wang, Kastha, Sch{\"{a}}fer,
  Dhurkunde, \& Cabero}]{Nitz2021a}
Nitz, A.~H., Capano, C.~D., Kumar, S., {et~al.} 2021, arXiv e-prints,
  \dodoi{10.3847/1538-4357/ac1c03}

\bibitem[{Nitz {et~al.}(2020)Nitz, Sch{\"{a}}fer, \& Canton}]{Nitz2020b}
Nitz, A.~H., Sch{\"{a}}fer, M., \& Canton, T.~D. 2020, The Astrophysical
  Journal Letters, 902, L29, \dodoi{10.3847/2041-8213/abbc10}

\bibitem[{Olejak \& Belczynski(2021)}]{Olejak2021a}
Olejak, A., \& Belczynski, K. 2021, The Astrophysical Journal Letters, 921, L2,
  \dodoi{10.3847/2041-8213/ac2f48}

\bibitem[{Olejak {et~al.}(2020)Olejak, Fishbach, Belczynski, Holz, Lasota,
  Miller, \& Bulik}]{Olejak2020}
Olejak, A., Fishbach, M., Belczynski, K., {et~al.} 2020, The Astrophysical
  Journal Letters, 901, L39, \dodoi{10.3847/2041-8213/abb5b5}

\bibitem[{Oliphant(2006)}]{numpy}
Oliphant, T.~E. 2006, {A guide to NumPy} (USA: Trelgol Publishing),
  \dodoi{10.5555/2886196}

\bibitem[{Olsen {et~al.}(2022)Olsen, Venumadhav, Mushkin, Roulet, Zackay, \&
  Zaldarriaga}]{Olsen2022}
Olsen, S., Venumadhav, T., Mushkin, J., {et~al.} 2022, arXiv e-prints.
\newblock \doarXiv{2201.02252}

\bibitem[{Patton {et~al.}(2022)Patton, Sukhbold, \& Eldridge}]{Patton2022}
Patton, R.~A., Sukhbold, T., \& Eldridge, J.~J. 2022, Monthly Notices of the
  Royal Astronomical Society, \dodoi{10.1093/mnras/stab3797}

\bibitem[{Paxton {et~al.}(2011)Paxton, Bildsten, Dotter, Herwig, Lesaffre, \&
  Timmes}]{Paxton2011}
Paxton, B., Bildsten, L., Dotter, A., {et~al.} 2011, The Astrophysical Journal
  Supplement Series, 192, 3, \dodoi{10.1088/0067-0049/192/1/3}

\bibitem[{Paxton {et~al.}(2013)Paxton, Cantiello, Arras, Bildsten, Brown,
  Dotter, Mankovich, Montgomery, Stello, Timmes, \& Townsend}]{Paxton2013}
Paxton, B., Cantiello, M., Arras, P., {et~al.} 2013, The Astrophysical Journal
  Supplement Series, 208, 4, \dodoi{10.1088/0067-0049/208/1/4}

\bibitem[{Paxton {et~al.}(2015)Paxton, Marchant, Schwab, Bauer, Bildsten,
  Cantiello, Dessart, Farmer, Hu, Langer, Townsend, Townsley, \&
  Timmes}]{Paxton2015}
Paxton, B., Marchant, P., Schwab, J., {et~al.} 2015, The Astrophysical Journal
  Supplement Series, 220, 15, \dodoi{10.1088/0067-0049/220/1/15}

\bibitem[{Paxton {et~al.}(2018)Paxton, Schwab, Bauer, Bildsten, Blinnikov,
  Duffell, Farmer, Goldberg, Marchant, Sorokina, Thoul, Townsend, \&
  Timmes}]{Paxton2018}
Paxton, B., Schwab, J., Bauer, E.~B., {et~al.} 2018, The Astrophysical Journal
  Supplement Series, 234, 34, \dodoi{10.3847/1538-4365/aaa5a8}

\bibitem[{Paxton {et~al.}(2019)Paxton, Smolec, Schwab, Gautschy, Bildsten,
  Cantiello, Dotter, Farmer, Goldberg, Jermyn, Kanbur, Marchant, Thoul,
  Townsend, Wolf, Zhang, \& Timmes}]{Paxton2019}
Paxton, B., Smolec, R., Schwab, J., {et~al.} 2019, The Astrophysical Journal
  Supplement Series, 243, 10, \dodoi{10.3847/1538-4365/ab2241}

\bibitem[{P{\'{e}}rez \& Granger(2007)}]{ipython}
P{\'{e}}rez, F., \& Granger, B.~E. 2007, IEEE Journals {\&} Magazines, 9, 21,
  \dodoi{10.1109/MCSE.2007.53}

\bibitem[{Price-Whelan {et~al.}(2018)Price-Whelan, Sipocz, G{\"{u}}nther, Lim,
  Crawford, Conseil, Shupe, Craig, Dencheva, Ginsburg, VanderPlas, Bradley,
  P{\'{e}}rez-Su{\'{a}}rez, de~Val-Borro, Aldcroft, Cruz, Robitaille, Tollerud,
  Ardelean, Babej, Bachetti, Bakanov, Bamford, Barentsen, Barmby, Baumbach,
  Berry, Biscani, Boquien, Bostroem, Bouma, Brammer, Bray, Breytenbach,
  Buddelmeijer, Burke, Calderone, Rodr{\'{i}}guez, Cara, Cardoso, Cheedella,
  Copin, Crichton, D{\'{A}}vella, Deil, Depagne, Dietrich, Donath, Droettboom,
  Earl, Erben, Fabbro, Ferreira, Finethy, Fox, Garrison, Gibbons, Goldstein,
  Gommers, Greco, Greenfield, Groener, Grollier, Hagen, Hirst, Homeier, Horton,
  Hosseinzadeh, Hu, Hunkeler, Ivezi{\'{c}}, Jain, Jenness, Kanarek, Kendrew,
  Kern, Kerzendorf, Khvalko, King, Kirkby, Kulkarni, Kumar, Lee, Lenz,
  Littlefair, Ma, Macleod, Mastropietro, McCully, Montagnac, Morris, Mueller,
  Mumford, Muna, Murphy, Nelson, Nguyen, Ninan, N{\"{o}}the, Ogaz, Oh, Parejko,
  Parley, Pascual, Patil, Patil, Plunkett, Prochaska, Rastogi, Janga, Sabater,
  Sakurikar, Seifert, Sherbert, Sherwood-Taylor, Shih, Sick, Silbiger,
  Singanamalla, Singer, Sladen, Sooley, Sornarajah, Streicher, Teuben, Thomas,
  Tremblay, Turner, Terr{\'{o}}n, van Kerkwijk, de~la Vega, Watkins, Weaver,
  Whitmore, Woillez, \& Zabalza}]{TheAstropyCollaboration2018}
Price-Whelan, A.~M., Sipocz, B.~M., G{\"{u}}nther, H.~M., {et~al.} 2018, The
  Astronomical Journal, 156, 123, \dodoi{10.3847/1538-3881/aabc4f}

\bibitem[{Qin {et~al.}(2018)Qin, Fragos, Meynet, Andrews, S{\o}rensen, \&
  Song}]{Qin2018}
Qin, Y., Fragos, T., Meynet, G., {et~al.} 2018, Astronomy {\&} Astrophysics,
  616, A28, \dodoi{10.1051/0004-6361/201832839}

\bibitem[{Qin {et~al.}(2019)Qin, Marchant, Fragos, Meynet, \&
  Kalogera}]{Qin2019}
Qin, Y., Marchant, P., Fragos, T., Meynet, G., \& Kalogera, V. 2019, The
  Astrophysical Journal, 870, L18, \dodoi{10.3847/2041-8213/aaf97b}

\bibitem[{Qin {et~al.}(2022{\natexlab{a}})Qin, Shu, Yi, \& Wang}]{Qin2022}
Qin, Y., Shu, X., Yi, S., \& Wang, Y.~Z. 2022{\natexlab{a}}, Research in
  Astronomy and Astrophysics, 22, 035023, \dodoi{10.1088/1674-4527/ac4ca4}

\bibitem[{Qin {et~al.}(2022{\natexlab{b}})Qin, Wang, Wu, Meynet, \&
  Song}]{Qin2022a}
Qin, Y., Wang, Y.-Z., Wu, D.-H., Meynet, G., \& Song, H. 2022{\natexlab{b}},
  The Astrophysical Journal, 924, 129, \dodoi{10.3847/1538-4357/ac3982}

\bibitem[{Robitaille {et~al.}(2013)Robitaille, Tollerud, Greenfield,
  Droettboom, Bray, Aldcroft, Davis, Ginsburg, Price-Whelan, Kerzendorf,
  Conley, Crighton, Barbary, Muna, Ferguson, Grollier, Parikh, Nair,
  G{\"{u}}nther, Deil, Woillez, Conseil, Kramer, Turner, Singer, Fox, Weaver,
  Zabalza, Edwards, Bostroem, Burke, Casey, Crawford, Dencheva, Ely, Jenness,
  Labrie, Lim, Pierfederici, Pontzen, Ptak, Refsdal, Servillat, \&
  Streicher}]{TheAstropyCollaboration2013}
Robitaille, T.~P., Tollerud, E.~J., Greenfield, P., {et~al.} 2013, Astronomy
  {\&} Astrophysics, 558, A33, \dodoi{10.1051/0004-6361/201322068}

\bibitem[{Rodriguez {et~al.}(2016)Rodriguez, Zevin, Pankow, Kalogera, \&
  Rasio}]{Rodriguez2016}
Rodriguez, C.~L., Zevin, M., Pankow, C., Kalogera, V., \& Rasio, F.~A. 2016,
  The Astrophysical Journal Letters, 832, 1, \dodoi{10.3847/2041-8205/832/1/L2}

\bibitem[{Rodriguez {et~al.}(2020)Rodriguez, Kremer, Grudi{\'{c}}, Hafen,
  Chatterjee, Fragione, Lamberts, Martinez, Rasio, Weatherford, \&
  Ye}]{Rodriguez2020}
Rodriguez, C.~L., Kremer, K., Grudi{\'{c}}, M.~Y., {et~al.} 2020, The
  Astrophysical Journal, 896, L10, \dodoi{10.3847/2041-8213/ab961d}

\bibitem[{Safarzadeh \& Hotokezaka(2020)}]{Safarzadeh2020c}
Safarzadeh, M., \& Hotokezaka, K. 2020, The Astrophysical Journal, 897, L7,
  \dodoi{10.3847/2041-8213/ab9b79}

\bibitem[{Sana {et~al.}(2012)Sana, {De Mink}, {De Koter}, Langer, Evans,
  Gieles, Gosset, Izzard, {Le Bouquin}, \& Schneider}]{Sana2012}
Sana, H., {De Mink}, S.~E., {De Koter}, A., {et~al.} 2012, Science, 337, 444,
  \dodoi{10.1126/science.1223344}

\bibitem[{Santoliquido {et~al.}(2021)Santoliquido, Mapelli, Giacobbo,
  Bouffanais, \& Artale}]{Santoliquido2021}
Santoliquido, F., Mapelli, M., Giacobbo, N., Bouffanais, Y., \& Artale, M.~C.
  2021, Monthly Notices of the Royal Astronomical Society, 502, 4877,
  \dodoi{10.1093/mnras/stab280}

\bibitem[{Schr{\o}der {et~al.}(2018)Schr{\o}der, Batta, \&
  Ramirez-Ruiz}]{Schroder2018}
Schr{\o}der, S.~L., Batta, A., \& Ramirez-Ruiz, E. 2018, The Astrophysical
  Journal Letters, 862, L3, \dodoi{10.3847/2041-8213/aacf8d}

\bibitem[{Spruit(2002)}]{Spruit2002}
Spruit, H.~C. 2002, Astronomy {\&} Astrophysics, 381, 923,
  \dodoi{10.1051/0004-6361}

\bibitem[{Steinle \& Kesden(2021)}]{Steinle2021}
Steinle, N., \& Kesden, M. 2021, Physical Review D, 103, 63032,
  \dodoi{10.1103/PhysRevD.103.063032}

\bibitem[{Stevenson {et~al.}(2019)Stevenson, Sampson, Powell,
  Vigna-G{\'{o}}mez, Neijssel, Sz{\'{e}}csi, \& Mandel}]{Stevenson2019}
Stevenson, S., Sampson, M., Powell, J., {et~al.} 2019, The Astrophysical
  Journal, 882, 121, \dodoi{10.3847/1538-4357/ab3981}

\bibitem[{Tagawa {et~al.}(2021)Tagawa, Haiman, Bartos, Kocsis, \&
  Omukai}]{Tagawa2021}
Tagawa, H., Haiman, Z., Bartos, I., Kocsis, B., \& Omukai, K. 2021, Monthly
  Notices of the Royal Astronomical Society, 507, 3362,
  \dodoi{10.1093/mnras/stab2315}

\bibitem[{Tanikawa {et~al.}(2022)Tanikawa, Yoshida, Kinugawa, Trani, Hosokawa,
  Susa, \& Omukai}]{Tanikawa2022b}
Tanikawa, A., Yoshida, T., Kinugawa, T., {et~al.} 2022, The Astrophysical
  Journal, 926, 83, \dodoi{10.3847/1538-4357/ac4247}

\bibitem[{Thorne(1974)}]{Thorne1974}
Thorne, K.~S. 1974, The Astrophysical Journal, 191, 507, \dodoi{10.1086/152991}

\bibitem[{{Van Der Walt} {et~al.}(2011){Van Der Walt}, Colbert, \&
  Varoquaux}]{numpy2}
{Van Der Walt}, S., Colbert, S.~C., \& Varoquaux, G. 2011, Computing in Science
  and Engineering, 13, 22, \dodoi{10.1109/MCSE.2011.37}

\bibitem[{van Son {et~al.}(2020)van Son, {De Mink}, Broekgaarden, Renzo,
  Justham, Laplace, Mor{\'{a}}n-Fraile, Hendriks, \& Farmer}]{VanSon2020}
van Son, L. A.~C., {De Mink}, S.~E., Broekgaarden, F.~S., {et~al.} 2020, The
  Astrophysical Journal, 897, 100, \dodoi{10.3847/1538-4357/ab9809}

\bibitem[{Vink \& de~Koter(2005)}]{Vink2005}
Vink, J.~S., \& de~Koter, A. 2005, Astronomy {\&} Astrophysics, 442, 587,
  \dodoi{10.1051/0004-6361:20052862}

\bibitem[{Vink {et~al.}(2001)Vink, de~Koter, \& Lamers}]{Vink2001}
Vink, J.~S., de~Koter, A., \& Lamers, H. J. G. L.~M. 2001, Astronomy {\&}
  Astrophysics, 369, 574, \dodoi{10.1051/0004-6361}

\bibitem[{Virtanen {et~al.}(2020)Virtanen, Gommers, Oliphant, Haberland, Reddy,
  Cournapeau, Burovski, Peterson, Weckesser, Bright, van~der Walt, Brett,
  Wilson, Millman, Mayorov, Nelson, Jones, Kern, Larson, Carey, Polat, Feng,
  Moore, VanderPlas, Laxalde, Perktold, Cimrman, Henriksen, Quintero, Harris,
  Archibald, Ribeiro, Pedregosa, van Mulbregt, Vijaykumar, Bardelli, Rothberg,
  Hilboll, Kloeckner, Scopatz, Lee, Rokem, Woods, Fulton, Masson,
  H{\"{a}}ggstr{\"{o}}m, Fitzgerald, Nicholson, Hagen, Pasechnik, Olivetti,
  Martin, Wieser, Silva, Lenders, Wilhelm, Young, Price, Ingold, Allen, Lee,
  Audren, Probst, Dietrich, Silterra, Webber, Slavi{\v{c}}, Nothman, Buchner,
  Kulick, Sch{\"{o}}nberger, {de Miranda Cardoso}, Reimer, Harrington,
  Rodr{\'{i}}guez, Nunez-Iglesias, Kuczynski, Tritz, Thoma, Newville,
  K{\"{u}}mmerer, Bolingbroke, Tartre, Pak, Smith, Nowaczyk, Shebanov, Pavlyk,
  Brodtkorb, Lee, McGibbon, Feldbauer, Lewis, Tygier, Sievert, Vigna, Peterson,
  More, Pudlik, Oshima, Pingel, Robitaille, Spura, Jones, Cera, Leslie, Zito,
  Krauss, Upadhyay, Halchenko, \& V{\'{a}}zquez-Baeza}]{scipy}
Virtanen, P., Gommers, R., Oliphant, T.~E., {et~al.} 2020, Nature Methods, 17,
  261, \dodoi{10.1038/s41592-019-0686-2}

\bibitem[{Vitale {et~al.}(2020)Vitale, Gerosa, Farr, \& Taylor}]{Vitale2020a}
Vitale, S., Gerosa, D., Farr, W.~M., \& Taylor, S.~R. 2020, arXiv e-prints.
\newblock \doarXiv{2007.05579}

\bibitem[{Webbink(1984)}]{Webbink1984}
Webbink, R. 1984, The Astrophysical Journal, 277, 355, \dodoi{10.1086/161701}

\bibitem[{Wong {et~al.}(2021)Wong, Breivik, Kremer, \& Callister}]{Wong2021a}
Wong, K.~W., Breivik, K., Kremer, K., \& Callister, T. 2021, Physical Review D,
  103, 83021, \dodoi{10.1103/PhysRevD.103.083021}

\bibitem[{Zaldarriaga {et~al.}(2018)Zaldarriaga, Kushnir, \&
  Kollmeier}]{Zaldarriaga2018}
Zaldarriaga, M., Kushnir, D., \& Kollmeier, J.~A. 2018, Monthly Notices of the
  Royal Astronomical Society, 473, 4174, \dodoi{10.1093/mnras/stx2577}

\bibitem[{Zevin(2021)}]{Zenodo_selection_effects}
Zevin, M. 2021, {Semianalytic VT},  Zenodo, \dodoi{10.5281/zenodo.5086359}

\bibitem[{Zevin {et~al.}(2020{\natexlab{a}})Zevin, Berry, Coughlin,
  Chatziioannou, \& Vitale}]{Zevin2020a}
Zevin, M., Berry, C. P.~L., Coughlin, S., Chatziioannou, K., \& Vitale, S.
  2020{\natexlab{a}}, The Astrophysical Journal, 899, L17,
  \dodoi{10.3847/2041-8213/aba8ef}

\bibitem[{Zevin {et~al.}(2020{\natexlab{b}})Zevin, Spera, Berry, \&
  Kalogera}]{Zevin2020b}
Zevin, M., Spera, M., Berry, C. P.~L., \& Kalogera, V. 2020{\natexlab{b}}, The
  Astrophysical Journal, 899, L1, \dodoi{10.3847/2041-8213/aba74e}

\bibitem[{Zevin {et~al.}(2021)Zevin, Bavera, Berry, Kalogera, Fragos, Marchant,
  Rodriguez, Antonini, Holz, \& Pankow}]{Zevin2021}
Zevin, M., Bavera, S.~S., Berry, C.~P., {et~al.} 2021, The Astrophysical
  Journal, 910, 152, \dodoi{10.3847/1538-4357/abe40e}

\end{thebibliography}
